\documentclass[pre,preprintnumbers,amsmath,amssymb,superscriptaddress,reprint]{revtex4} 

\usepackage[dvipdfmx]{graphicx}
\usepackage{dcolumn}
\usepackage{bm,amsmath}
\usepackage{here}

\renewcommand{\vec}[1]{{\bm #1}}

\begin{document}
\title{Turbulence, cascade and singularity in a generalization of the Constantin-Lax-Majda equation}
\author{Takeshi \surname{Matsumoto}}
\email{takeshi@kyoryu.scphys.kyoto-u.ac.jp}
\affiliation{%
Division of Physics and Astronomy,
Graduate School of Science,
Kyoto University,
Kyoto, 606-8502, Japan
}
\author{Takashi \surname{Sakajo}}
\email{sakajo@math.kyoto-u.ac.jp}
\affiliation{%
Department of Mathematics,
Graduate School of Science,
Kyoto University,
Kyoto, 606-8502, Japan}
\date{\today}
\begin{abstract}
We study numerically a Constantin-Lax-Majda-De Gregorio model generalized by 
Okamoto, Sakajo and Wunsch, which is a model of fluid turbulence in one dimension with
an inviscid conservation law.
In the presence of the viscosity and two types of the large-scale forcings, we show that 
turbulent cascade of the inviscid invariant, which is not limited to quadratic quantity,
occurs and that properties of this model's
turbulent state are related to singularity of the inviscid case by adopting standard
tools of analyzing fluid turbulence.
\end{abstract} 

\maketitle

\section{Introduction}
Fostering a number of simpler nonlinear partial differential equations (PDE) 
is
a defining feature of the Navier-Stokes (NS) or Euler equations. 
This is perhaps because 
a reduced equation is more insightful and direct in understanding
one particular phenomenon than the whole NS equations which include
countless facets of fluid phenomena.
The Korteweg-de Vries equation derived via the water-wave equation from the Euler equations 
is a prominent example for understanding a peculiar behavior of the shallow water wave, 
which is now called solitary wave.

Reaching a good reduced model is not at all 
limited to systematic derivations from the NS or Euler equations. 
Phenomenological modeling of them
with one-dimensional (1D) PDE or a set of ordinary differential equations
has been proved to be fruitful. Famous examples include the Burgers' equation \cite{burg},
the Constantin-Lax-Majda (CLM) equation \cite{clm} and
the shell models of turbulent cascade \cite{Bi}.

The major interest behind these models is in 
statistical laws of incompressible high-Reynolds number 
turbulence, 
a putative singular solution of the incompressible NS or Euler equations
and 
a possible relation between them (see e.g., \cite{f, mb}).
By statistical laws, we mean those of homogeneous isotropic turbulence
such as the scaling laws of the energy spectrum and of the structure functions, 
with the turbulent cascade of the energy or other inviscid conserved quantity.
Since these problems are known to be one of the toughest in physics
and mathematics, approach from a simple model is indispensable.
The influential CLM eq. yields the analytic solution of the vorticity analogue becoming 
infinite in a finite time \cite{clm}. 
However it does not have a turbulent solution with viscosity (see e.g., \cite{saka}).
There may not be commonly accepted reduced PDE models suitable for studying
those points.
Nevertheless, we here mention two recent studies to develop such models. 

Zikanov, Thess and Grauer introduced a nonlocal generalization 
of the 1D Burgers' equation. 
They showed that the solution has 
the energy spectrum $E(k) \propto k^{-5/3}$, which is consistent with the Kolmogorov scaling,
and 
that the scaling exponent of 
the $p$-th order velocity structure function,
$\langle [u(x + r, t) - u(x, t)]^p\rangle \propto r^{\zeta_p}$ \cite{ztg}
is without intermittency, namely $\zeta_p = p / 3$
(here $\langle \cdot \rangle$ denotes an ensemble average).
In their model the degree of the nonlocality can be changed by one parameter.
The above result is obtained for the maximally nonlocal case. 
For an intermediately nonlocal case, they found that 
the scaling exponent $\zeta_p$ deviates from $p/3$ in the quantitatively 
same way as the three dimensional (3D) incompressible turbulence \cite{ztg}.

Recently, Luo and Hou numerically found a potentially singular 
solution to the 3D axisymmetric Euler flow confined in a cylindrical surface,
where the vorticity grew by $10^{8}$ times larger \cite{lh3d, lh3d2}.
To understand the nature of this, 1D PDE models have been developed 
by Luo and Hou \cite{lh3d} and by Choi, Keselev and Yao \cite{cky}.
It is proven that a solution to each model
starting from a smooth initial condition does blow up in a finite time 
\cite{chklsy}.

In the same spirit of the two models with an emphasis on the statistical 
laws and the possible role of the singularity, 
we here study a generalization, proposed by Okamoto, Sakajo and Wunch \cite{oswn},
of the Constantin-Lax-Majda-De Gregorio (gCLMG) equation \cite{dg1} 
with the viscosity and a forcing term $f(x, t)$ 
\begin{eqnarray}
 \partial_t \omega + a u \partial_x u = \omega \partial_x u + \nu \partial_{xx} \omega + f.
\label{gclmg}
\end{eqnarray}
Here $\omega(x, t)$ is a scalar modeling of the vorticity in three dimensions
and the velocity analogue is expressed with $u(x, t) = -(-\partial_{xx})^{-1}\omega$
and $\partial_x u = H(\omega)$ which is the Hilbert transform of $\omega$.
The Hilbert transform first considered in these 1D modelings \cite{clm} 
is one of the key ingredients, which was used also in the models
we mentioned \cite{ztg, lh3d}.
Notice that the velocity is no longer incompressible in 1D. 
A historical background of the gCLMG eq. (\ref{gclmg}) can be found 
in \cite{ms, chklsy}.
The parameter $a$ in front of the advection term introduced in \cite{oswn}
enables the equation to have a conserved quantity in the inviscid
($\nu = 0$) and unforced ($f = 0$) setting. 
Specifically, for $a \le -1$, it is easily shown that 
\begin{eqnarray}
C_{a} =  \frac{1}{-a}\int \omega^{-a}(x, t) dx
\label{cq}
\end{eqnarray}
is a conserved quantity if there is no input or output on the boundary \cite{oswn}
(If $a$ is not integer, we take $|\omega|^{-a}$ in the integrand. Furthermore if $a$ is odd,
$\int |\omega|^{-a} dx/(-a)$ is also a conserved quantity).
We notice here that negative $a$ has no physical origin and that the Galilean
invariance is lost for $a < 0$.
However the inviscid conservation law leaves possibility
of turbulent cascade. Indeed for the case of $a = -2$, where Eq.(\ref{cq})
coincides dimensionally with the enstrophy, it has been numerically shown that the enstrophy 
cascade takes place \cite{ms}. 
It may appear paradoxical that, for $a = -2$, we have two-dimensional (2D) turbulence analogue 
from the model (\ref{gclmg}) with the vortex stretching term that is the essential ingredient of
the 3D vorticity equation. 
This suggests that, regardless of the form of the equation, the conservation law matters most.
In this paper, from a mathematical and theoretical view point, we extend 
our previous study of the gCLMG eq. \cite{ms} (which was limited to $a = -2$)
to general negative $a$'s. The $a = -2$ case in \cite{ms} is the baseline of our analysis.

We now summarize findings of the previous study \cite{ms} and state
the plan of the present paper.
In \cite{ms}, the gCLMG eq. was numerically
studied in a periodic interval. It is observed that 
a turbulent state occurs as a statistically steady state 
if the large-scale forcing $f(x, t)$ is random
and that, if the forcing is deterministic, a solution becomes stationary.
The turbulent state exhibited the cascade of $C_{-2}$ (enstrophy cascade) 
and the energy spectrum close to that of the 2D enstrophy-cascade turbulence 
$k^{-3}$ but with a measurable deviation from it in the inertial range. 
Interestingly, the stationary solution had the energy spectrum which is
indistinguishable from the turbulent spectrum in the inertial range.
The vorticity structure functions of the turbulent
state at high even orders were possibly characterized with 
negative scaling exponents, indicating infinite vorticity
as $\nu \to 0$. Also the nonlinear stationary solution as we decreased the viscosity $\nu$
suggested infinite vorticity with the finite enstrophy dissipation rate.
Lastly the phase-space orbit of the turbulence state normalized by the stationary
solutions showed a peculiar self-similarity.

In this paper, we show numerically for general negative $a$'s that the same
above holds. Furthermore we analyze in detail
the cascade of the inviscid invariant (\ref{cq}) in the turbulent state, 
the profile of the stationary solution with $\nu \to 0$ 
and compare the scaling of the viscous case with the inviscid case.

The organization of the paper is as follows. 
In Sec.\ref{s:random}, we study the turbulent state under the random forcing.
Specifically, we characterize it with the energy spectrum and analyze
the cascade with the filtering flux method. 
We also consider  the K\'arm\'an-Howarth Monin relation and the vorticity structure 
functions.
In Sec.\ref{s:stat}, we study the nonlinear stationary state under the deterministic 
forcing. 
Specifically, we consider the energy spectrum and the vorticity profile and
then compare the the energy spectrum to that of the inviscid solution. 
In Sec.\ref{s:selfsimilar}, we show the self-similarity of the phase-space orbit of the
turbulent solution normalized by the stationary solution. 
A summary and concluding discussion are given in Sec.\ref{s:sc}.

\section{\label{s:random}Turbulence under the random forcing}

Throughout the paper we consider the gCLMG eq. (\ref{gclmg}) in a periodic interval
of length $2\pi$.
We hence use the Fourier spectral method for numerical simulation. 
We set the vorticity Fourier mode of the zero wavenumber to zero initially.
The dealiasing is done with the two-third method. The time stepping scheme is 
the forth-order Runge-Kutta method. 
It is known that the round-off noise grows in the spectral simulation
of the gCLMG eq. with the double precision. 
To suppress this, we use the same spectral filter as \cite{oswn}. 
Namely, if the absolute value of the vorticity Fourier modes is smaller than 
$10^{-12}$, we set it to zero at each time step.

First, we set the large-scale forcing to be random. 
Specifically, we set the Fourier mode of the forcing to Gaussian random variable
without temporal correlation having the following mean and variance
\begin{eqnarray}
 \langle \widehat{f}(k, t) \rangle &=& 0,\\
 \langle \widehat{f}(k', t')  \widehat{f}(k, t)\rangle &=& 2\sigma_f^2 \delta_{k', k} \delta(t' - t).
\end{eqnarray}
To make the forcing effective in a large scale, we set non-zero $\widehat{f}$ only for 
the wavenumbers $k = \pm 1$. We set $\sigma_f = 1.0\times10^{-2}$ leading to
the average enstrophy-input rate $2\sigma_f^2 = 2.0\times 10^{-4}$.
The initial condition of the simulation is $\omega = 0$.

Next, we present the vorticity profile and the energy spectrum 
for a wide range of $a$'s as an overview of gCLMG turbulence. 
After that, by limiting to a smaller range of $a$'s, we study its property in more 
detail.

\subsection{Appearance of the vorticity and the energy spectrum}

\begin{figure} 
\includegraphics[scale=0.6]{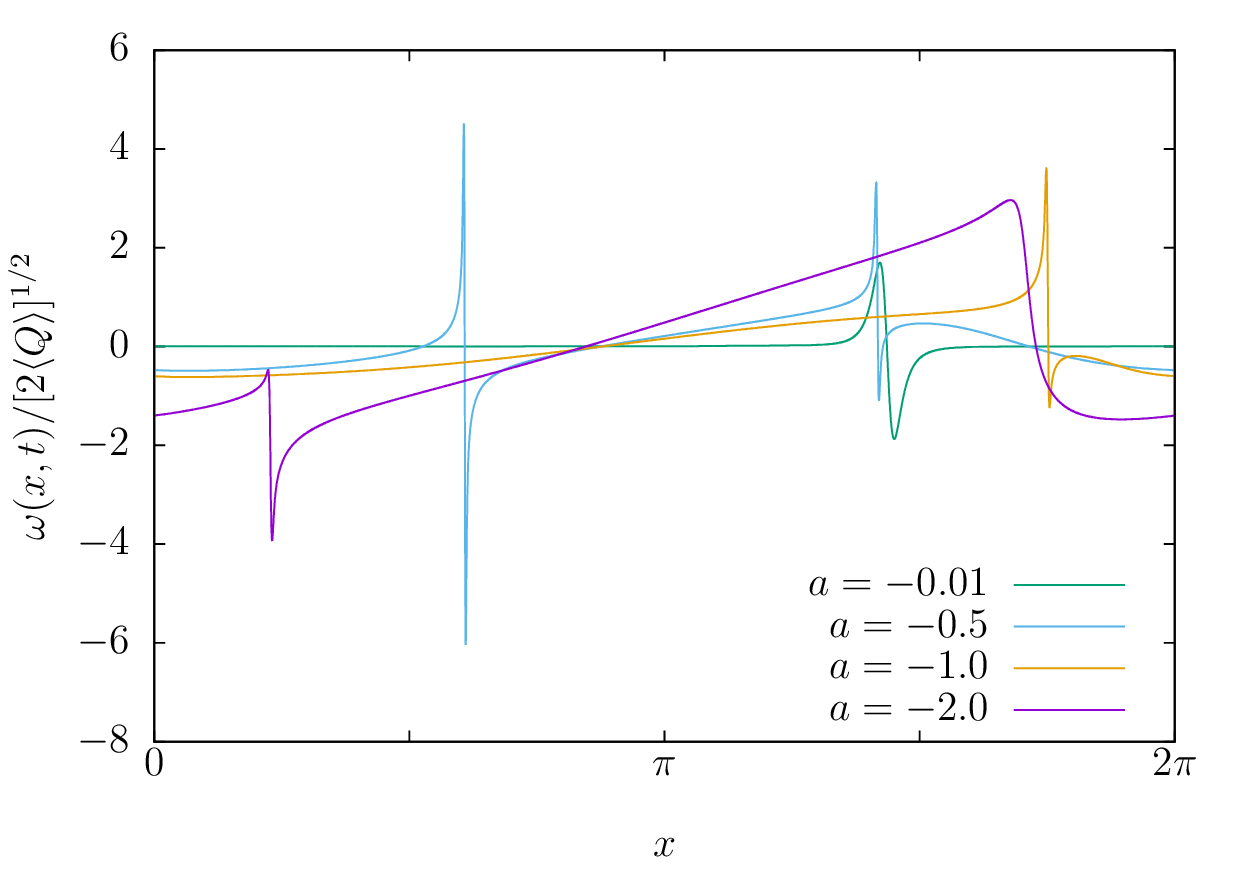}
\includegraphics[scale=0.6]{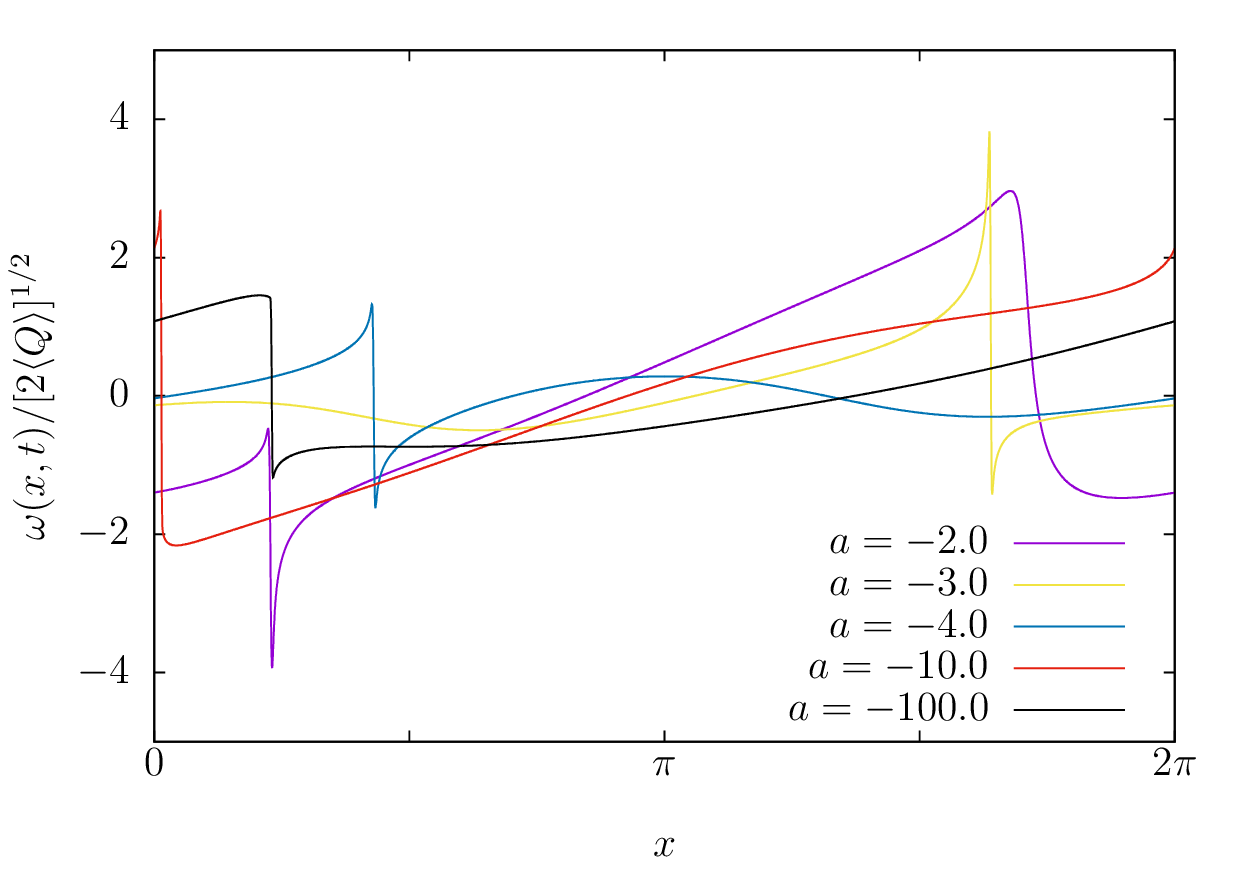}
\caption{\label{prof} The vorticity snapshots for various $a$'s normalized
with the enstrophy-based vorticity, $(2\langle Q \rangle)^{1/2}$.
The spatial resolution is $2^{13}$ grid points.
The kinematic viscosity is $\nu = 2.5\times 10^{-5}$ with exceptions
for $a = -0.01$ ($\nu = 3.2\times 10^{-3}$) and $a = -100$ ($\nu = 1.0\times10^{-4}$).
The time step for the integration is $\Delta t = 2.5\times10^{-4}$
with exceptions for $a = 100$ ($\Delta t = 1.25\times10^{-4}$).}
\end{figure}

In Fig.\ref{prof}, we plot vorticity snapshots in statistically steady states
for $a = -0.01, -0.5, -1.0, -2.0, -3.0, -4.0, -10.0$ and $-100.0$.
They are normalized by the temporally averaged enstrophy
\begin{eqnarray}
 \langle Q \rangle
  = \left\langle \frac{1}{2\pi} \int_0^{2\pi} \frac{1}{2}\omega^2(x, t) dx \right\rangle
  = \left\langle \sum_k \frac{1}{2}|\widehat{\omega}(k, t)|^2 \right\rangle.
\end{eqnarray}
The vorticity is characterized with one or two pulses for small $|a|$ 
and with shocks for large $|a|$, which are formed at a velocity null point with
negative velocity gradient. Otherwise the solution is very much smooth.  
Roughly the pulse is made by the stretching term 
of the gCLMG eq. as in the CLM eq. but the blowup is avoided primarily by the negative
advection term. Owing to the forcing and the viscous terms, the system reaches
a statistically steady state. The pulses or shocks move around and sometimes merge. 
The pulse-like structure for small $|a|$ resembles the analytical
blow-up solution of the CLM eq. 
For the cases shown in Fig.\ref{prof}, we observe that the enstrophy
obeys $\langle Q \rangle \propto (-a)^{-1}$ (although $\nu$ is not the same for all the nine cases). 
This is consistent to the blow-up of vorticity of the viscous CLM eq. (see e.g., \cite{saka}).

\begin{figure} 
\includegraphics[scale=0.6]{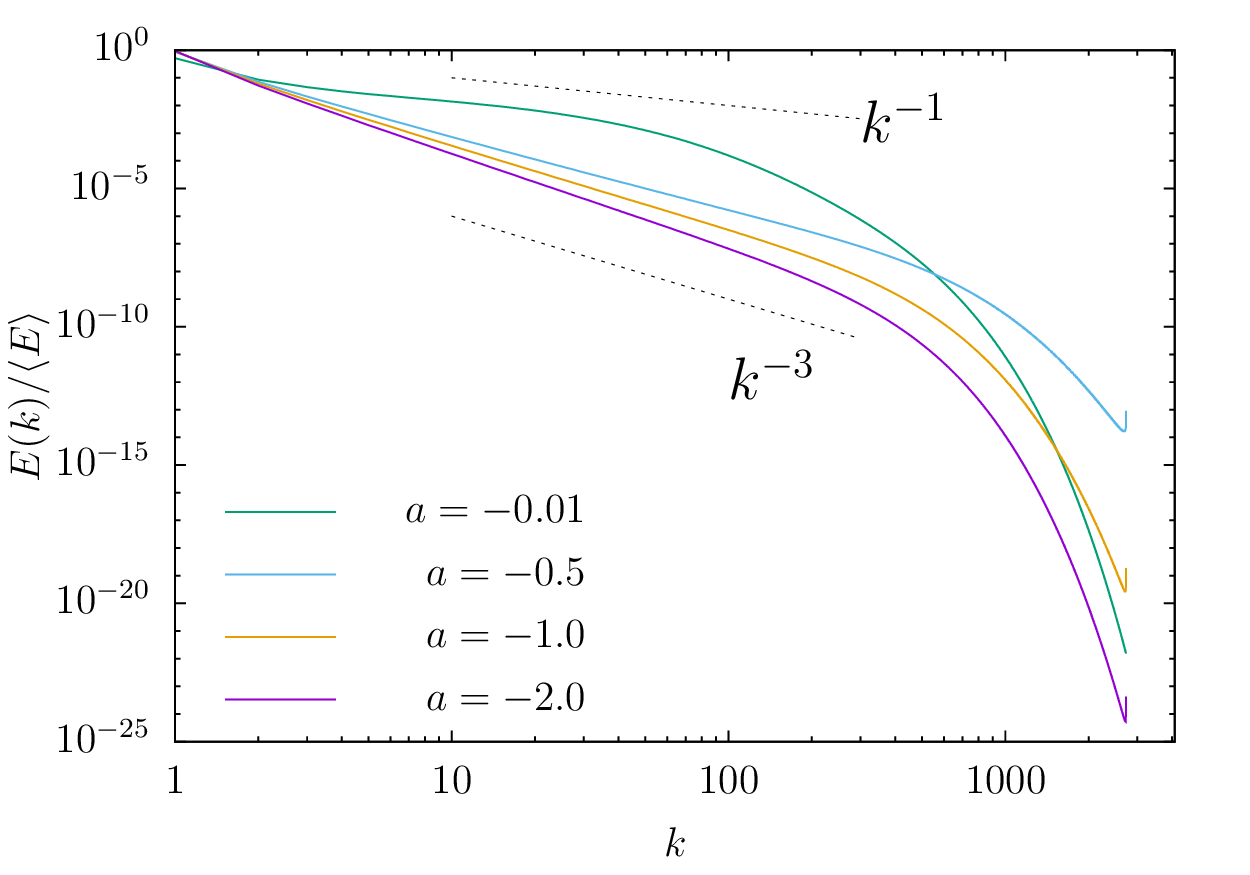}
\includegraphics[scale=0.6]{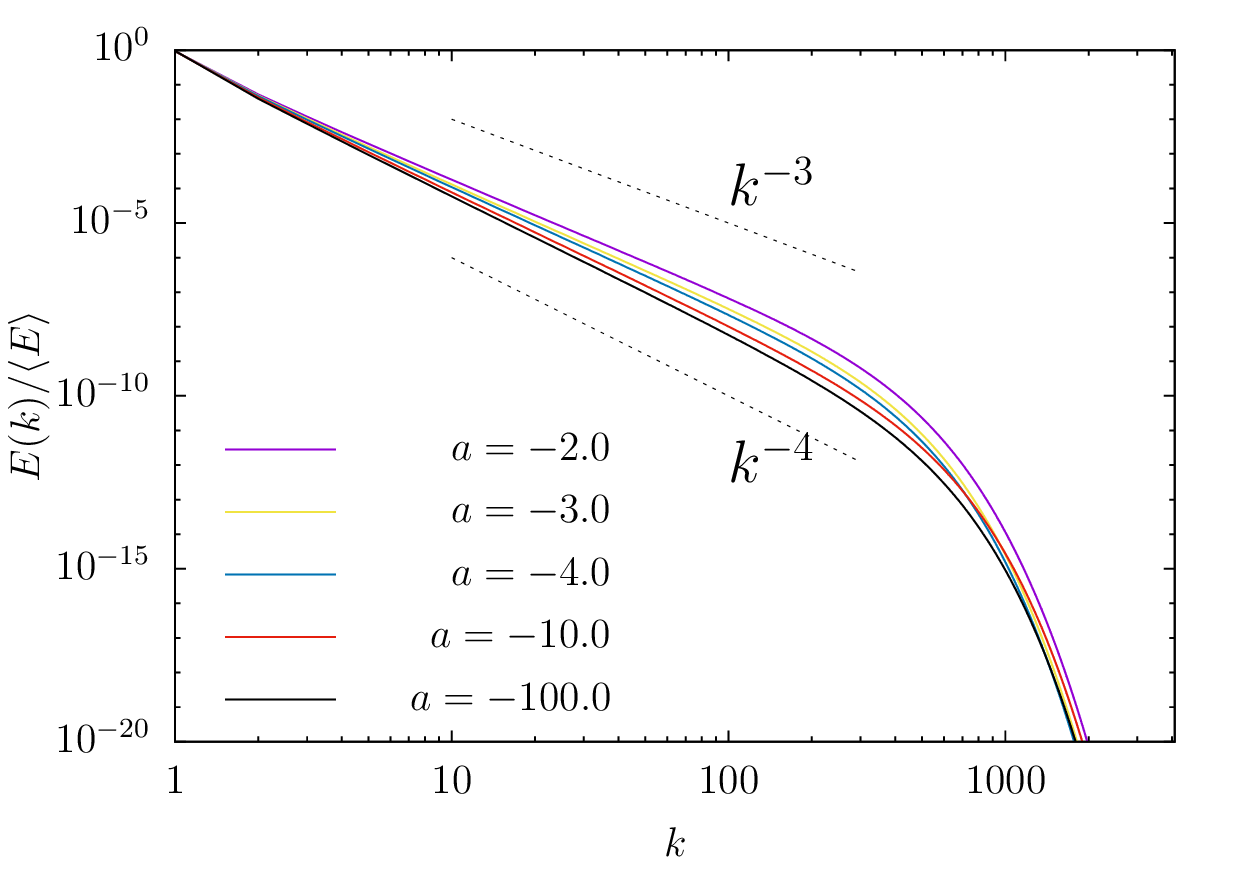}
\caption{\label{spc} The time-averaged energy spectra for various $a$'s normalized
with the energy, $\langle E \rangle$. The time average is taken over $9\times 10^3$
equi-spaced instances between $10^4 \le t \le 10^5$.
Notice that the inertial range for the $a=-0.01$ case 
is narrower since its viscosity is $100$ times larger than other cases. The power laws $k^{-1}, k^{-3}$ and $k^{-4}$ are meant for a guide.}
\end{figure}
In Fig.\ref{spc}, we plot the corresponding energy spectra 
\begin{eqnarray}
 E(k) = \left\langle \sum_{k \le |k'| < k + 1} \frac{1}{2}|\widehat{u}(k', t)|^2 \right\rangle,
\end{eqnarray}
which are normalized with the temporally averaged energy
\begin{eqnarray}
 \langle E \rangle
  = \left\langle \frac{1}{2\pi} \int_0^{2\pi} \frac{1}{2} u^2(x, t) dx \right\rangle
  = \left\langle \sum_k \frac{1}{2}|\widehat{u}(k, t)|^2 \right\rangle.
\end{eqnarray}
There are two ranges which are analogous to the inertial range and 
the dissipation rage in the NS turbulence.
If we fit $E(k)$ in the inertial range  with a power-law scaling $k^{-q}$,
the scaling exponent $q$ varies probably
from $0$ ($a \to 0$) to around $4$ ($a \to -\infty$). 
The former is again consistent with a blowup solution to the CLM eq., 
seemingly having a flat ($k^0$) energy spectrum. 
The latter limit is consistent with the shock like, or step-function like, 
vorticity profile for large $|a|$.

Now we present Kolmogorov-type dimensional analysis about the scaling exponent of the energy 
spectra.
Notice that the inviscid conservation of $C_{a}$ is not guaranteed for $-1 < a < 0$.
We first assume that the inviscid invariant (\ref{cq}) is cascading down 
to smaller scales and that its ``dissipation rate'',
\begin{eqnarray}
 \beta_a = (-a - 1) \nu \frac{1}{2\pi} \int_0^{2\pi} \omega^{-a-2}(\partial_x \omega)^2 dx,
\label{betaa}
\end{eqnarray} 
determines the inertial range quantity for $a \le -1$ ($\beta_{-2}$ is dimensionally
the same as the enstrophy dissipation rate of the NS turbulence).
Since $\beta_a$ has the dimension [(time)$^{a-1}$], the inertial-range spectrum
behaves as
\begin{eqnarray}
 E(k) \propto \beta_a^{\frac{3}{1 - a}} k^{-3},
\label{spcform}
\end{eqnarray}
which can be obtained by application of the Kraichnan-Leith-Bachelor
argument on the 2D enstrophy-cascade NS turbulence \cite{k67,l68,b69}.
However Eq.(\ref{spcform}) does not agree well with the numerical result plotted in 
in Fig.\ref{spc} even if we omit the cases with $-1 < a \le 0$ since they do not have
the inviscid conservative quantity.
More precisely, the numerical result shows $a$-dependence of the spectrum such that
$E(k)$ seems to take the power law $k^{-q}$ with exponent $2 \le q \le 4$ 
by choosing some $a < 0$.
Nevertheless $E(k)$ seems concentrating around $k^{-3}$ 
in the intermediate wavenumber range.

One way to understand the discrepancy from $k^{-3}$ is
the logarithmic correction that was first proposed by Kraichnan \cite{k71} 
for the 2D enstrophy-cascade NS turbulence. If we apply his derivation
to the gCLMG turbulence, the logarithmic correction takes the form
\begin{eqnarray}
 E(k) \propto \beta_a^{\frac{3}{1 - a}} k^{-3} \left[\log\left(\frac{k}{k_f}\right)\right]^{-\frac{1}{1 - a}},
\label{loge}
\end{eqnarray}
where $k_f$ is the wavenumber in which the forcing is added. 
Here we make a wildly heuristic assumption that the flux of $C_{-a}$  in the Fourier space
can be expressed with $\Omega(k) [k^3 E(k)]^{-a/2}$ where $\Omega(k)$ is the non-local frequency
$\Omega(k)^2 \sim \int_{k_f}^{k} p^2 E(p) dp$.
Obviously Eq.(\ref{loge}) with $a = -2$ coincides with the log-corrected spectrum of 
the 2D enstrophy-cascade NS turbulence. 
As observed in \cite{ms}, for $a = -2$ the log-corrected form, $k^{-3}\log^{-1/3}(k/k_f)$ does 
not agree with $E(k)$. The same is true for other $a$'s as we will see later.

What can be inferred from behavior of the $E(k)$ then?  
Is the cascade of $C_{-a}$ in the gCLMG turbulence 
just a coincidence for certain $a$'s?
For small $a$'s ($a \ll -1$) and large $a$'s ($a\sim -0$), the assumption of the cascade may be invalid
since $E(k)$ is rather distinct from $k^{-3}$.
Indeed it may appear strange that the high-order quantity, such as
$C_{-10}$ or $C_{-100}$, determines $E(k)$ which is the second 
order quantity of $\widehat{u}$. 
Therefore we study in detail the cases of $a$'s in which $E(k)$ is around $k^{-3}$.
Specifically, we take five cases, $a = -1.0, -1.5, -2.0, -3.0$ and $-4.0$ 
($a = -1.5$ is taken as a representative of the fractional cases).

\begin{figure} 
\includegraphics[scale=0.6]{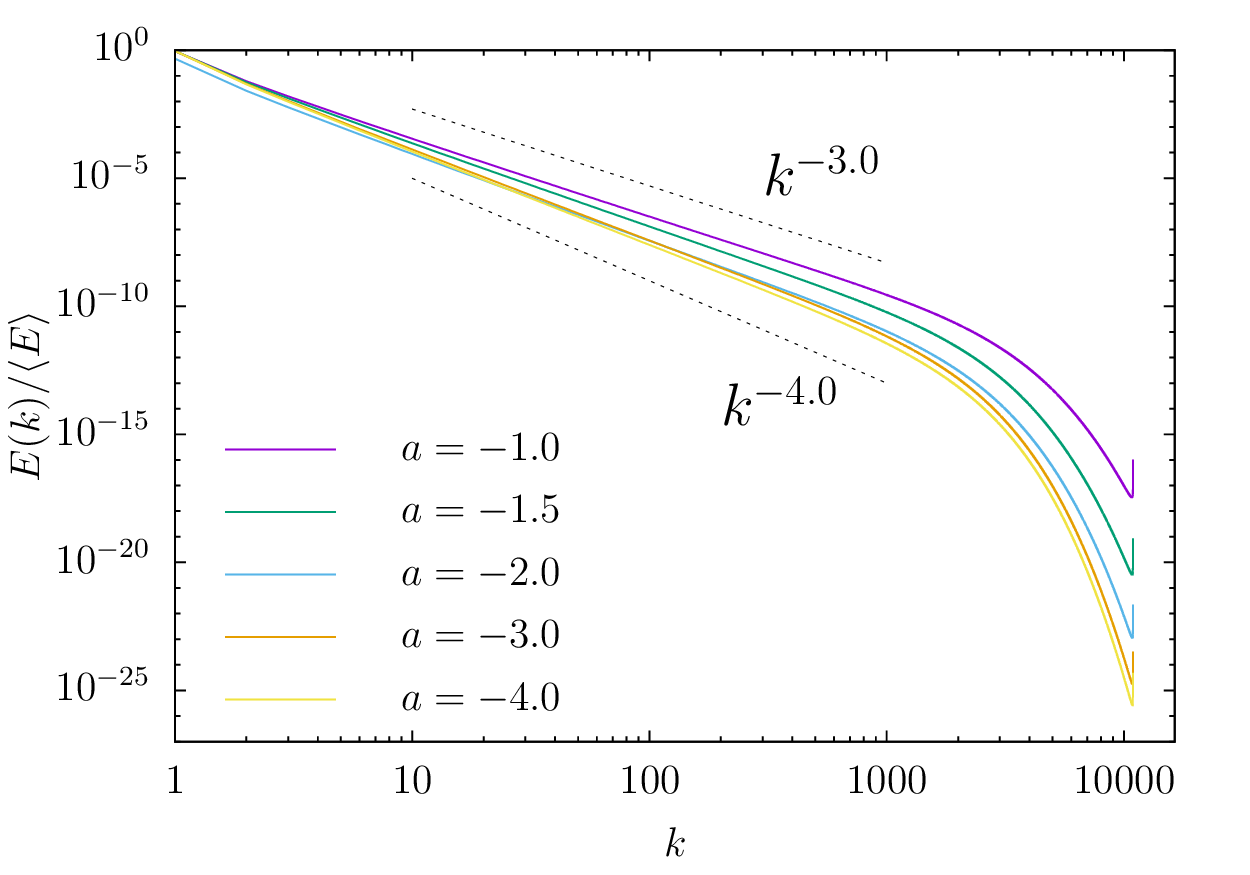}
\caption{\label{spc3} The time-averaged energy spectra for various $a$'s normalized
with the energy, $\langle E \rangle$. Here the viscosity is $\nu =  1.6\times 10^{-6}$
and the number of grid points is $2^{15}$.
The time average is taken over $3150$
equi-spaced instances between $700 \le t \le 7000$ from three different realizations of the random forcing.
The scaling laws, $k^{-3}$ and $k^{-4}$, are shown as a guide.}
\end{figure}

First, we check behavior of $E(k)$ as decreasing $\nu$. 
The energy spectra with a smaller viscosity are shown in Fig.\ref{spc3}.
We observe that the ``inertial-range behavior'' of each $a$ 
extends to the larger wavenumber region than Fig.~\ref{spc} without 
changing the wavenumber dependence. In particular, the energy spectrum $E(k)$ 
is close to $k^{-3.0}$ for $a = -1.0$ and to $k^{-4.0}$ for $a = -4.0$. 
Comparing to Eq.(\ref{spcform}),
this dependence of $E(k)$ on the parameter $a$ indicates that turbulent
cascade of the inviscid conservative quantity $C_{-a}$ is unlikely and 
that the enstrophy cascade for $a =-2$ analyzed previously in \cite{ms} 
is just coincidental. 
In the next subsection we numerically analyze directly whether or not
a scale-wise nonlinear transport of $C_{-a}$ is considered to be turbulence
cascade with a spatial-filter method \cite{filter}.

\subsection{Analysis of the cascade}

Working in the periodic domain, the most convenient method to analyze
the cascade is the transfer function or the flux in the Fourier space,
if the cascade quantity is quadratic. 
For the case of $a = -2$, the enstrophy flux in the the Fourier space was
used to show the cascade of the enstrophy \cite{ms}. 
Here, for general $a$'s where the quantity is no longer quadratic, 
we adopt a more versatile method introduced in \cite{filter} to investigate
whether or not the turbulent cascade of the conservative quantity $C_{-a}$ occurs.

This method uses a low-pass spatial filter with a filtering scale $\ell$
in the physical space. 
Let us write the filter function with $g_\ell(x)$. The filtered 
quantity of a function $A(x, t)$ is then expressed as 
\begin{eqnarray}
 \overline{A_\ell}(x, t) = \int_0^{2\pi} A(x', t) g_\ell(x' - x) dx'.
\end{eqnarray}
Now the filtered gCLMG eq. can be written as
\begin{eqnarray}
\partial_t \overline{\omega_\ell} 
 + a \overline{u}_\ell
 \partial_x \overline{\omega}_\ell
 =
 \overline{\omega}_\ell \partial_x \overline{u}_\ell 
 + \sigma_\ell
 + \nu \partial_{xx} \overline{\omega}_\ell
 + \overline{f}_\ell,
\end{eqnarray}
where the vorticity input from the scales smaller than $\ell$ (the subgrid scales) 
is
\begin{eqnarray}
 \sigma^{(a)}_\ell 
= 
 - a \partial_x \left[\overline{u \omega}_\ell - \overline{u}_\ell \overline{\omega}_\ell \right]
 + (1 + a)  \left[\overline{\omega (\partial_x u)}_\ell - \overline{\omega}_\ell \partial_x \overline{u}_\ell\right].
\end{eqnarray}
This leads to the equation of the low-pass filtered 
$(-a)$-th power of the vorticity  as
\begin{eqnarray}
 &&
  \partial_t \left(\frac{1}{-a}\overline{\omega}_\ell^{-a} \right)
- \partial_x \left( \overline{u}_\ell \overline{\omega}_\ell^{-a - 1} + \nu \overline{\omega}_\ell^{-a - 1} \partial_x \overline{\omega}_\ell \right) \nonumber \\
&& = -Z^{(a)}_\ell 
  - \nu (-a - 1) \overline{\omega}_\ell^{-a - 2} (\partial_x \overline{\omega}_\ell)^2
 + \overline{\omega}^{-a - 1}_\ell \overline{f}_\ell.
\end{eqnarray}
Here $Z^{(a)}_\ell(x, t)$ is the flux of the grid-scale moment $\overline{\omega}_\ell^{-a}$
being transferred to smaller scales than $\ell$, which 
is expressed as
\begin{eqnarray}
 Z^{(a)}_\ell(x, t) =  \overline{\omega}^{-a - 1}_\ell \sigma^{(a)}_\ell.
\end{eqnarray}
With  $Z^{(a)}_\ell(x, t)$, we can analyze the cascade in a precise way \cite{filter}.
By cascade, it is understood that the space-time average of $Z^{(a)}_\ell(x, t)$,
which we denote $\langle Z^{(a)}_\ell \rangle$, becomes independent of the filter 
scale $\ell$ in some range of $\ell$. If such a range of $\ell$ exists, we here 
call it inertial range. Notice that we assume spatial homogeneity and statistical
steadiness of the flux, $Z^{(a)}_\ell(x, t)$. 
It is known that the expressions of the flux is not unique.
This non-uniqueness does not matter since we consider the spatial average of the flux.

\begin{figure}
\includegraphics[scale = 0.6]{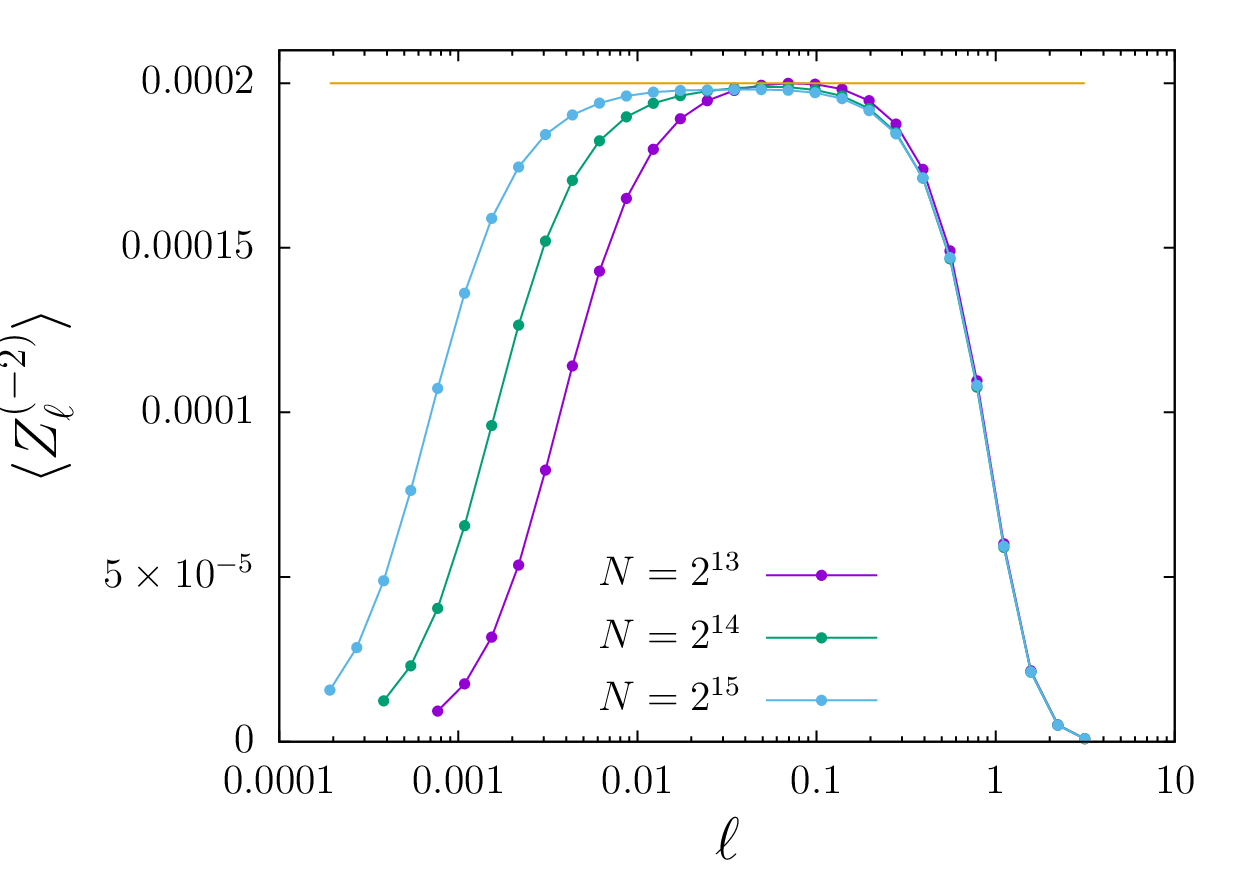}
\includegraphics[scale = 0.6]{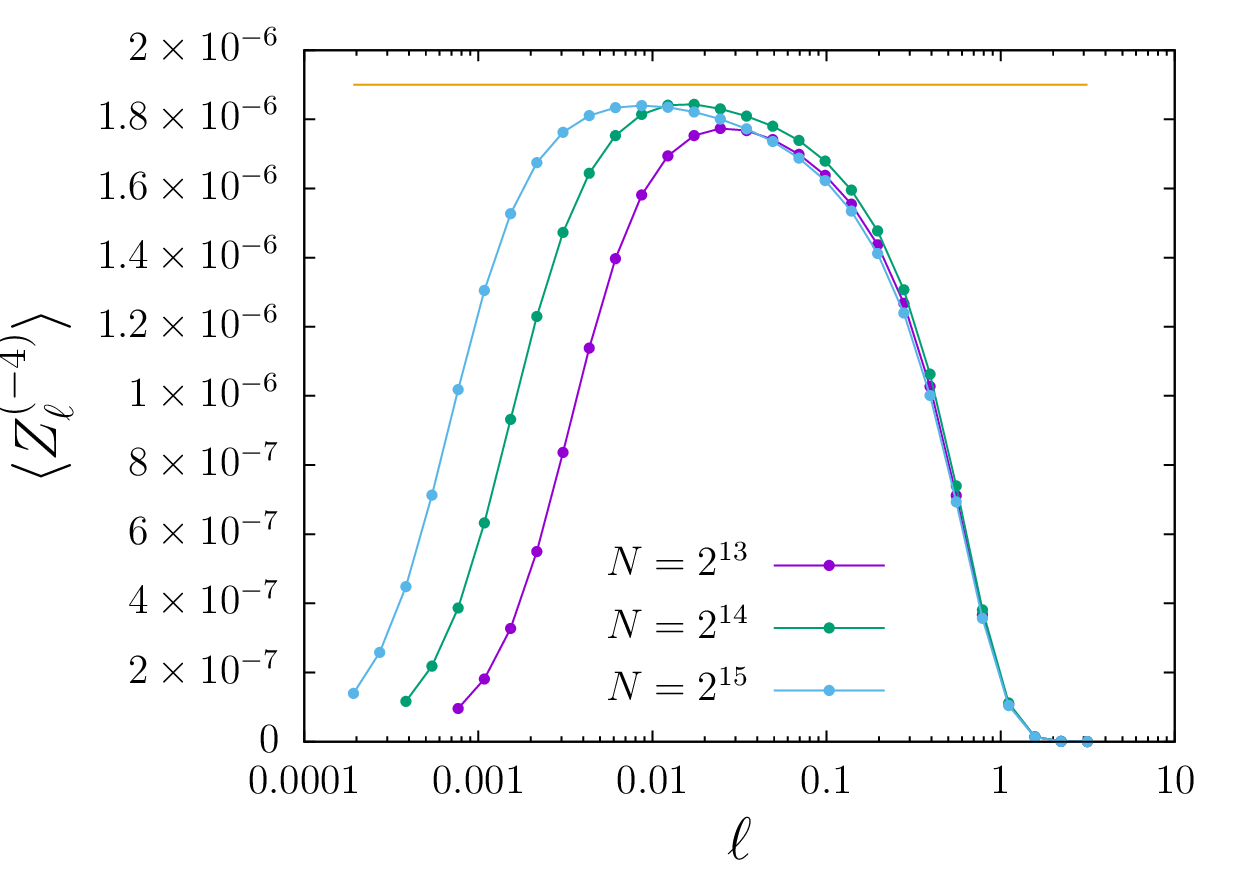}
 \caption{\label{fluxeven} Averaged flux $\langle Z^{(a)}_\ell \rangle$
 of the inviscid conservative quantity $C_a$  for $a = -2$ (left) and $a = -4$ (right).
 The horizontal line indicates the average dissipation rate $\langle \beta_{a} \rangle$.
 The viscosity and the resolution ($\nu, N$) are
 $(2.5\times 10^{-5}, 2^{13}), (6.25\times 10^{-6}, 2^{14})$ and $(1.5625\times10^{-6}, 2^{15})$.
 The method of time average is the same as in Figs.\ref{spc} and \ref{spc3}.
 For the $N = 2^{14}$ case, the time average is taken over equi-spaced 3700 snapshots from $4000 \le t \le 41000$.}
\end{figure}

In Fig.\ref{fluxeven} we show $\langle Z^{(a)}_\ell \rangle$ for even $a$ cases,
where the inviscid conservative quantities $C_{-a}$ are positive definite. 
As the filter function $g_\ell(x)$,
we use the Gaussian filter $g_\ell(x) = \frac{1}{\sqrt{2\pi}\ell}\exp[-(x/\ell)^2/2]$.
For $a = -2$, there is a plateau that amounts to the $\ell$-independent flux.
This implies that the cascade of the enstrophy, $C_{-2}$, occurs,
as indicated with the equivalent flux in the Fourier space in \cite{ms}.
While for $a = -4$ such a plateau is not well developed in comparison. 
Instead of plateau, the flux in the intermediate range is a mildly decreasing
function of $\ell$. The variation shown in Fig.\ref{fluxeven} may suggest $\langle Z^{(-4)}_\ell \rangle \propto -\log(\ell)$.
This indicates that $C_{-4}$ does not cascade at least for the ranges of $\nu$ considered here. 
Nevertheless, if we decrease $\nu$ furthermore, a plateau may appear in small scales. 
Hence the cascade is not completely ruled out for $a = -4.0$.

\begin{figure} 
\includegraphics[scale = 0.6]{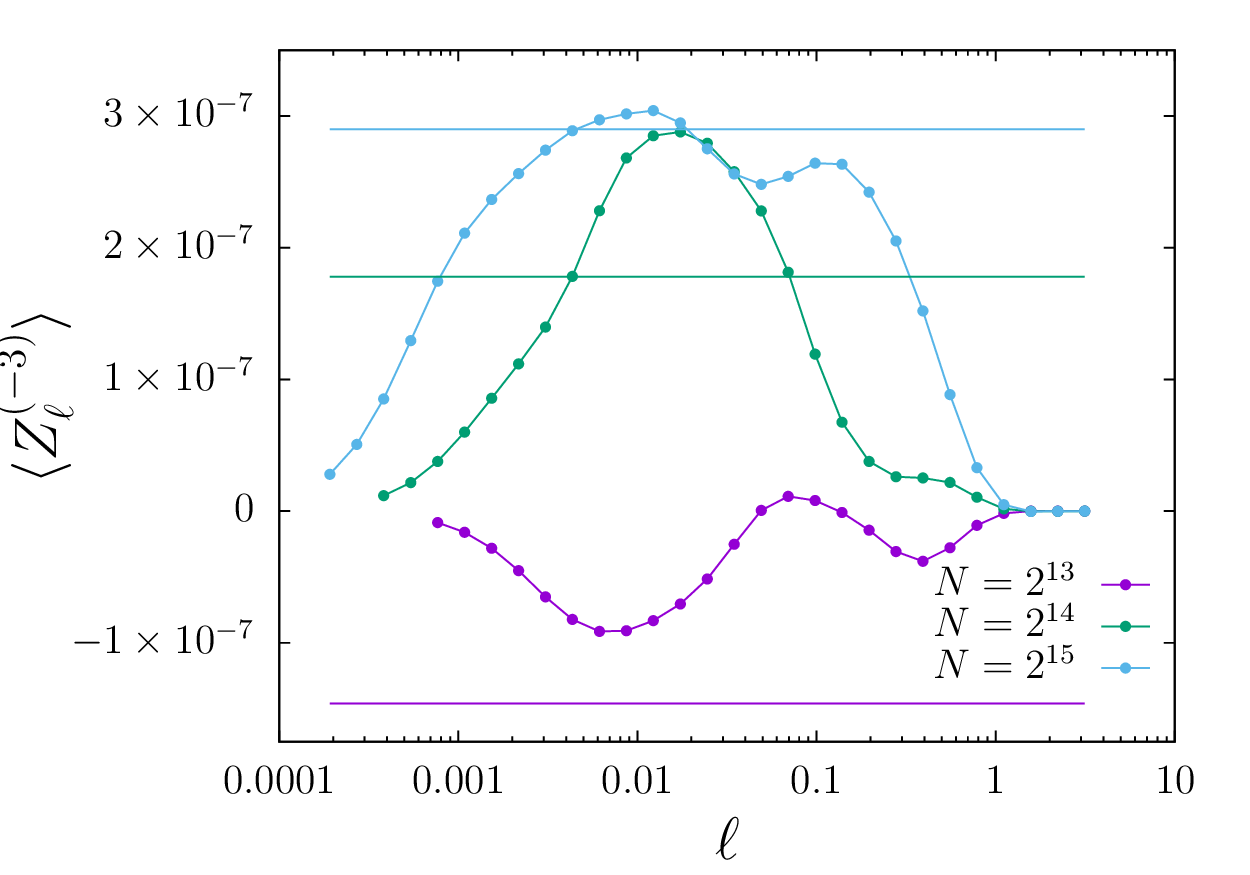}
 \caption{\label{fluxodd} Same as Fig.~\ref{fluxeven} but for $a = -3$.
For the $N = 2^{15}$ and $2^{14}$ cases, $C_{-3} > 0$ while $C_{-3} < 0 $ for the $N = 2^{13}$ case.
The horizontal line indicates the average dissipation rate $\langle \beta_{a} \rangle$ for each case.} 
\end{figure}

For an odd $a$ case, $a = -3$, the flux is shown in Fig.\ref{fluxodd}. 
The wiggly variation in the intermediate range of $\ell$ indicates evidence
against the cascade. This wiggles may be caused because $C_{-3}$ and $\beta_{-3}$ are not sign 
definite and hence fluctuation effects are strong.
There may be cascade for smaller $\nu$ but the way of extension
of the possible inertial range shown in Fig.\ref{fluxodd} 
is not convincing.
Thus we do not have a numerical evidence for the cascade of $C_{-3}$. 
If we can control the input rate of $C_{-3}$ with the forcing, a clearer result
may be obtained. 
For the other odd case, $a = -1$,
the average flux $\langle Z^{(a)}_\ell \rangle$ is by definition zero
since the nonlinearity vanishes in the equation of $C_{-1}$. 
Therefore the cascade of $C_{-1}$ is not possible.
However recall that the energy spectrum for $a = -1$ is broad 
and close to a power law $k^{-3}$ in Fig.\ref{spc3}. At least for $a = -1$, the 
spectrum has nothing to do with the cascade of $C_{-1}$.

\begin{figure} 
\includegraphics[scale = 0.6]{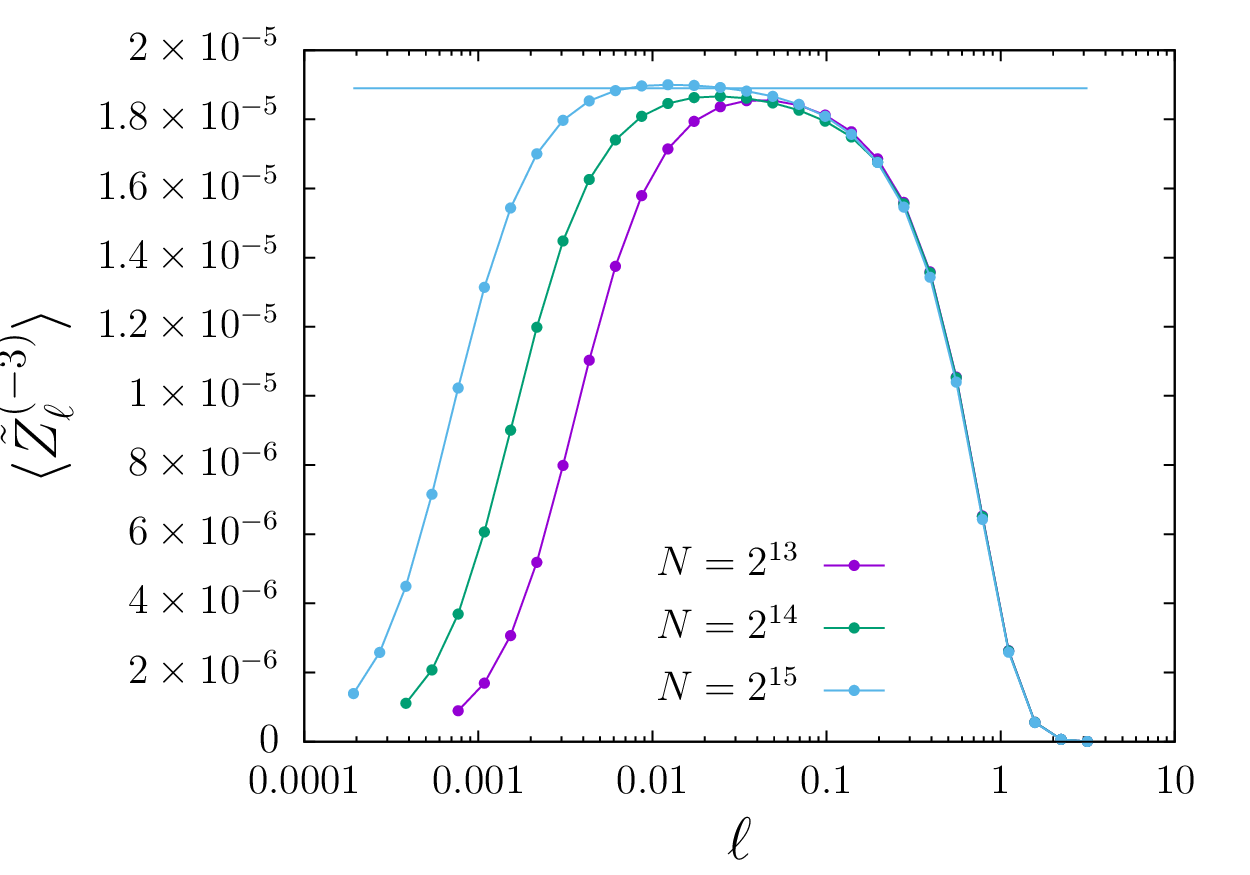}
\includegraphics[scale = 0.6]{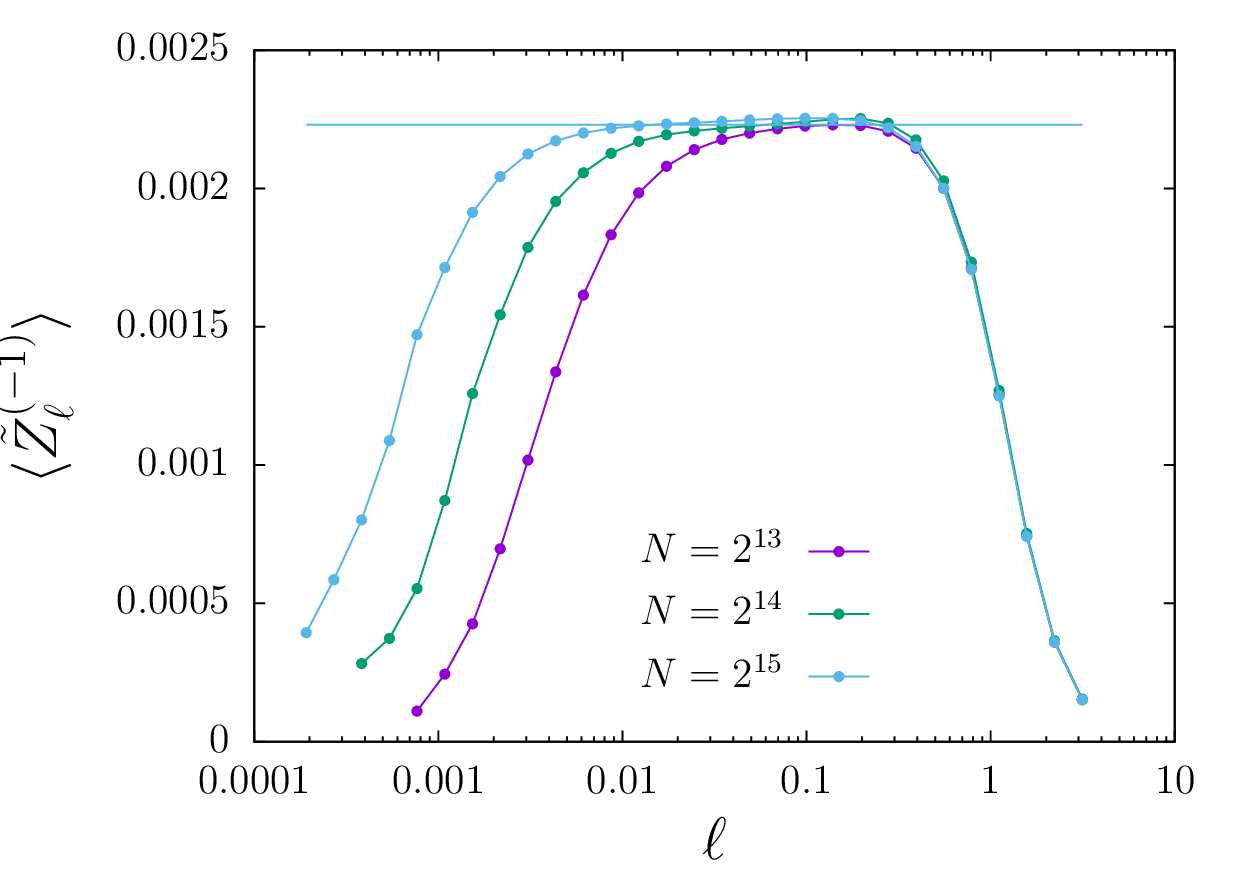}
\caption{\label{fluxabs31}Averaged flux $\langle \tilde{Z}^{(a)}_{\ell} \rangle$
of the conservative quantity $\tilde{C}_{a}$ for  $a = -3$ (left) and $a = -1$ (right).
The horizontal line indicates the averaged dissipation rate $\langle \tilde{\beta}_{a} \rangle$
measured in the simulation with $2^{15}$ grid points.} 
\end{figure}

Now we move to a different form of the inviscid conservation law.
For negative odd integer $a$, the absolute $(-a)$-th moment
\begin{eqnarray}
 \tilde{C}_{a} = \frac{1}{-a}\int_0^{2\pi} |\omega(x, t)|^{-a} dx
\label{abmo}
\end{eqnarray}
is also an inviscid conserved quantity. The dissipation rate
of $\tilde{C}_{a}$ can be defined as
\begin{eqnarray}
\tilde{\beta}_{a} = -\nu \frac{1}{2\pi}\int_0^{2\pi} {\rm sgn}(\omega) |\omega|^{-a - 1} \partial_x^2 \omega dx,
\end{eqnarray}
where ${\rm sgn}(\omega)$ denotes the sign of $\omega$. 
The corresponding flux can be obtained via the equation of $|\overline{\omega}_\ell|^{-a}$. 
One expression is 
\begin{eqnarray}
\tilde{Z}^{(a)}_{\ell} 
= 
{\rm sgn}(\overline{\omega}_\ell) 
|\overline{\omega}_\ell|^{-a - 1}
\left[
 a \partial_x (\overline{u \omega}_\ell)
-(a + 1) \overline{\omega (\partial_x u)}_\ell
\right].
\end{eqnarray}
For $a = -1$ and $-3$ cases, the averaged flux $\langle \tilde{Z}^{(a)}_\ell \rangle$
is shown in Fig.\ref{fluxabs31}. 
Comparing with Fig.\ref{fluxodd}, the flux of the absolute
third order moment appears quite different for $a = -3$ and looks similar to 
the flux $\langle Z^{(4)}_\ell \rangle$ for $a = -4$. We begin to see the plateau for the smallest
$\nu$ case. For $a = -1$ case, a well-developed plateau is seen. 
From this, the cascade of the absolute moment is plausible for $a = -1$ and $-3$. 
Is this consistent with the dimensional analysis of the energy spectrum, Eq.(\ref{spcform}), 
provided that the dissipation rate is now $\tilde{\beta}_a$ ?
As seen in Fig.\ref{spc3}, for $a = -1$, it may be consistent since $E(k)$ is close to $k^{-3}$.
While $a = -3$, it is not since $E(k)$ is closer to $k^{-4}$. This point will be revisited
with a stationary solution under a deterministic forcing in the next section. 

\begin{figure} 
\includegraphics[scale = 0.6]{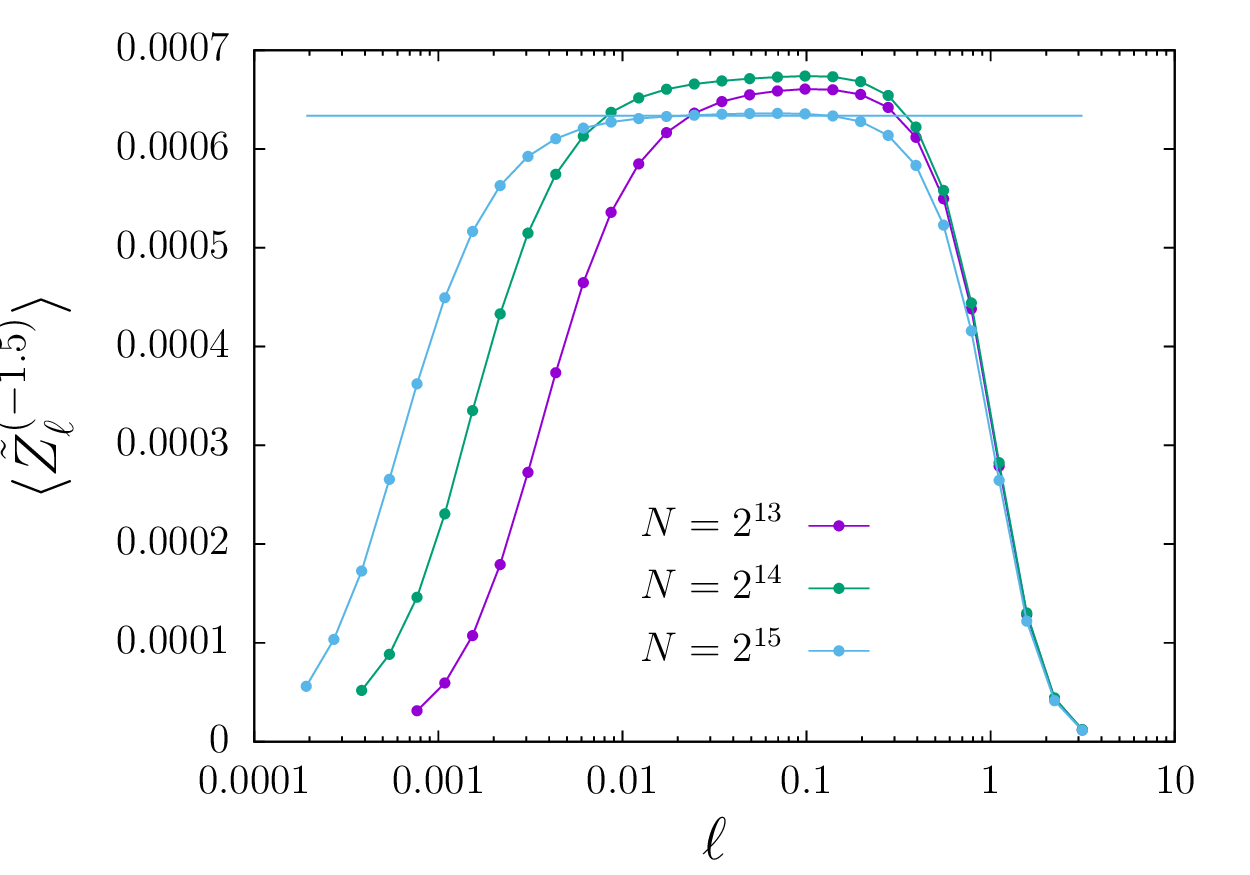}
 \caption{\label{fluxhalf}Averaged flux $\langle \tilde{Z}^{(-1.5)}_{\ell} \rangle$.
The horizontal line indicates the averaged dissipation rate $\langle \tilde{\beta}_{-1.5} \rangle$
measured in the simulation with $2^{15}$ grid points.} 
\end{figure}

As a non-integer value, we here take $a = -1.5$ as a representative case.
We plot the flux of the absolute
moment $\langle \tilde{Z}^{(-1.5)}_\ell \rangle$ in Fig.\ref{fluxhalf}. 
It shows a well-developed plateau as in the previous case $a = -1.0$. 
The difference in the plateau values among the three resolutions
is large. However they are consistent with values of the dissipation rate $\tilde{\beta}_{-1.5}$.
Although this cascade of $a = -1.5$ indicated by the plateau implies $E(k) \propto k^{-3}$ dimensionally,
the measured $E(k)$ presented in Fig.\ref{spc3} has a slight but measurable deviation 
from the dimensional result.

In summary of this cascade analysis,
we observe that, for $a = -1.0, -1.5, -2.0$ and $-3.0$, the cascade of
$\tilde{C}_{a}$ is indicated by the plateau of the averaged 
flux $\langle \tilde{Z}^{(a)}_\ell \rangle$ 
and that, for $a  = -4.0$, the indication of the cascade becomes 
weaker. 
Therefore the inertial range, in the sense of the range of scales where the flux 
becomes constant, is likely to exist at least for $a \ge -3.0$. 
At the same time we see the systematic change of the energy spectrum $E(k)$ in the inertial range 
as a function of $a$ in Figs.\ref{spc}--\ref{spc3} 
This is not consistent with the dimensional  result, Eq.(\ref{spcform}) although the change is 
around $k^{-3}$ except for extreme values of $a$. 
Our view on the discrepancy between the cascade analyzed here 
and the dimensional form of $E(k)$ is that the variation around $k^{-3}$ can be understood as a non-dimensional 
correction such as the logarithmic correction proposed by Kraichnan for 2D enstrophy-cascade 
turbulence. This point will be studied in the next section with the stationary solution under the deterministic forcing.
For extreme values of $a$'s (close to 0 and smaller than $-3$), the behavior of $E(k)$ may be inferred
from the corresponding limits,
such as the CLM eq. for $a \to -0$ and 
the advection equation for $a \to -\infty$, not from the cascade.

\subsection{K\'arm\'an-Howarth-Monin relation and dissipative weak solution}

\begin{figure} 
\includegraphics[scale=0.6]{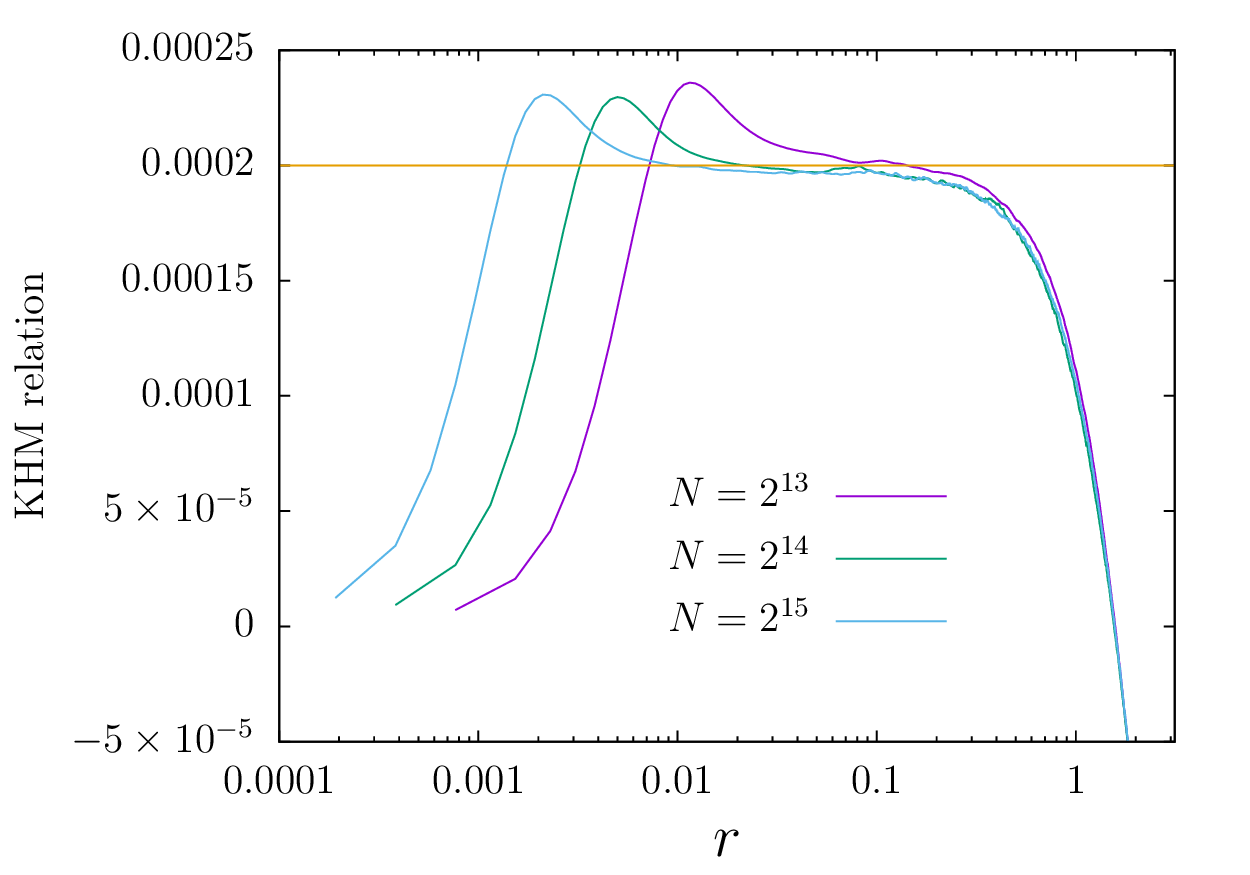}
\caption{\label{khmfig}The K\'arm\'an-Howarth-Monin relation of the gCLMG turbulence for $a = -2.0$.
The horizontal line is the averaged dissipation rate $\langle \beta_{-2} \rangle$ (Eq.(\ref{betaa})) 
calculated in the simulation with $N = 2^{15}$ grid points.
Both terms on the right hand side of Eq.(\ref{khm}) have the same order of magnitude in the inertial range.}
\end{figure}

Having obtained an evidence of the cascade of $\tilde{C}_{-a}$ for certain $a$'s, 
we now consider a gCLMG analogue of the K\'arm\'an-Howarth-Monin (KHM) relation
of the NS turbulence \cite{f}. 
For the case of $a = -2.0$, an expression of the KHM relation for the gCLMG turbulence 
is 
\begin{eqnarray}
&& B_{-2}(r) 
= - \partial_r \langle \omega(x) \omega(x') [u(x') - u(x)] \rangle \nonumber \\
&&  + \frac{1}{2} \langle \omega(x)\omega(x') [\partial_x u(x) + \partial_{x'} u(x')] \rangle,
\label{khm}
\end{eqnarray}
where $x' = x + r$.  
This $B_{-2}(r)$ is interpreted as the enstrophy flux across scale $r$, which 
is similar to $\langle Z_r^{(-2)} \rangle$. 
In other words, the KHM relation is yet another device to look at the cascade. 
Although it is closely related to the filtering flux used previously,
the KHM relation does not involve spacial filtering.
Here we assume that the correlation functions on the right hand side of Eq.(\ref{khm})
are homogeneous (depending only on $r$) and in a statistically steady state. 
Due to the compressibility,
expression of the KHM relation is not unique.
Furthermore, it cannot be expressed in terms of divergence of certain products of 
the velocity and vorticity increments. 
This means that the analogue is not like the 4/5-law of the energy
cascade of the 3D NS turbulence (see, e.g., \cite{f}) or the 2-law of the enstrophy
cascade of the 2D NS turbulence (e.g., \cite{g98}).
A numerical confirmation of the KHM relation (\ref{khm}) is shown in 
Fig.\ref{khmfig}. 
The flux $B_{-2}(r)$ in the inertial range is $r$ independent and 
close to the enstrophy dissipation rate $\langle \beta_{-2} \rangle$,
again demonstrating the cascade of the enstrophy $C_{-2}$.

This fact has an interesting consequence on dissipation without viscosity, namely
gCLMG analogue of the Onsager's conjecture 
(for the Onsager's conjecture on the 3D Euler equations, see, e.g., \cite{esrmp,ds}). 
By following the Duchon-Robert formalism \cite{dr}, let us now consider a weak solution 
of the inviscid and unforced gCLMG eq. with $a = -2.0$.
With a $\epsilon$-scale mollifier $\varphi_\epsilon(x' - x)$,
the regularized vorticity $\omega_{\epsilon} = \int \omega(x') \varphi_\epsilon(x' - x) dx'$
of the weak solution obeys the equation
\begin{equation}
 \partial_t \omega_\epsilon = 2 \partial_x (u \omega)_\epsilon  - (\omega \partial_x u)_\epsilon.
\end{equation}
The local enstrophy budget equation becomes
\begin{eqnarray}
&& \partial_t \frac{1}{2}(\omega \omega_\epsilon) - \partial_x (u \omega \omega_\epsilon) 
  = - \frac{1}{2}\omega 
\big[
-2 \partial_x (u \omega)_\epsilon 
+ \partial_x(u\omega_\epsilon) \nonumber \\
&& 
+ u \partial_x \omega_\epsilon 
+ (\omega \partial_x u)_\epsilon \big].
\end{eqnarray}
This motivates us to put the right hand side as ``dissipation''
\begin{equation}
 D^{(-2)}_\epsilon(\omega)
 =  
\frac{1}{2}\omega 
\big[
-2 \partial_x (u \omega)_\epsilon 
+ \partial_x(u\omega_\epsilon)
+ u \partial_x \omega_\epsilon 
+ (\omega \partial_x u)_\epsilon \big].
\end{equation}
If $D^{(-2)}_\epsilon(\omega) > 0$ as $\epsilon \to 0$, the weak solution 
to the inviscid gCLMG eq. can be called \textit{dissipative}. 
Under what conditions it becomes dissipative is an interesting question.
Unlike the 3D Euler case, condition for $D^{(-2)}_\epsilon(\omega) = 0$ (or $ >  0$) 
may not be characterized by the H\"older exponent of $\omega$.
Another formal analysis leads to an expression of $D^{(-2)}_\epsilon(\omega)$ as
\begin{eqnarray}
&& \int dr~ 
\omega(x)
\omega(x') 
\bigg[(u(x') - u(x)) 
(\partial_r \varphi_\epsilon(r)) \nonumber \\
&& + 
\frac{1}{2}
\varphi_\epsilon(r)
(\partial_{x'} u(x') + \partial_x u(x)) 
\bigg]
 = D^{(-2)}_\epsilon(\omega),
\end{eqnarray}
which is an equivalent of the KHM relation of the weak solution.

For the general $a$, the formal expression of the KHM flux of the gCLMG turbulence 
is
\begin{eqnarray}
&&B_{a}(r) = 
-\partial_r  \langle \omega^{-a-1}(x) \omega(x') [u(x') - u(x)] \rangle \nonumber \\
&& - \frac{1}{a} \langle \omega^{-a-1}(x) \omega(x') [\partial_{x} u(x) - (a + 1) \partial_{x'} u(x')] \rangle,
\end{eqnarray}
and the ``dissipation'' of the weak solution to the inviscid, unforced gCLMG eq.
is 
\begin{eqnarray}
&& D^{(a)}_\epsilon(r)
= 
\frac{1}{-a}
\omega^{-a - 1}
\big[
a\partial_x (u \omega)_\epsilon
+ \partial_x ( u \omega_\epsilon) \nonumber \\
&& 
- (a + 1) u \partial_x \omega_\epsilon
- (a + 1) (\omega \partial_x u)_\epsilon
\big].
\end{eqnarray}
Consequently, the KHM relation of the weak solution can be
\begin{eqnarray}
&&\int dr~ 
\omega^{-a - 1}(x)
\omega(x')
\bigg[
(u(x') - u(x))
(\partial_r \varphi_\epsilon(r))  \nonumber \\
&& 
- \frac{1}{a}
\varphi_\epsilon(r) 
(\partial_x u(x) - (1 + a) \partial_{x'} u(x'))
\bigg]
 = D^{(a)}_\epsilon (\omega).
\end{eqnarray}
This can also be interpreted as a law which holds local in time
and space without assuming the homogeneity, isotropy and 
statistical steadiness.

\subsection{Vorticity structure function}

Now, coming back to the gCLMG turbulent solution under the random forcing, 
we look into the $p$-th order structure function of the vorticity 
\begin{eqnarray}
 S^{(a)}_p(r) = \langle [\omega(x + r, t) - \omega(x, t)]^p \rangle
\end{eqnarray}
and its logarithmic local slope
\begin{eqnarray}
 \frac{\log S^{(a)}_p(r + \Delta r)  - \log S^{(a)}_p(r)}{\log (r + \Delta r) - \log r}.
\end{eqnarray}
The purpose here is to analyze the possible presence of the (non-power law)
correction to the simple scaling law.
Recall that, if the energy spectrum is a power law with an exponent $\gamma$, 
namely, $E(k) \propto k^{-\gamma}$, the structure function can be predicted 
as $S^{(a)}_p(r) \propto r^{(3 - \gamma)p/2}$ with the dimensional analysis. 
As we discussed at the end of the previous section, the correction to
$E(k) \propto k^{-3} ~(\gamma = 3)$ can be observed with the second order 
vorticity structure function. Here we focus on even orders,  $p = 2, 4$ and $6$.

\begin{figure} 
\includegraphics[scale=0.6]{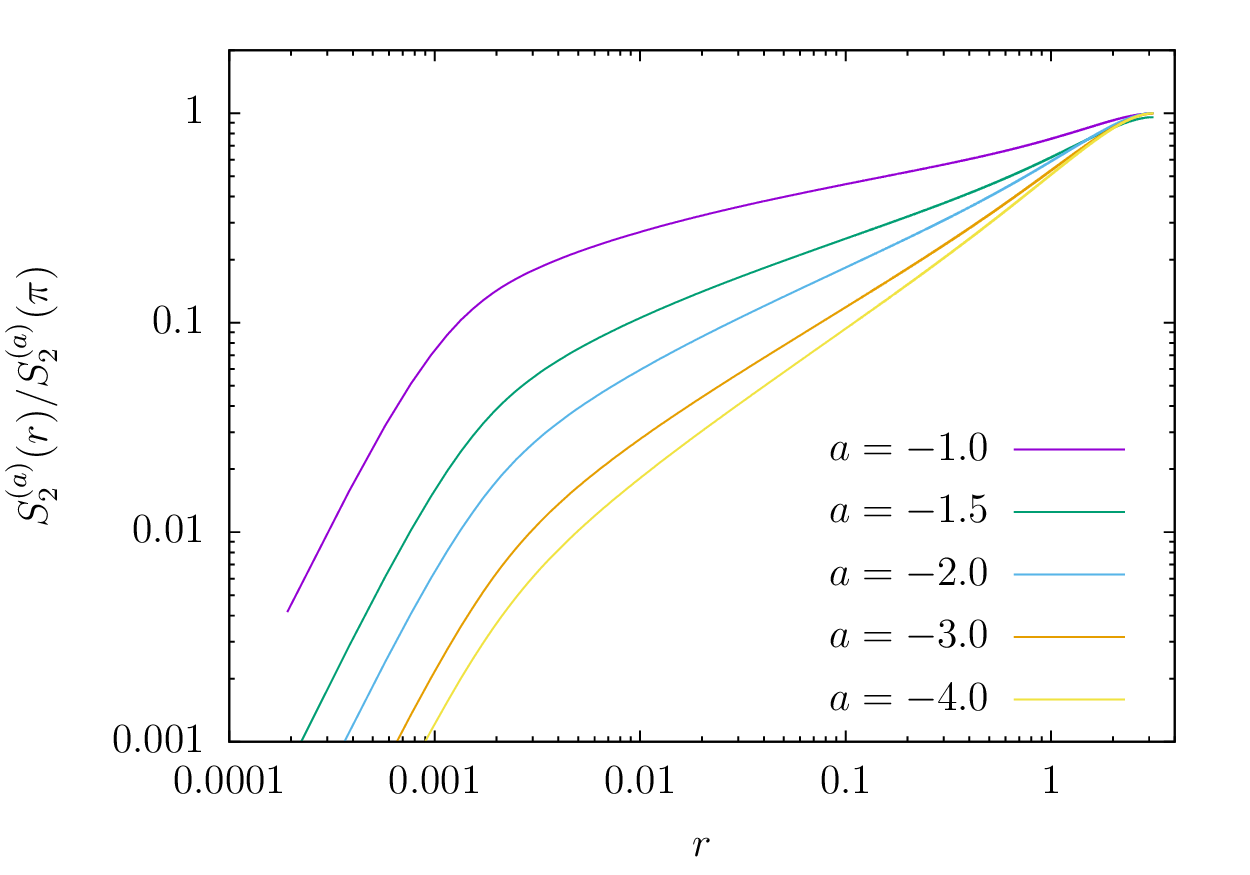}
\includegraphics[scale=0.6]{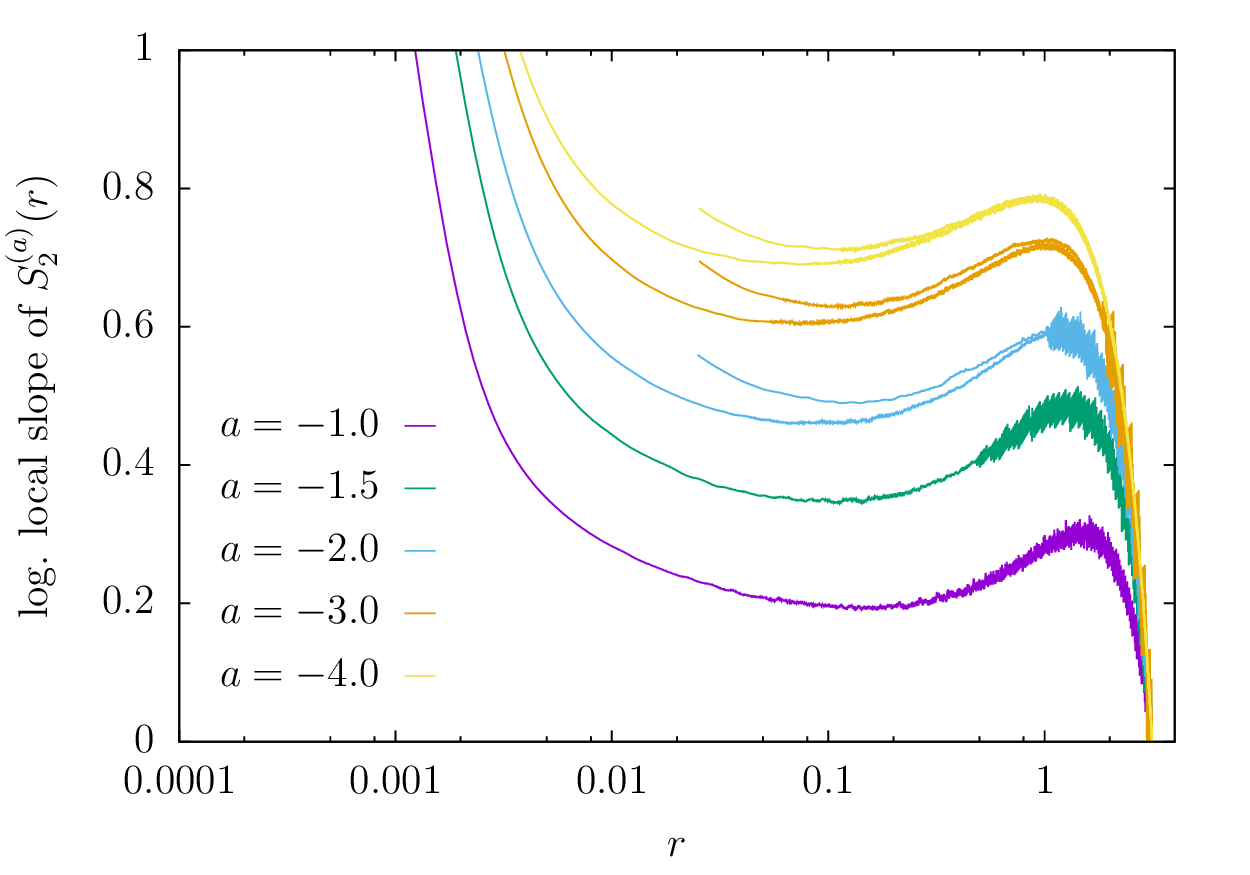}
 \caption{\label{stf2}Second order structure function of the vorticity (left) and its logarithmic local slope (right) for various $a$'s ,
 which are  calculated with the same data sets used in Figs.\ref{spc3}--\ref{fluxhalf}. Shorter curves for $a = -2.0, -3.0$ and $-4.0$ are data with the larger $\nu$ and the smaller $2^{14}$ grid points.}
\end{figure}

The second order structure function is shown in Fig.\ref{stf2}. 
For $a = -1.0$ and $-1.5$, the local slope does not become
flat for the range $0.01 \le r \le 0.1$ where the absolute flux $\tilde{Z}^{(a)}_r$
is reasonably flat (the inertial range). 
This indicates that $S^{(a)}_2(r)$ is not characterized with a power-law scaling, 
which supports a non-dimensional correction to the energy spectrum $E(k) \propto k^{-3}$.
For $a = -2.0, -3.0$ and $-4.0$, the local slope has a local minimum around $r \approx 0.1$, 
which may become a flat region with smaller $\nu$. 
However comparing with a different $\nu$ case, we see that the minimum cannot be interpreted
as the beginning of such a flat region. If it is the beginning, the minimum value is unchanged as we change $\nu$.
This indicates that $S^{(a)}_2$ is not a simple power law also for $a = -2.0, -3.0$ and $-4.0$.

As we increase the order to $p = 4$ and $6$, 
a visible feature is a local maximum in the dissipative range.
This is a reflection of the vorticity pulses shown in Fig.\ref{prof} since
the location of the maximum corresponds roughly to the width of the pulse. 
In Fig.\ref{prof}, as increasing $a$, we observe that the height of the pulse becomes larger and that
the width becomes smaller. 
The observation is consistent with 
the fact that the maximum of the fourth order structure function appears only for large $a$ cases, $a = -1.0$ and $-1.5$
and also with the fact that the location of the maximum in $S^{(a)}_6(r)$ shifts to smaller $r$ as we increase $a$.

\begin{figure} 
\includegraphics[scale=0.6]{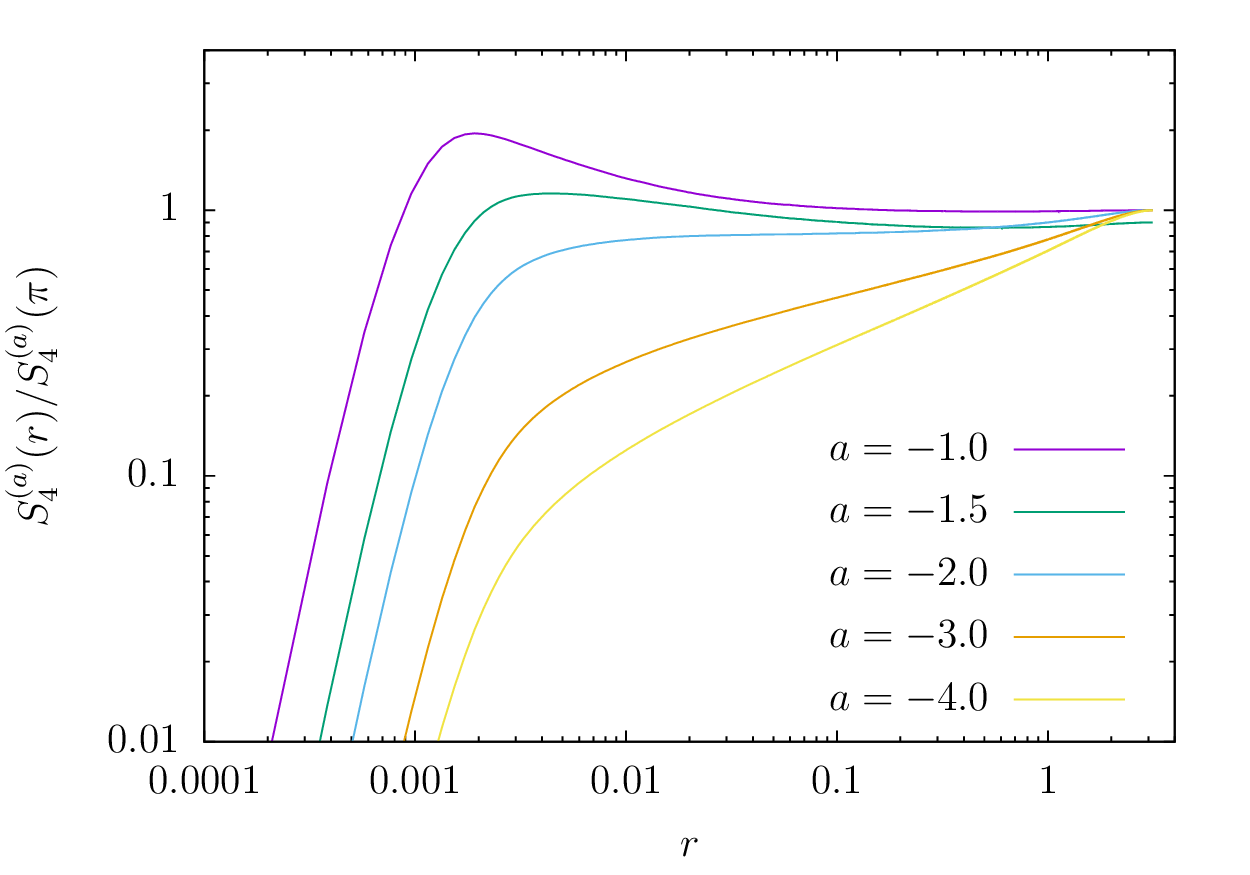}
\includegraphics[scale=0.6]{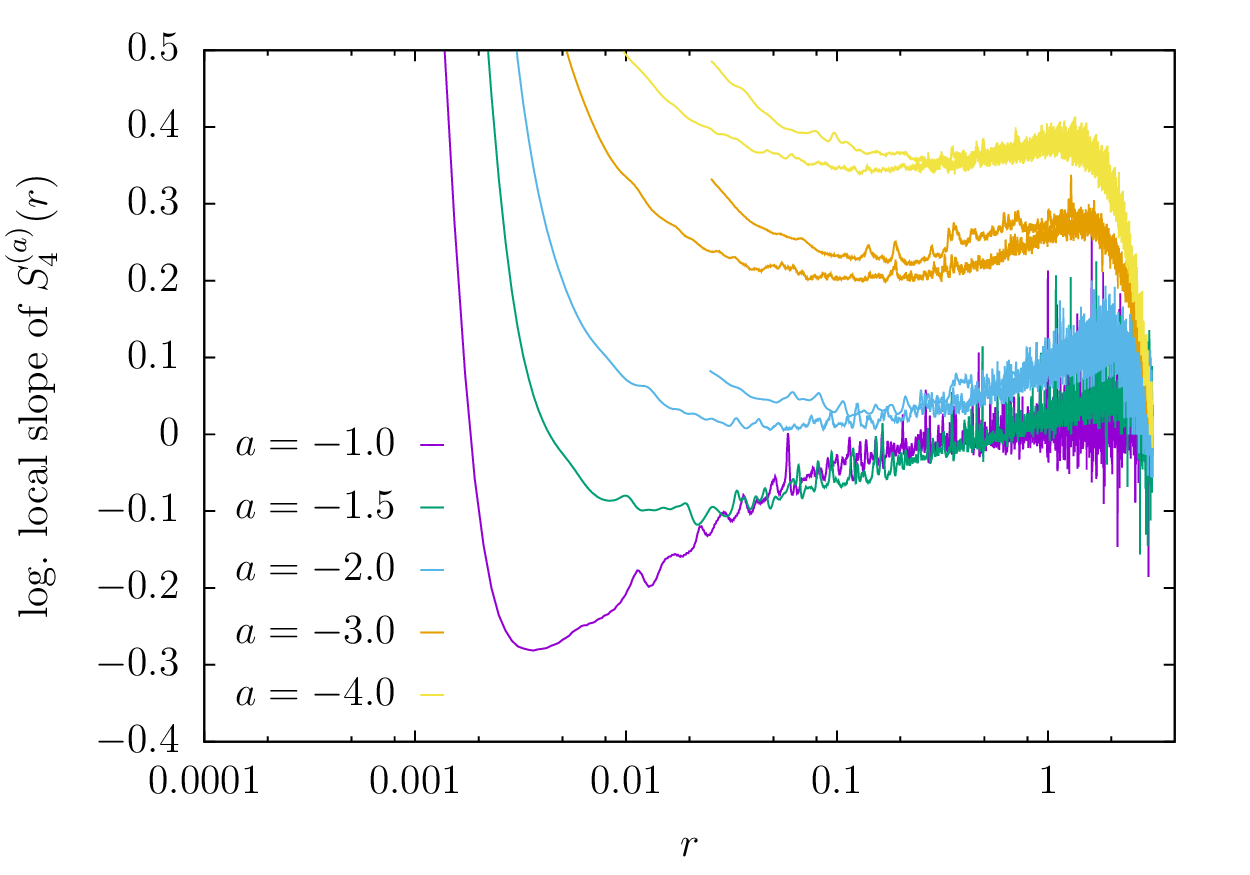}
\caption{\label{stf4}Same as Fig.\ref{stf2} but for $p = 4$.}
\end{figure}

\begin{figure} 
\includegraphics[scale=0.6]{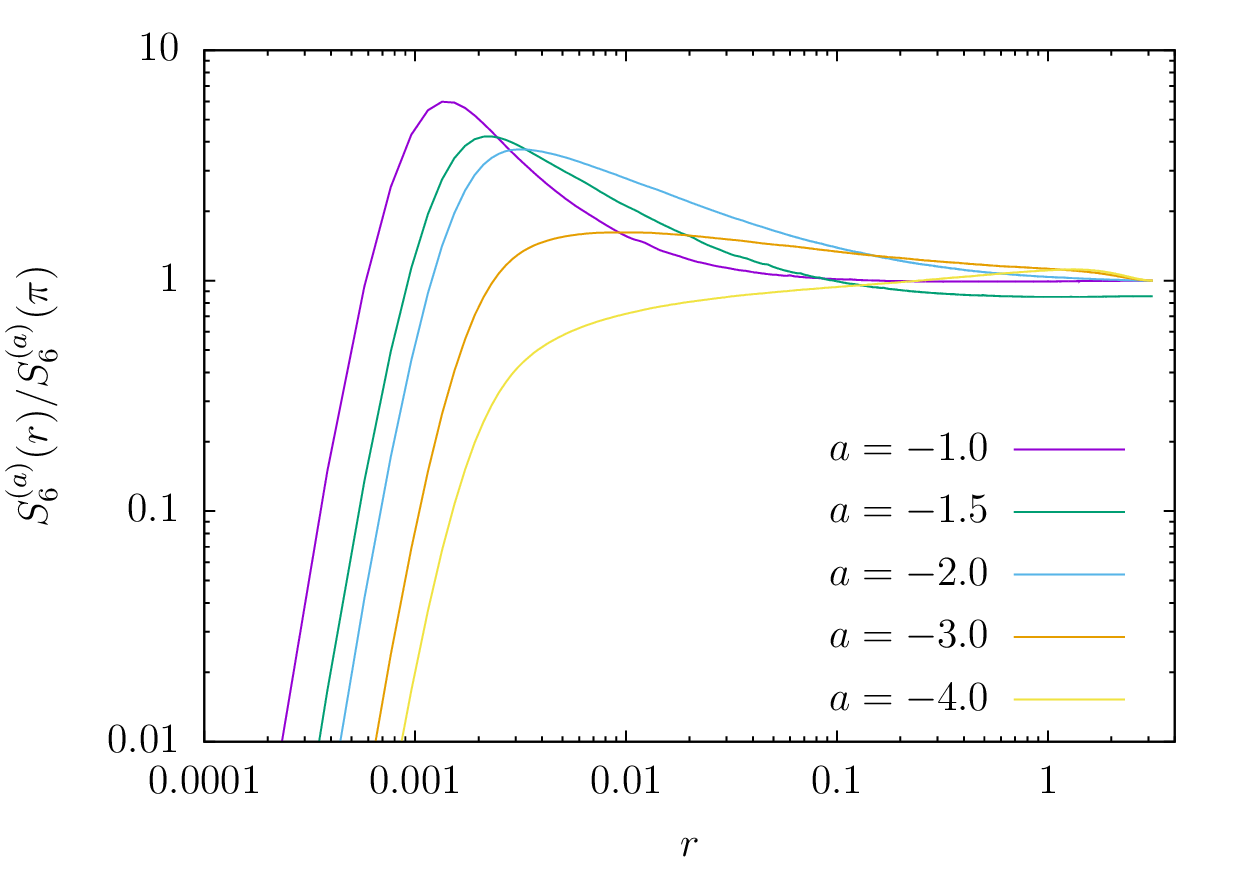}
\includegraphics[scale=0.6]{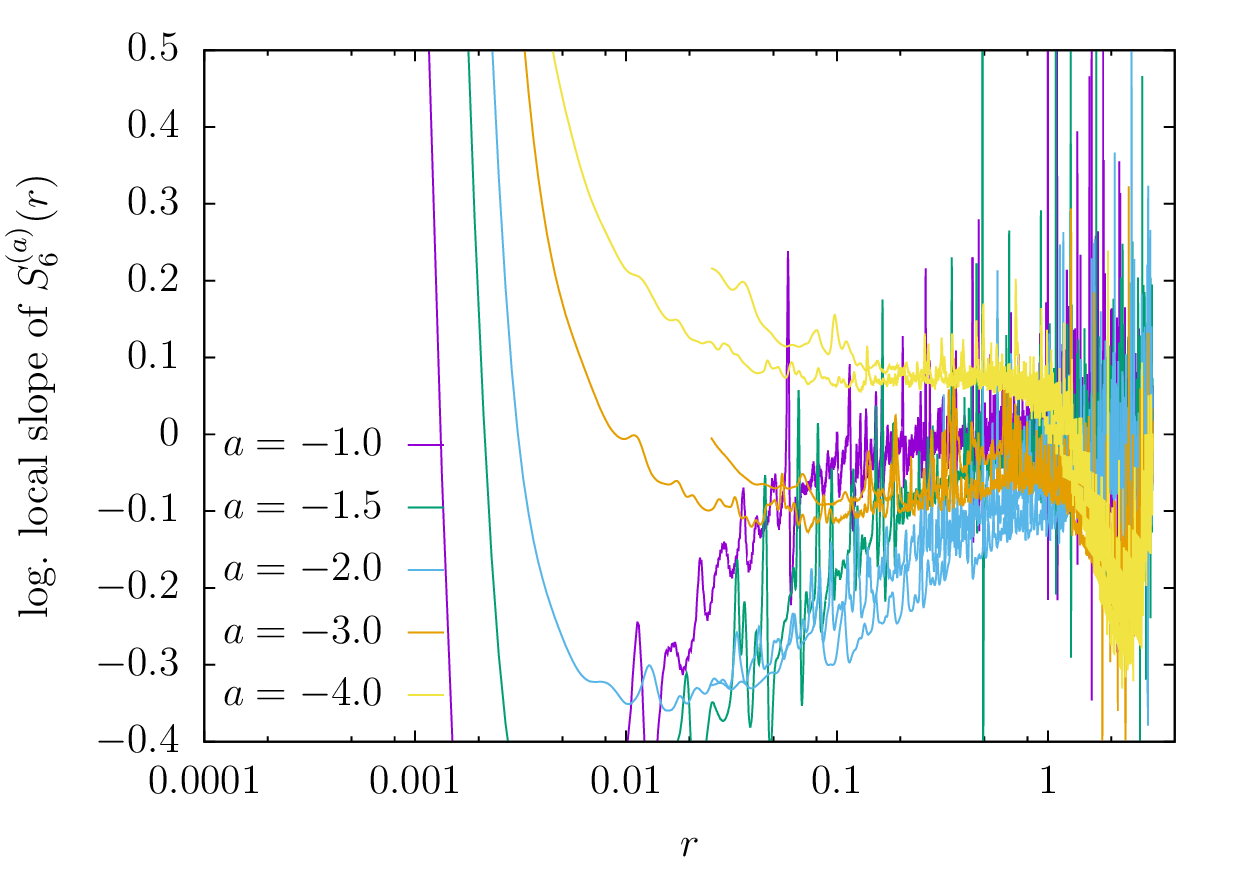}
\caption{\label{stf6}Same as Fig.\ref{stf2} but for $p = 6$.}
\end{figure}

Inevitably, for $p = 4$ and $6$, the local slope becomes noisier since
the high order structure functions are affected by rare events.
Nevertheless in Figs.\ref{stf4} and \ref{stf6},  
a power-law scaling is absent for $a = -1.0$ and $-1.5$
within the inertial range $0.01 \le r \le 0.1$.
In contrast, for $a = -2.0, -3.0$ and $-4.0$,
the local slope appears to be flatter than that of the second order case,
which may suggest a power-law scaling. But $S^{(-2.0)}_6(r)$ is an exception.
However the change from the larger $\nu$ data casts doubt on 
the scaling behavior.

Now let us assume that $S^{(a)}_4(r)$ and $S^{(a)}_6(r)$ are 
power-law functions for $a = -2.0, -3.0$ and $-4.0$. 
Then the scaling exponent of $S^{(a)}_6(r)$ is smaller than that of $S^{(a)}_4(r)$.
This decrease of the exponent of the even-order vorticity structure function implies that the vorticity 
is not bounded as $\nu \to 0$ \cite{f}. 
Moreover the possible scaling exponent of $S^{(a)}_6(r)$ for $a = -2.0$ and $-3.0$ 
is negative.

With the structure function, which is a standard tool to probe
scaling property of turbulence, 
we observed that, from the second order structure function, 
the possible correction to the dimensional-analysis scaling of the gCLMG turbulence 
is not a power-law type. 
From the higher order structure functions, we obtained an indication that 
the vorticity becomes infinite as $\nu \to 0$. Both points are next studied
with a different large-scale forcing.

\section{\label{s:stat}Nonlinear stationary solution under the deterministic forcing}
In this section, 
we change the large-scale forcing to a deterministic and static forcing
\begin{eqnarray}
 f(x, t) = C_0 \sin x,
\end{eqnarray}
where we set $C_0 = -0.1$.
As observed previously for $a = -2.0$ \cite{ms}, 
with the deterministic forcing we obtain a nonlinear stationary 
solution for other $a$'s using the same time stepping method as in the previous section.
Although the numerical solution is not not at all turbulent,
it has interesting properties from which 
we can get insights on the gCLMG turbulence realized under the random 
forcing as we will see.

The numerical method for the stationary solutions is as follows.
Starting from the zero initial condition with $2^{13}$ grid points,
we run the simulation up to $t = 50$ with the time step $\Delta t = 1.25 \times 10^{-4}$.
In this simulation, the 12 digits of the palinstrophy, $\sum_{k} k^4 |\widehat{u}(\vec{k}, t)|^2$, 
stays the same in $30 \le t \le 50$ for the $a = -1.0$ case (whose energy spectrum is the shallowest). 
Then we take the resultant Fourier modes as the initial condition 
of the next simulation with the doubled grid points and the one-half of the time step.
The duration of each simulation of a given resolution is up to $t=50$.
For $a = -1.0$, the viscosity is set to $\nu = 1.0\times 10^{-4}$ with $2^{13}$ grid points
and $\nu = 2.5\times 10^{-5} \times 2^{14-m}$ with $2^{m}$ grid points ($m = 14, \ldots, 18$). 
For $a = -1.5, -2.0, -3.0$ and $-4.0$, it is set to $\nu = 1.0\times 10^{-4} \times 4^{13-m}$ with $2^{m}$ grid points  
($m = 13, \ldots, 18$). 
With the largest number of grid points, $2^{18}$, 
the palinstrophy for the case $a = -1.0$ stays the same value in the 11 digits for $15 < t \le 50$.

\subsection{The energy spectrum}

\begin{figure} 
\includegraphics[scale=0.6]{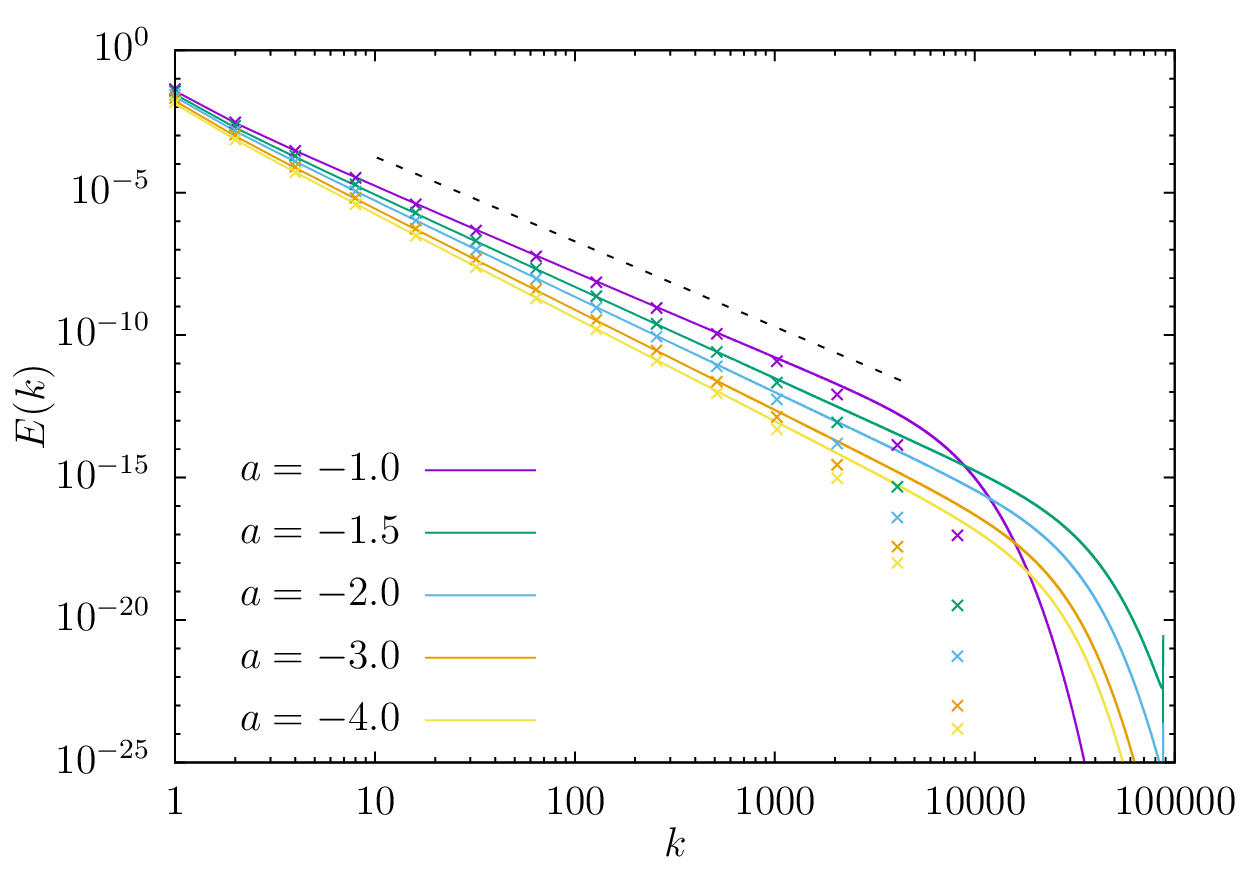}
 \caption{\label{specstat} The energy spectrum $E(k)$ of the stationary solution
 with the largest number of the grid points $2^{18}$.
 The crosses are the spectrum data of the randomly forced case with $2^{15}$ grid points,
 which are shown in Fig.\ref{spc3} but suitably shifted.
 The dotted line is an expression of the spectrum, Eq.(\ref{e}), for $a = -1.0$ with
 parameters shown in Table \ref{numbers}.}
\end{figure}

As shown in Fig.\ref{specstat}, 
the stable nonlinear stationary solution for each $a$ has the energy spectrum
that is indistinguishable in the inertial range 
from the turbulent one realized with the random forcing.
This implies that the possible correction $E(k)$ to the $k^{-3}$ scaling 
can be obtained through analysis of the stationary solution with the deterministic 
forcing. However we do not succeed in such a theoretical analysis so far.

Instead, as we have done in \cite{ms} for the $a =-2.0$ case, we do
a curve fitting to empirically measure the correction. Now we
fit the logarithmic local slope of $E(k)$ with a functional form 
\begin{equation}
 \frac{\log \frac{E(k + \Delta k)}{E(k)}}
  {\log \frac{k + \Delta k}{k}}
  \simeq 
k \frac{d}{dk} \log E(k)
  = -c_0 - c_1 \left[\log \mbox{$\left(\frac{k}{k_f}\right)$}\right]^\delta,
\label{a}
\end{equation}
where $k_f$ is the forcing wavenumber ($k_f = 1$ here).
Notice that $c_0 = 3$ and $\delta = -1$ correspond to the form of the Kraichnan's logarithmic correction, $E(k) \propto k^{-3} [\log(k / k_f)]^{-c_1}$.
For $\delta \ne -1$, integration of Eq.(\ref{a}) leads 
to the expression of the energy spectrum,
\begin{equation}
E(k) \propto k^{-c_0} \exp\left\{- c_2 \left[\log \mbox{$\left(\frac{k}{k_f}\right)$} \right]^{\theta}\right\},
\label{e}
\end{equation}
where $c_2 = c_1 / (1 + \delta)$ and $\theta = 1 + \delta$.

\begin{figure} 
\includegraphics[scale=0.6]{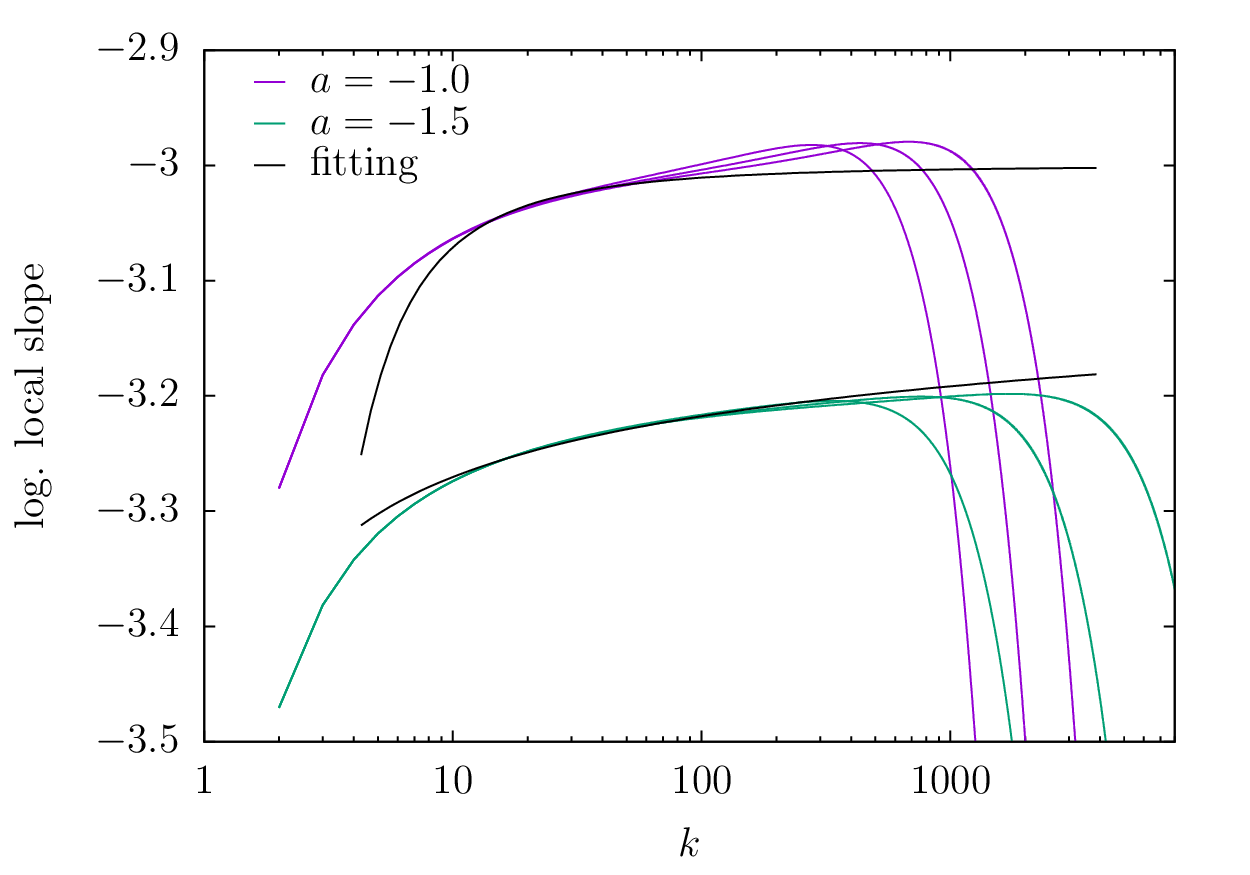}
\includegraphics[scale=0.6]{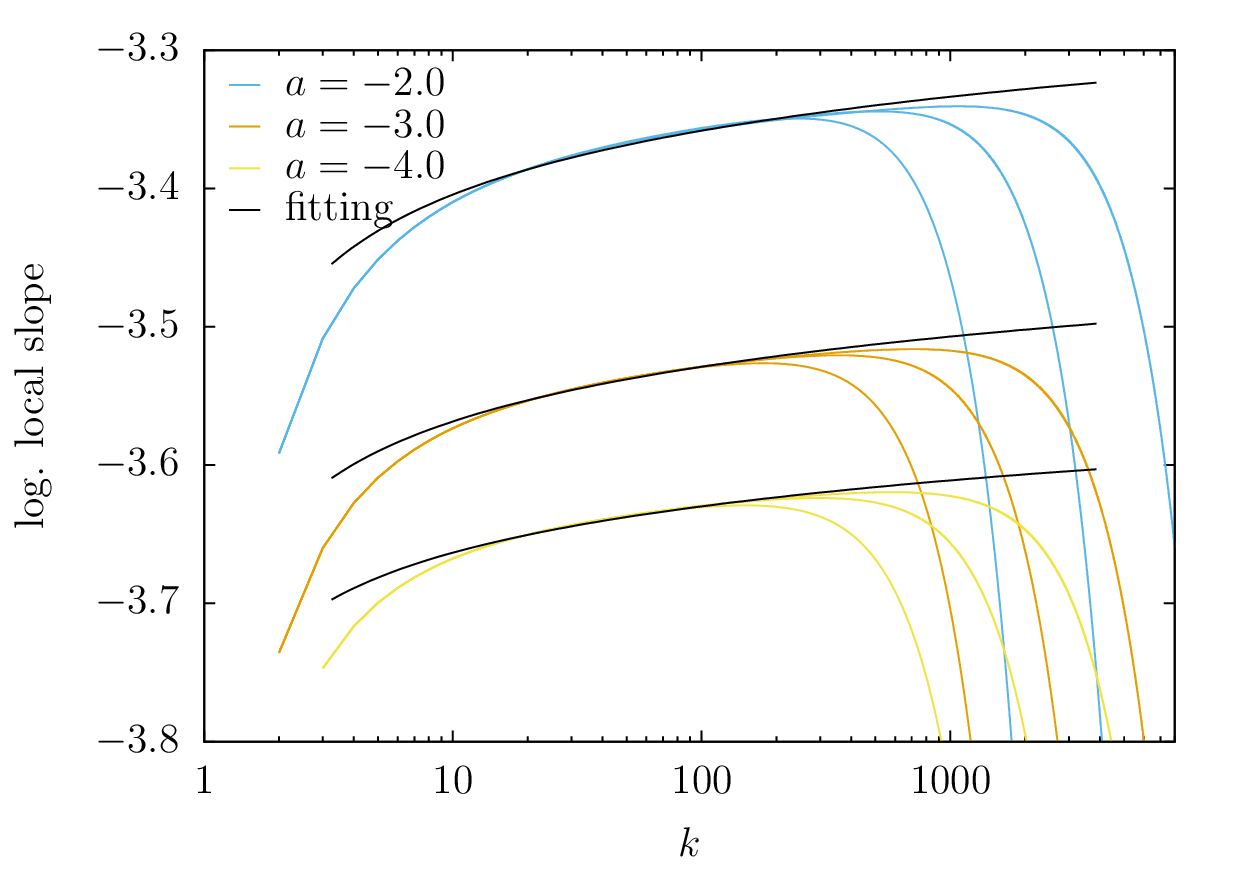}
 \caption{\label{speccook} Logarithmic local slope of the energy spectra
 of the stationary solutions with three different resolutions
 ($2^{16}, 2^{17}, 2^{18}$ grid points). The fitting function is
 Eq.(\ref{a}) with $c_0 = 3.0$.}
\end{figure}

In the fitting, we fix the first parameter as $c_0 = 3$ by assuming that 
the power-law part of $E(k)$ is given by Eq.(\ref{spcform}) since we have
obtained an evidence for the cascade of $C_{a}$ with $a \ge -3$.
For the fitting range, we take a range of $k$ in which the curves of different resolutions overlap
in Fig.\ref{speccook}, namely, $20 \le k \le 200$.
A least square fitting yields parameter values shown in Table \ref{numbers}.
The fitting with $\delta = -1$, which corresponds to the logarithmic corrected 
form, Eq.(\ref{loge}), does not yield a better fit for every $a$. 
For $a = -4.0$, $\theta \approx 1$ suggests that $E(k)$ is simply proportional 
to $k^{-(c_0 + c_2)} = k^{-3.763}$, which is consistent with the non-cascade of the inviscid conservative
quantity. 
The spectrum seems to be well parametrized with the form Eq.(\ref{e}). 
Although Eqs.(\ref{a}) and (\ref{e}) are purely empirical,
the functional form with $c_0 = 3$ and $\theta = -1$ can be obtained 
theoretically with the incomplete self-similarity analysis, which is given in Appendix A.

\begin{figure} 
\includegraphics[scale=0.6]{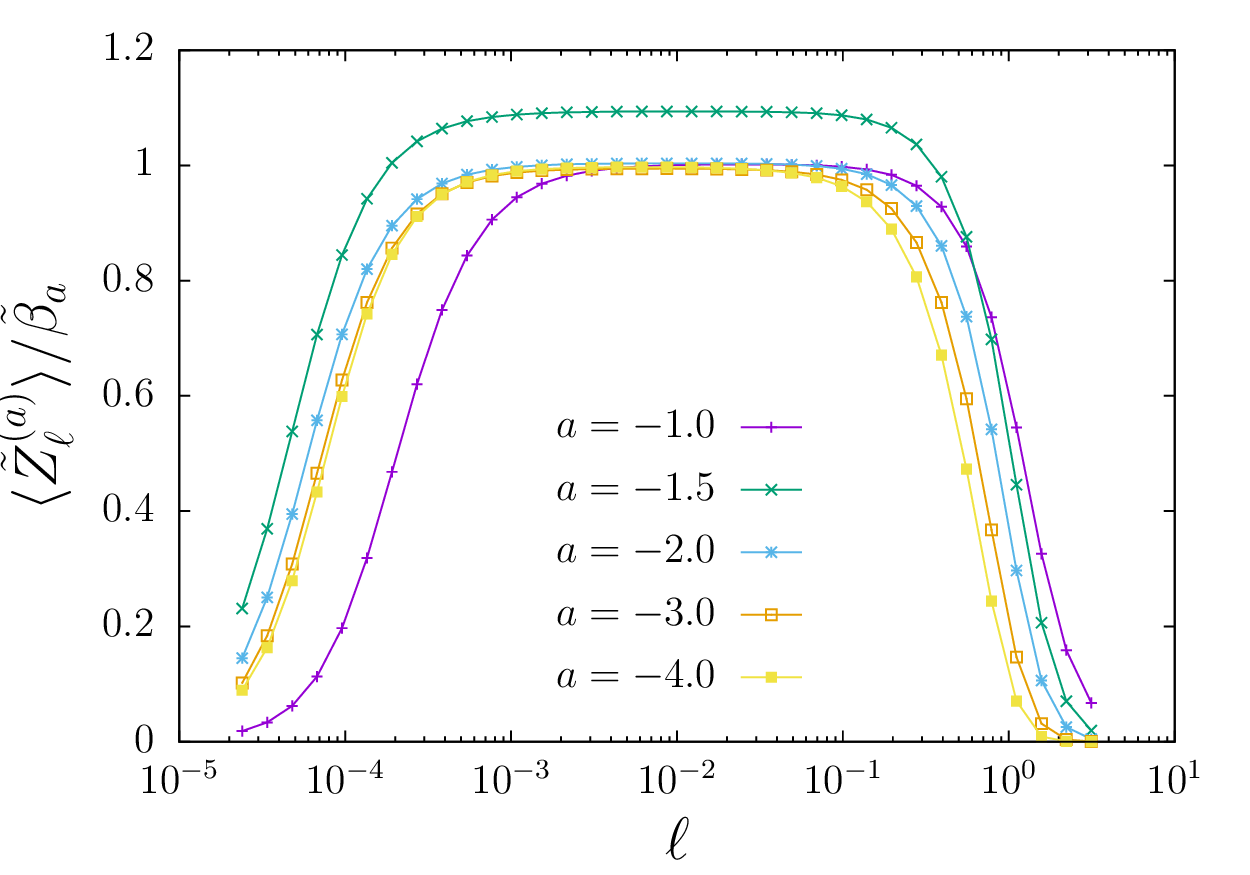}
\caption{\label{statcascade}Spatially averaged flux $\langle \tilde{Z}_\ell^{(a)}\rangle$ for the stationary solution normalized with the dissipation rate $\tilde{\beta}_a$. Here the number of grid points is $2^{18}$. Recall that the $a = -1.0$ case has a larger viscosity.}
\end{figure}

We now formally calculate the spatially averaged flux, $\langle \tilde{Z}_\ell^{(a)} \rangle$,
of the stationary solution. In spite of the inhomogeneity, we do this in order to 
look at nature of the nonlinear equilibrium.
The averaged flux as a function of $\ell$ exhibits the $\ell$-independent range
as shown in Fig.\ref{statcascade}. This supports the assumption of $c_0 = -3.0$ made
in the fitting of the energy spectrum.
For $a = -1.5$ case, the value of the plateau is about 10\% larger than 
the dissipation rate $\tilde{\beta}_{-1.5}$ of the stationary solution (notice that such discrepancy is
not seen in the randomly forced case, see Fig.\ref{fluxhalf}).
In other cases the differences are less than 1\%.
Apart from this discrepancy, we observe that the steady state is maintained with the constant flux, 
which is again a similarity to the turbulent solution.

\begin{table}
 \begin{tabular}{c|cc|cc}
  $a$   & $c_1$    & $\delta$    & $c_2$      & $\theta$  \\ \hline
 $-1.0$ & $0.699$  &  $-2.74$   &   $0.402$ &  $-1.74$  \\
 $-1.5$ & $0.351$  &  $-0.313$   &   $0.511$ &  $0.687$  \\
 $-2.0$ & $0.468$  &  $-0.175$   &   $0.567$  &  $0.825$  \\
 $-3.0$ & $0.620$  &  $-0.104$   &   $0.691$  &  $0.896$  \\
 $-4.0$ & $0.706$  &  $-0.0747$  &   $0.763$  &  $0.925$  \\
 \end{tabular}
 \caption{\label{numbers}Parameter values in Eq.(\ref{a}) determined
 by a least square fit.
 The values of $c_2$ and $\theta$ are calculated with
 $c_2 = c_1 / (1 + \delta), \theta = 1 + \delta$.
 For the $a = -2.0$ case, the values here are different
from those obtained in \cite{ms} ($c_1 = 0.442, \delta = -0.138$) due to difference 
in the fitting range.}
\end{table}

\subsection{\label{vp}Vorticity pulse and blowup of the inviscid limit}

\begin{figure} 
\includegraphics[scale = 0.6]{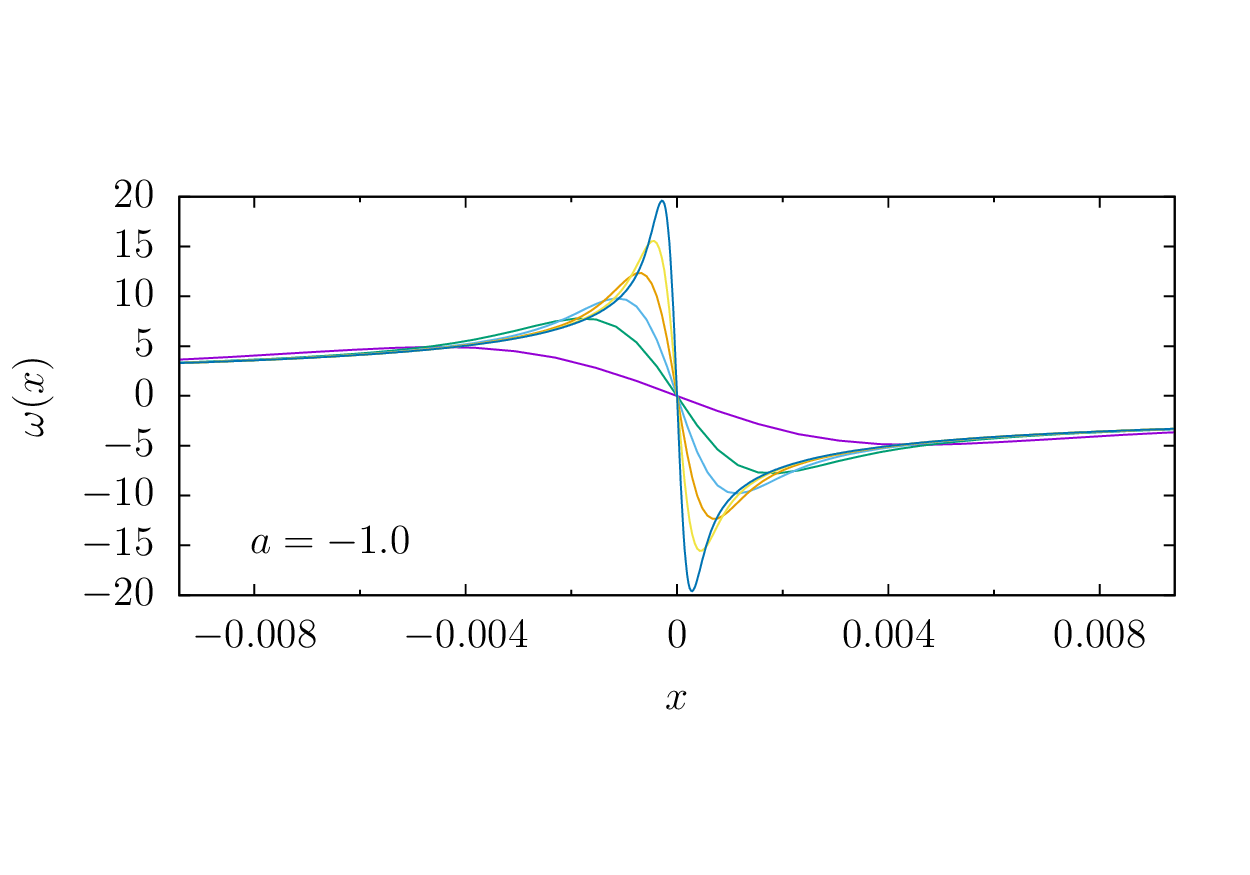}
\includegraphics[scale = 0.6]{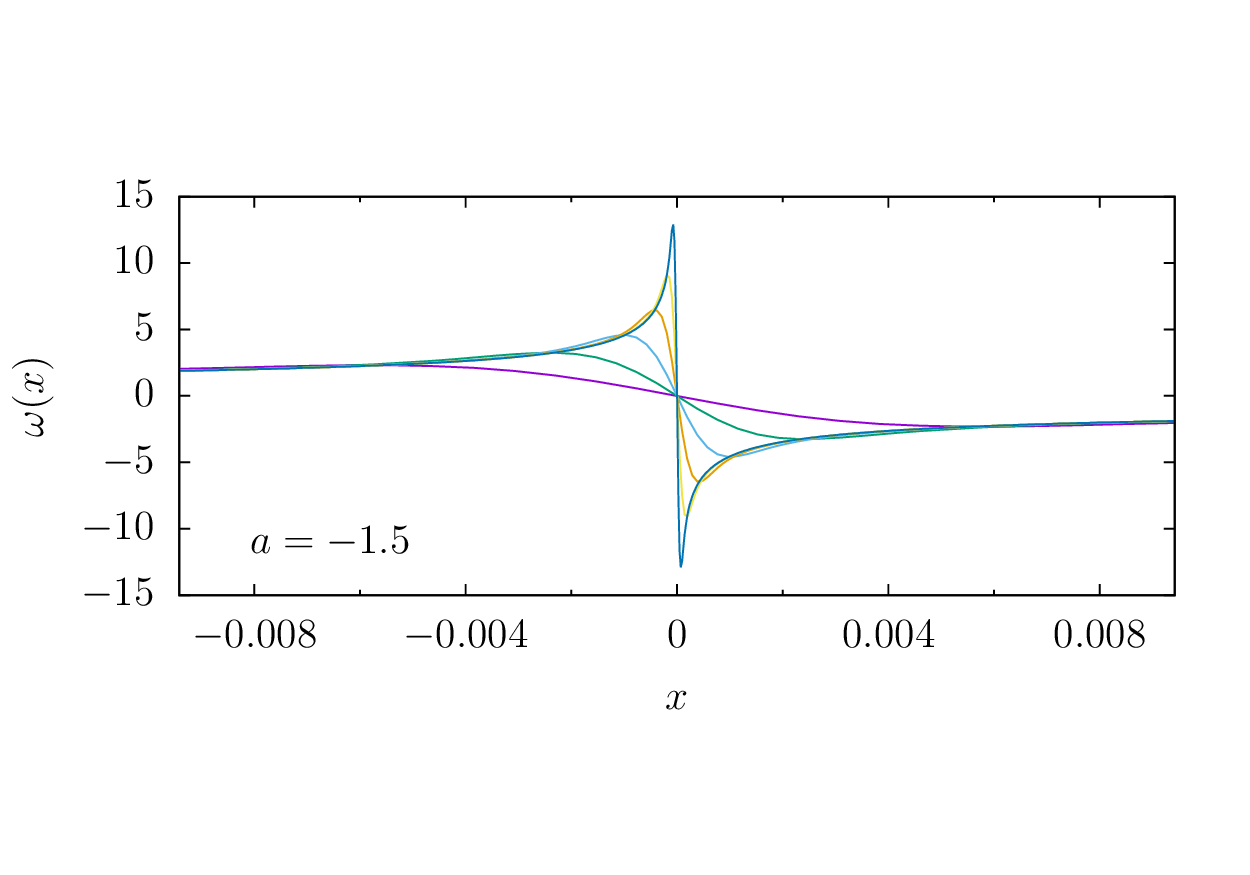}
\includegraphics[scale = 0.6]{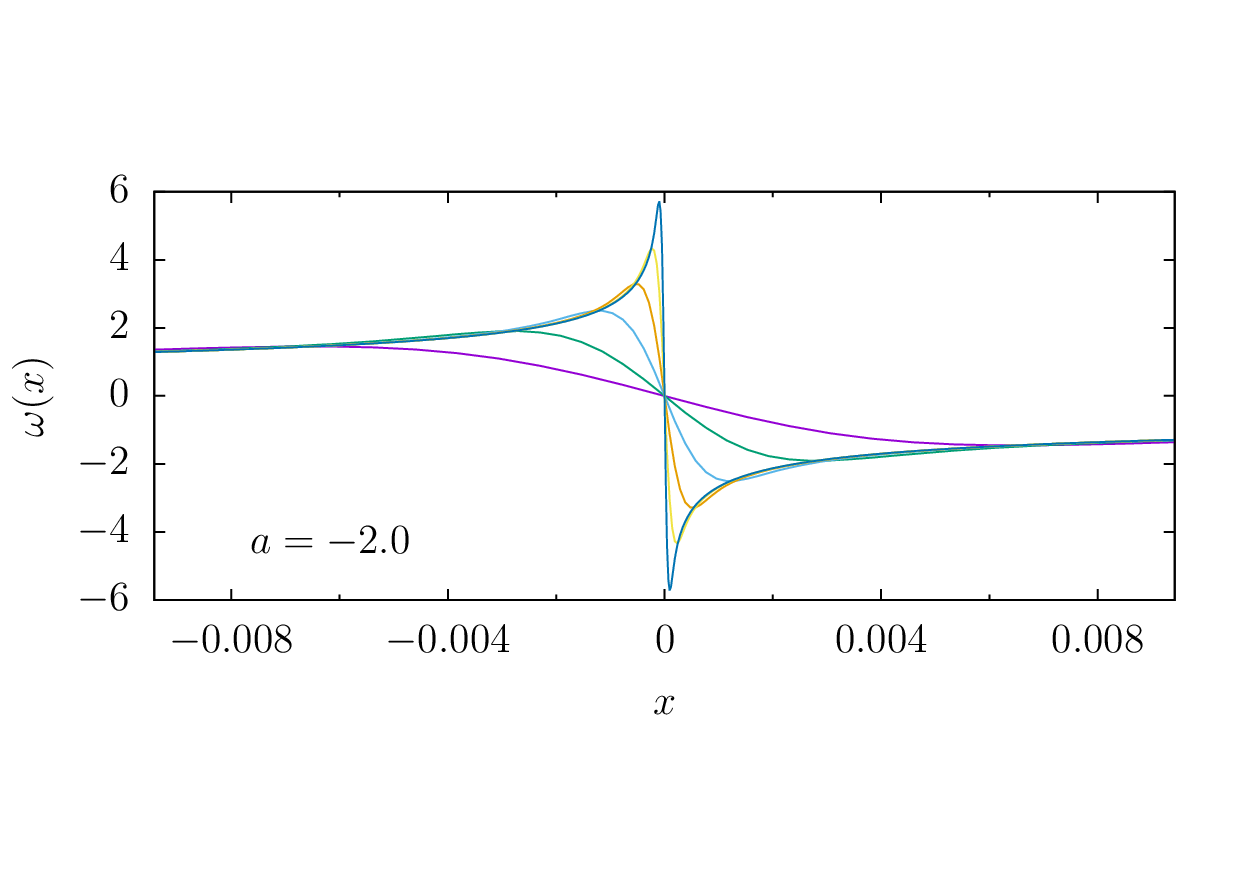}
\includegraphics[scale = 0.6]{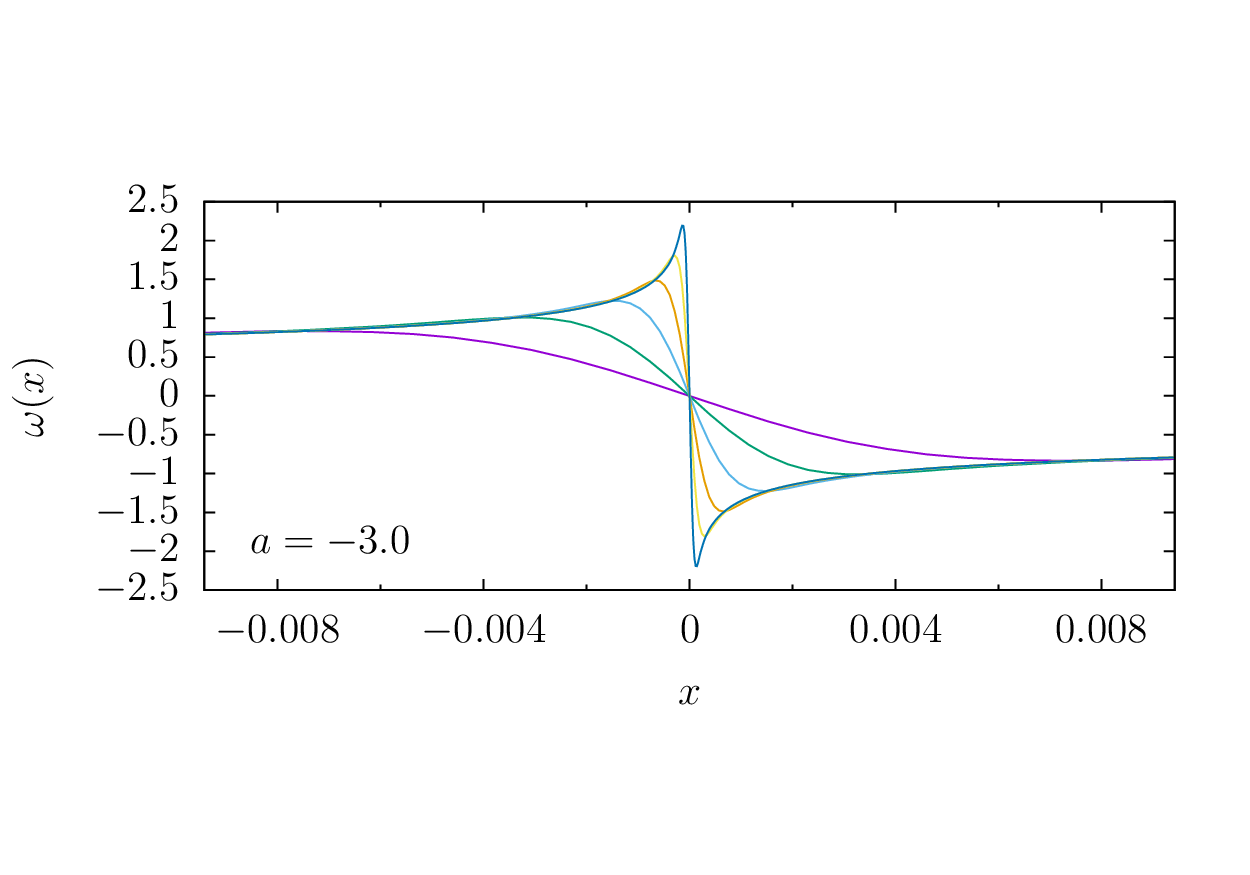}
\includegraphics[scale = 0.6]{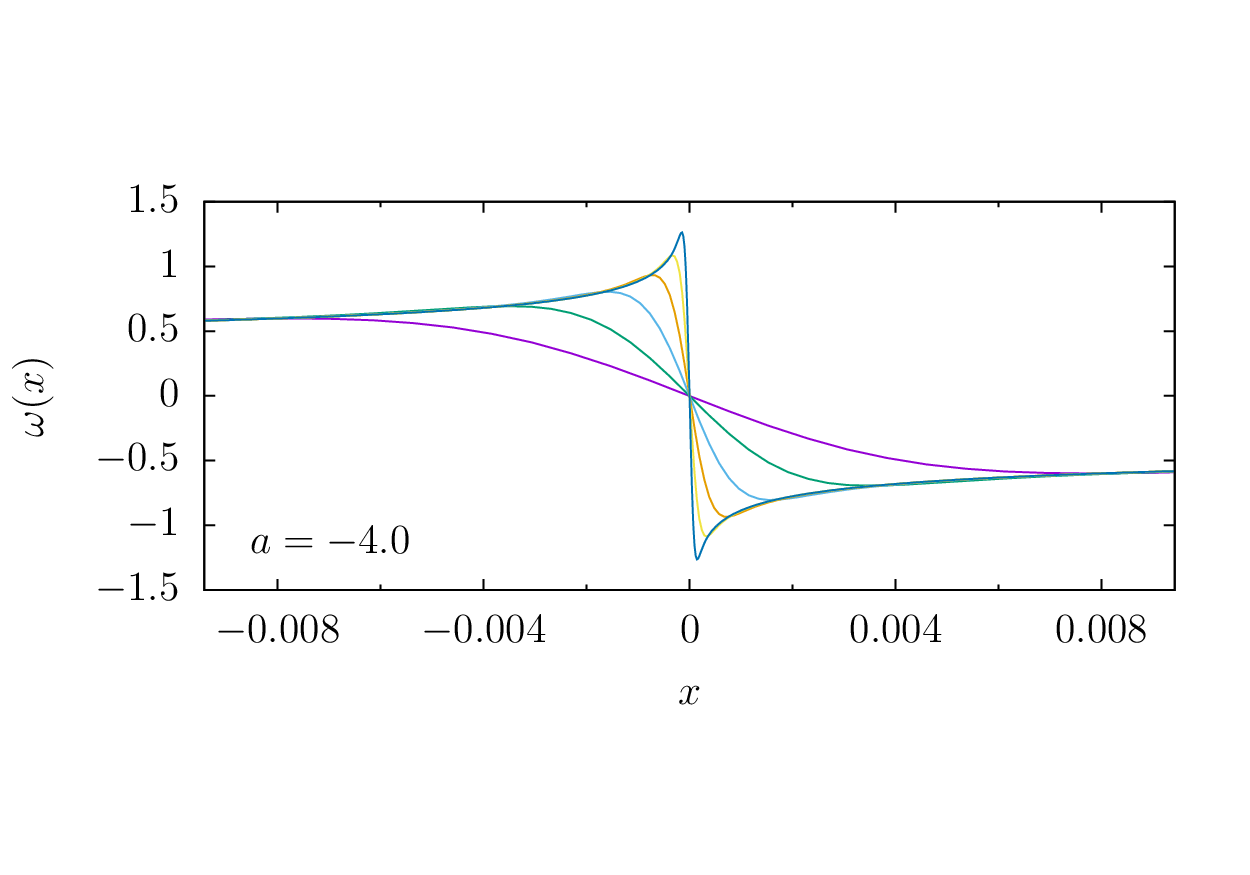}
\caption{\label{profraw} The vorticity pulse of the stationary solution around the origin. Each panel has profiles of six different values of the viscosity $\nu$.}
\end{figure}

The vorticity profile of the stationary solution for each $a$ consists
of a single pulse around the origin. A magnified view of the pulse
for each $a$ is shown in Fig.\ref{profraw}. The energy spectrum 
of the stationary solution studied in the previous subsection is a result
of this single pulse. The question is now whether we can relate 
the form Eq.(\ref{e}) in the inertial range with the pulse profile in the physical space. 
Probably we cannot do so because the width of the pulse belongs to the dissipation range. 
Rather it is likely to be related with the profile far from the 
pulse, namely how the vorticity decreases from the pulse peaks.

\begin{figure} 
\includegraphics[scale = 0.6]{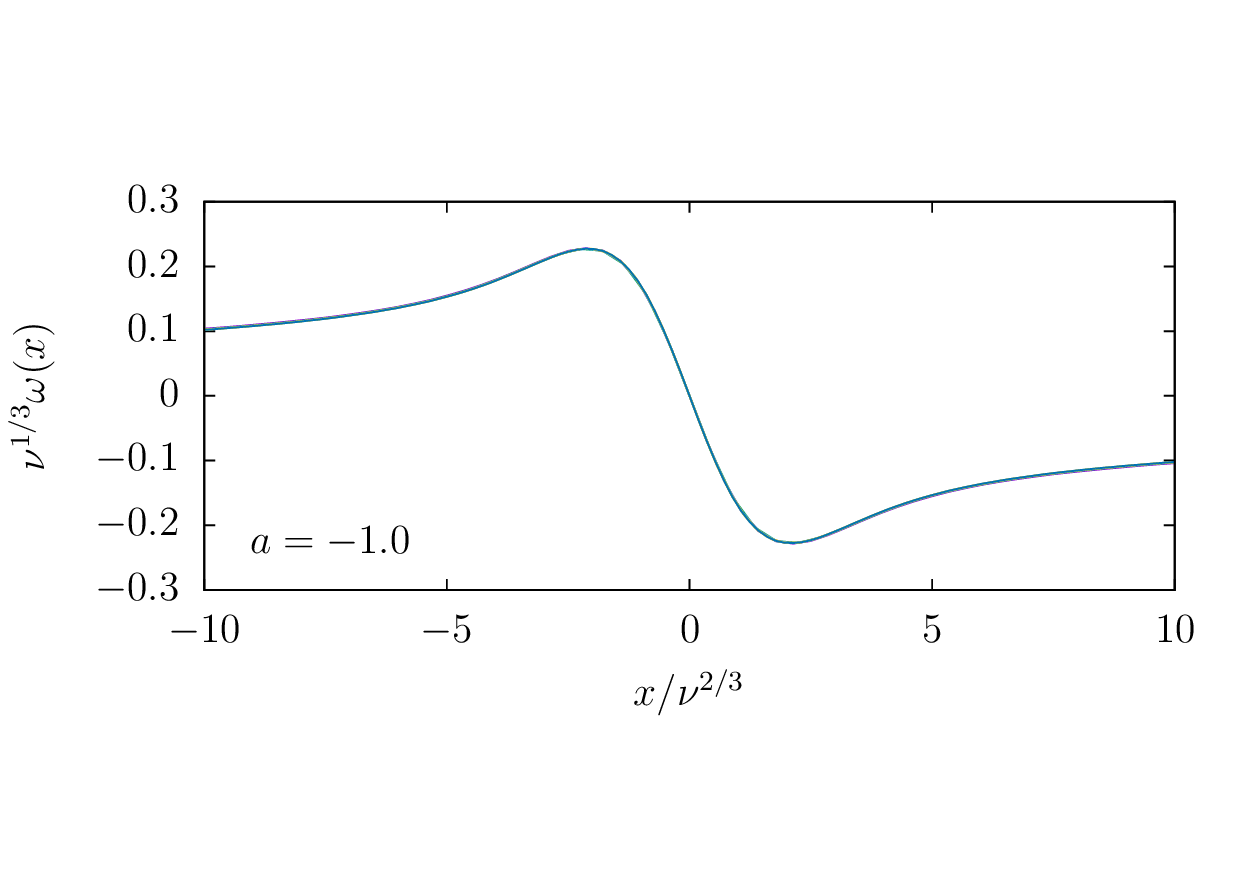}
\includegraphics[scale = 0.6]{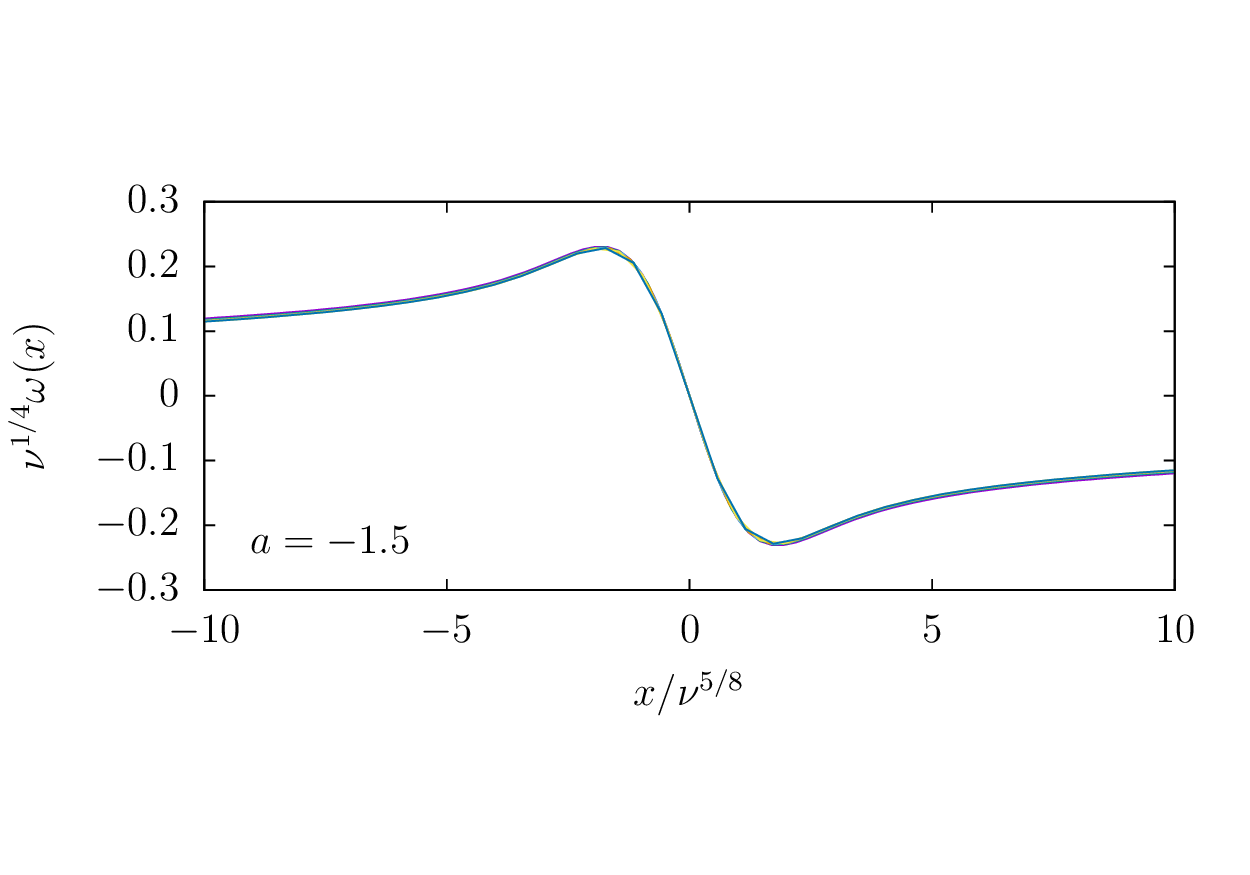}
\includegraphics[scale = 0.6]{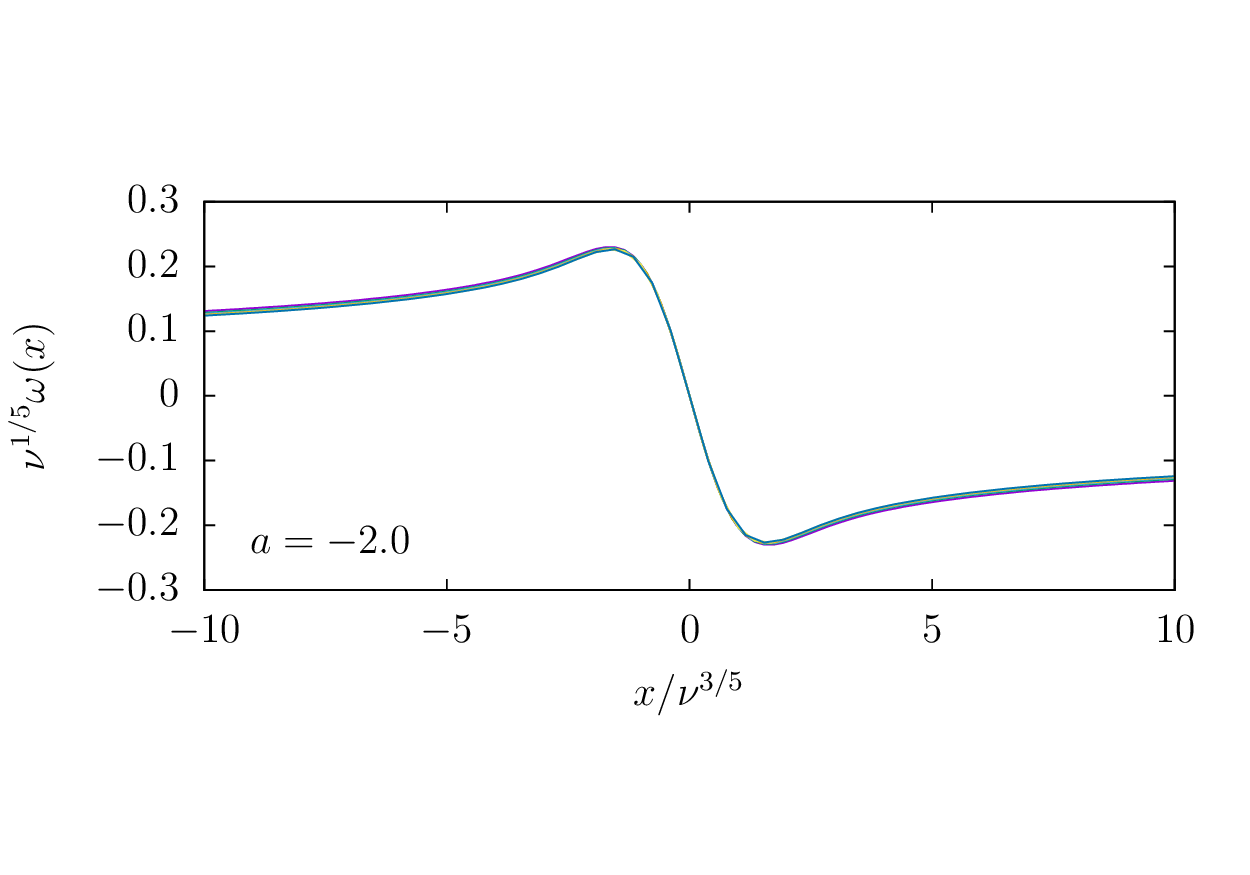}
\includegraphics[scale = 0.6]{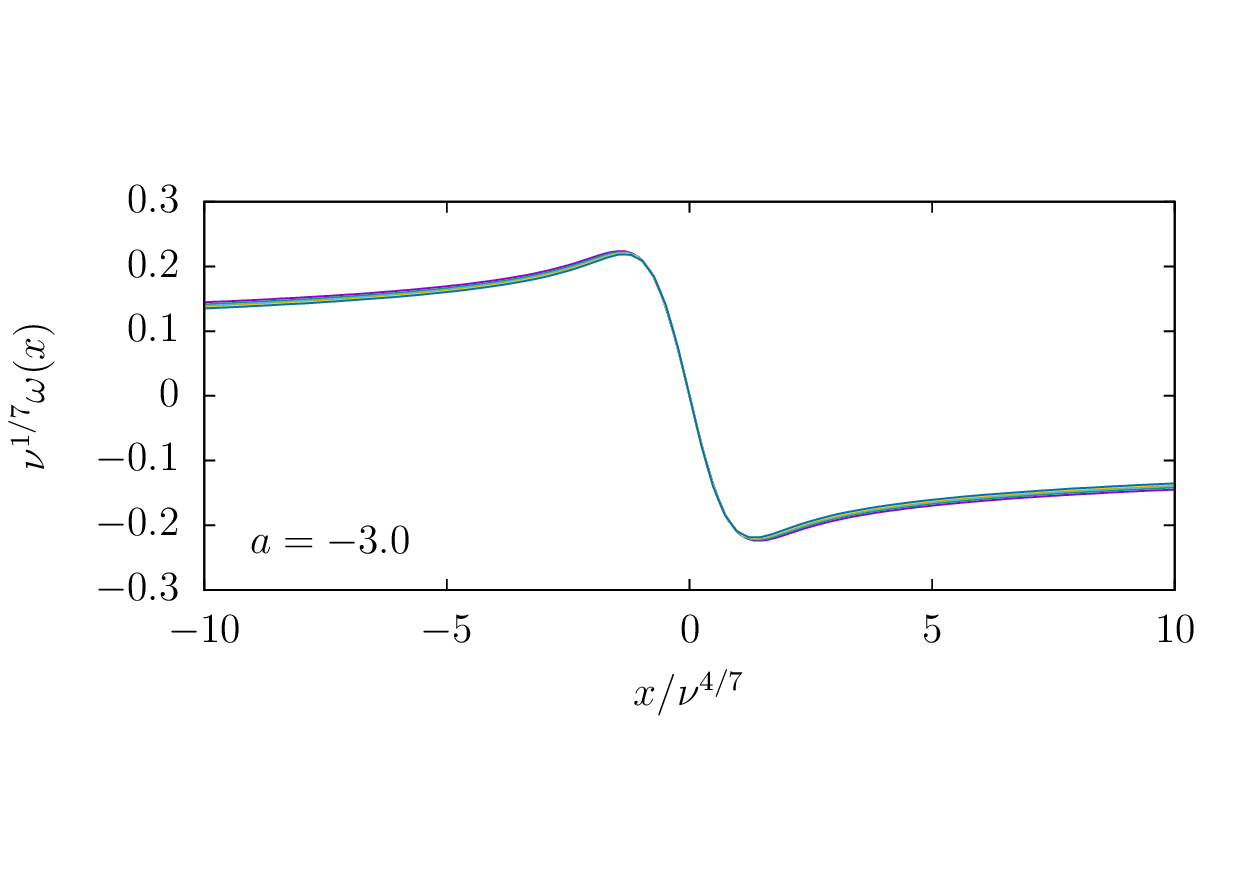}
\includegraphics[scale = 0.6]{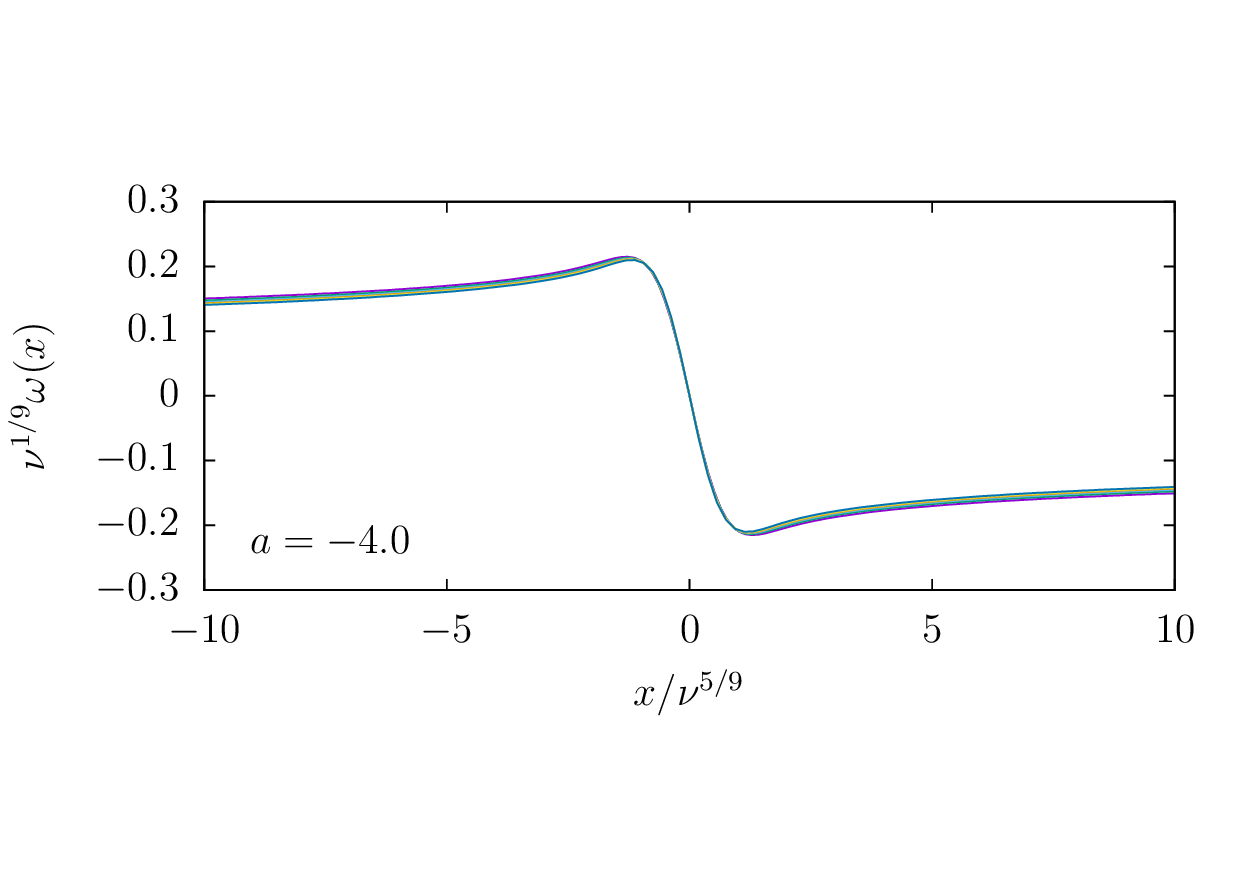}
\caption{\label{profscale} Scaled vorticity pulse of the stationary solution around the origin, $\nu^{-\alpha}\omega(\nu^{\mu} x)$,
with $\alpha = - 1/(1 - 2a)$ and $\mu = (1 - a) / (1 - 2a)$. Each panel has profiles of six different values of the viscosity $\nu$.}
\end{figure}

Nevertheless, we study now the pulse profile closely because
it is related to the constant ($\nu$-independent) dissipation rate $\beta_{a}$.
As observed in Fig.\ref{profraw}, the height and width of the
pulse is affected by the viscosity $\nu$. These profiles with
various values of the viscosity are found to collapse to a single curve when
the vorticity $\omega(x)$ is scaled as $\nu^{-\alpha}\omega(\nu^{\mu} x)$,
where $\alpha = - 1/(1 - 2a)$ and $\mu = (1 - a) / (1 - 2a)$. 
The scaled profiles are shown in Fig.\ref{profscale}.
This viscous scaling implies that, for $a < 0$, the vorticity becomes 
infinite,  $\omega(x) \to \nu^{\alpha} ~(\alpha < 0)$, as $\nu \to 0$.

Now we argue that the exponents of the viscous scaling can 
be determined with a boundary-layer type analysis as $\nu \to 0$.
Let us first assume that the width of the pulse is given by $\delta = \nu^{\mu}$.
In the stretched coordinate $X = x / \delta$,
we assume that the solution is
\begin{equation}
 \omega(x) = \nu^{\alpha} \Omega(X), \quad u(x) = \nu^{\gamma} U(X).
\end{equation}
Then the stationary gCLMG eq. with the stationary forcing, 
\begin{equation}
 a u \partial_x \omega = \omega \partial_x u + \nu \partial_x^2 \omega + C_0 \sin x,
\end{equation}
becomes 
\begin{equation}
a\frac{\nu^{\alpha + \gamma}}{\delta} U \partial_X \Omega 
= 
\frac{\nu^{\alpha + \gamma}}{\delta} \Omega \partial_X U  
+ \frac{\nu^{1 + \alpha}}{\delta^2}  \partial_X^2 \Omega 
+ C_0 \sin (\delta X).
\label{beq}
\end{equation}
Here we assume that the dominant balance in Eq.(\ref{beq}) holds
between the nonlinear term and the viscous term, resulting in
\begin{equation}
 \gamma + \mu = 1.
\label{er1}
\end{equation}
We further assume the scaling relation $\omega(x) \sim u(x) / \delta$, which
gives
\begin{equation}
 \alpha - \gamma + \mu = 0.
\label{er2}
\end{equation}
(the same scaling relation can be obtained through analysis of the Hilbert 
transform $\partial_x u(x) = H(\omega)(x)$).
Lastly, we assume that the dissipation rate of the invisid conserved 
quantity is $\nu$ independent as $\nu \to 0$. 
The dissipation rate inside the pulse may be written as
\begin{eqnarray}
\beta_{a} \sim
\nu^{1 - \alpha a - \mu}
(-a - 1)
\int_{-\infty}^{\infty}
 \Omega^{-a - 2}
(\partial_X \Omega)^2
 dX.
\end{eqnarray}
Hence we have a relation
\begin{equation}
 1 - a \alpha - \mu = 0.
\label{er3}
\end{equation}
The three relations, Eqs.(\ref{er1}), (\ref{er2}) and (\ref{er3}),
yield the exponents as a function of $a$:
\begin{equation}
 \alpha = -\frac{1}{1 - 2a}, \quad
 \gamma = -\frac{a}{1 - 2a}, \quad
 \mu =  \frac{1 - a}{1 - 2a}.
\label{ve}
\end{equation}
These exponents indeed scale well the numerical solutions as seen
in Fig.\ref{profscale} 
(not only $\omega(x)$ but also $u(x)$ and $H(\omega)(x)$, figures
not shown).
The last assumption of the independence of the dissipation rate on $\nu$
is not trivial \cite{f,e250} but the result suggests that it
is plausible including for the $a = -4.0$ case, where 
the turbulent cascade does not take place with the random forcing. 

The width of the stationary pulse scales with the viscosity as 
$\delta \propto \nu^{\mu} = \nu^{(1 - a)/(1 - 2a)}$. 
In contrast, the Kolmogorov-Kraichnan dissipation length scale of the gCLMG turbulence is
$\eta_a \propto \beta_a^{1/[2(a - 1)]} \nu^{1/2}$, 
which is the unique combination of the dissipation rate $\beta_a$ and the viscosity $\nu$ 
having the dimension of length.
The two viscous length scales have different scaling exponents of $\nu$ except for $a \to -\infty$. 
This difference can be due to the fact that we determine the viscous scaling
by considering only the ``boundary layer''. It does not involve matching with
the ``outer layer'' which corresponds to the inertial range. Recall that matching between
the inertial and the dissipation ranges is the way to obtain the Kolmogorov-Kraichnan dissipation 
scale.

To see which viscous length scales is more relevant with respect to the spectrum
in the dissipation range, we scale the enstrophy spectra of the randomly forced cases
with $1/\delta$ and $1/\eta_a$ for different $\nu$'s (for the energy spectra, good collapse 
is not obtained for both viscous scales).
A better collapse in the dissipation range for the enstrophy spectra
is observed with $1/\delta$. This indicates that $\delta$ is more relevant in the dissipation range 
than $\eta_a$ for the turbulent cases under the random forcing.
Additionally we observe numerically that $E(k)$ of the stationary solution 
in the dissipation range decreases exponentially with the form
$\exp(- c k))$, with some constant $c$.

The ``inner solution'' shown in Fig.\ref{profscale} is a solution
to the nonlinear and nonlocal equations
\begin{eqnarray}
  a U \partial_X \Omega 
&=& 
 \Omega \partial_X U  
+  \partial_X^2 \Omega,
\label{ieqlimit}
\\
U(X) &=& 
\frac{1}{\pi} 
\int_{-\infty}^{\infty} 
 \Omega(Y) 
\log |X - Y| dY,
\label{bolimit}
\end{eqnarray}
which we are not able to solve so far. 
Recall that the solution to the equations above 
may not be sufficient to determine the inertial-range properties
which require the outer solution.

\subsection{Energy spectra of inviscid blowup solution and stationary solution}
Given the indication of the blowup of the vorticity of the stationary solution
as $\nu \to 0$, comparison with the inviscid solution is of next interest.
It is proven in \cite{cc10} that an inviscid solution of the gCLMG eq. 
without a forcing term blows up in a finite time for $a < 0$ in an unbounded domain.

Now we compare the energy spectrum of the stationary solution to 
that of the inviscid gCLMG eq. without any forcing term starting
from the initial condition $\omega(x) = 0.1 \sin x$. Notice that 
the inviscid-limit case $\nu \to 0$ is different from the inviscid 
case $\nu = 0$.

The inviscid and forceless gCLMG eq. is numerically solved with the same spectral 
method as in Sec.\ref{s:random}. 
The inviscid solution does not become stationary and its Fourier modes in ever higher
wavenumbers are generated in the course of time.
With the finite resolution we hence should stop the numerical simulation at some
time before the Fourier modes at the largest truncation wavenumber becomes 
larger than the filtering threshold (recall that we set the vorticity Fourier
modes to zero if their magnitudes are smaller than the threshold value $10^{-12}$).

\begin{figure} 
\includegraphics[scale=0.6]{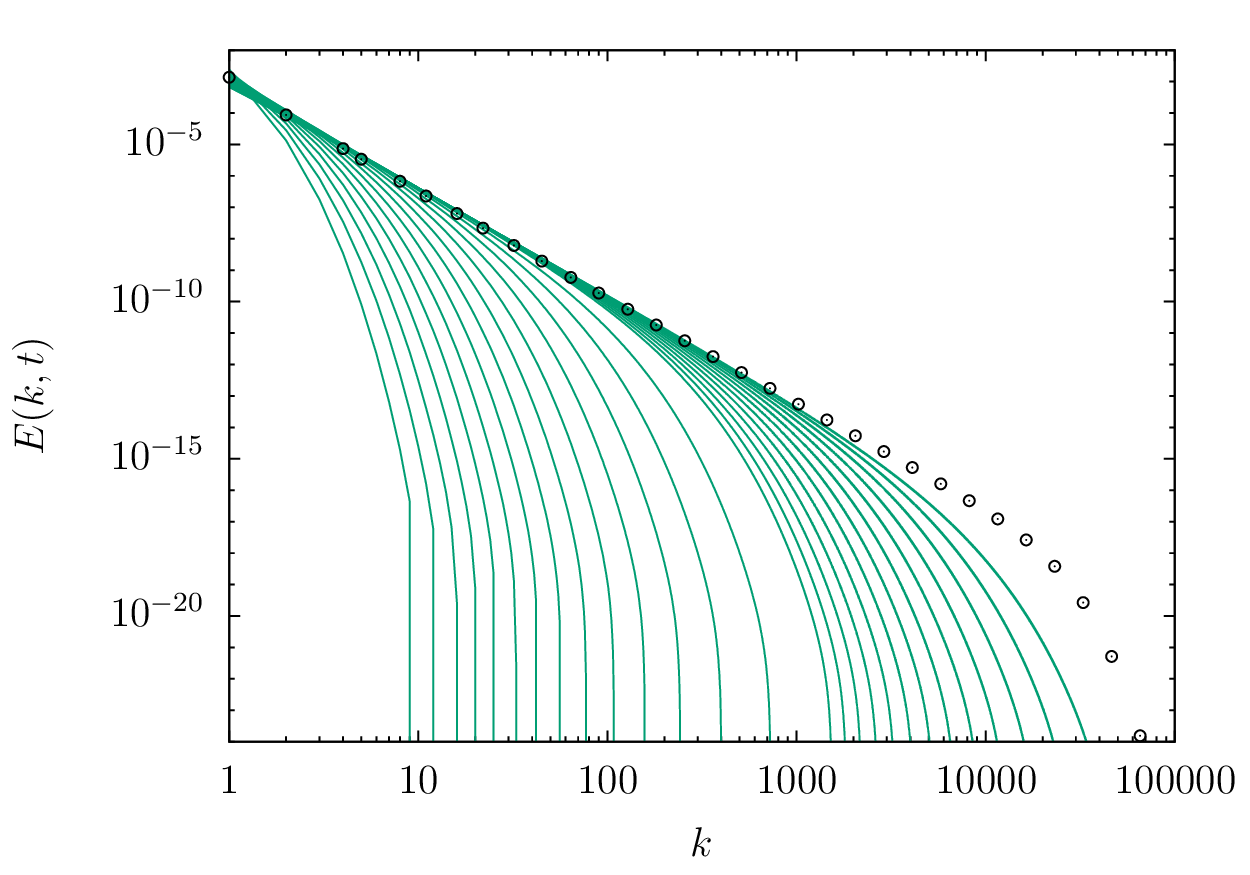} 
 \caption{\label{invspcevolv} Time evolution of the energy spectrum $E(k, t)$ of the inviscid and unforced gCLMG eq. with $a = -2.0$
 at $t = 1.0, 1.5, 2.0, \ldots, 7.5$ and $8.0, 8.1, 8.2, \ldots, 9.2$.
The circles are the energy spectrum of the viscous stationary solution under the deterministic forcing 
for $a = -2.0$ calculated with $2^{18}$ grid points (the same one shown in Fig.\ref{specstat}).
The inviscid numerical solution is calculated with $2^{18}$ grid points and the time step $\Delta t = 7.81\times10^{-6}$. }
\end{figure}

\begin{figure} 
\includegraphics[scale=0.6]{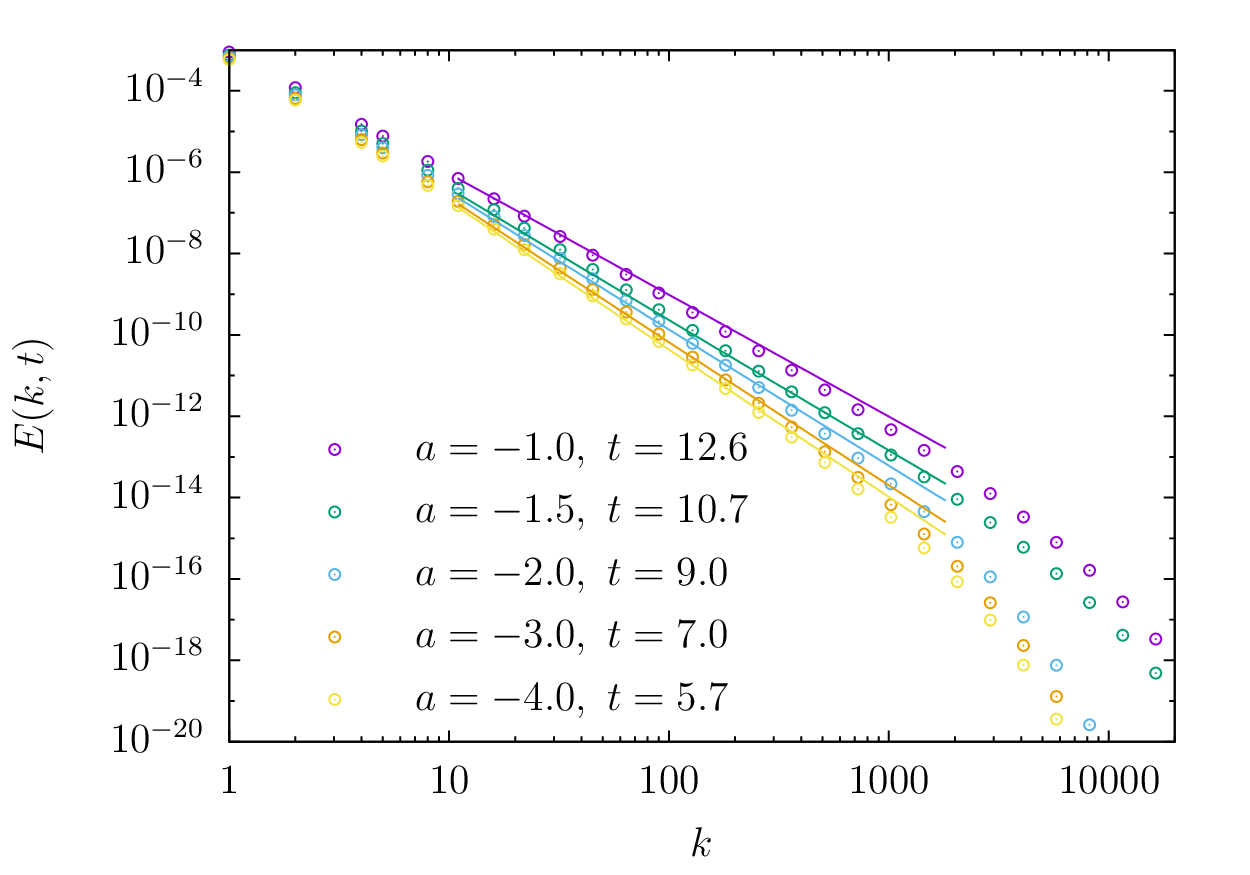}
 \caption{\label{invspc} Instantaneous energy spectrum $E(k, t)$ of the inviscid and unforced gCLMG eq. for various $a$'s.
Here the inviscid numerical solution is calculated with $2^{18}$ grid points and the time step $\Delta t = 7.81\times10^{-6}$.
The solid line corresponds to the parametrization of $E(k)$ of the stationary solution, Eq.(\ref{e}),
with the parameter values in Table \ref{numbers}.}
\end{figure}

The time evolution of the energy spectrum obtained numerically is shown
in Fig.\ref{invspcevolv}. 
The functional form of $E(k, t)$ does not change 
in the intermediate wavenumber range which corresponds to the inertial range of
the turbulent solution.
It is remarkable that the functional form of $E(k, t)$
in this range is quite close to that of the stationary solution
for each $a$ case as seen in Fig.\ref{invspc}. 
This implies that the inviscid $E(k, t)$ in the intermediate range is the same form 
as that of the randomly forced case in the inertial range as well.
Such an agreement is not found in numerical solutions to the Euler and 
the Navier-Stokes equations in the 3D space (see, e.g.,\cite{bb}).
As long as we run the simulation,  $E(k, t)$ in the dissipation range decreases 
exponentially.

In the physical space, the vorticity profile of the invisid solution looks quite similar to that of the stationary
solution with the deterministic forcing except that the inviscid pulse becomes sharper and sharper 
as the time elapses. Its scaling analysis is done in Appendix B.


\section{\label{s:selfsimilar}Self-similarity of the phase-space orbit}

We showed that, depending on the large-scale forcing, the gCLMG eq. has two
classes of solutions: the turbulent one under the random forcing and the stationary
one under the deterministic static forcing. The resemblance of the energy spectra
of the two described in the previous section indicates that  the turbulent
solution is somehow fluctuating around the stationary solution. 
This point is now examined
through a visualization of the phase-space orbit. 
As in \cite{ms}, we consider the following 3D projection of the phase space:
\begin{eqnarray}
\vec{X}(\vec{\kappa}, t) &=& (X_1(k_1, t), X_2(k_2, t), X_3(k_3, t)),\\
 X_j(k_j, t) &=& \frac{{\rm Im}~\tilde{\omega}_T(k_j, t)}{{\rm Im}~\tilde{\omega}_S(k_j, t)} - 1 \quad (j = 1, 2, 3),
\end{eqnarray}
where we take two triplets of the wavenumbers $\vec{\kappa} = (k_1, k_2, k_3)$, 
in the inertial range as powers of two, $\vec{\kappa}_1 = (4, 8, 16)$ and $\vec{\kappa}_2 = (32, 64, 128)$.
Here ${\rm Im}~\tilde{\omega}_T(k_j, t)$ denotes the imaginary part of the vorticity Fourier coefficient
of the turbulent solution normalized by the square root of the mean energy, 
$\tilde{\omega}_T(k, t) = \widehat{\omega}(k, t) / \langle E \rangle^{1/2}$. 
Notice that the counterpart of the stationary solution, $\tilde{\omega}_S(k, t)$, is purely imaginary.

\begin{figure} 
\centerline{%
\includegraphics[scale=0.11]{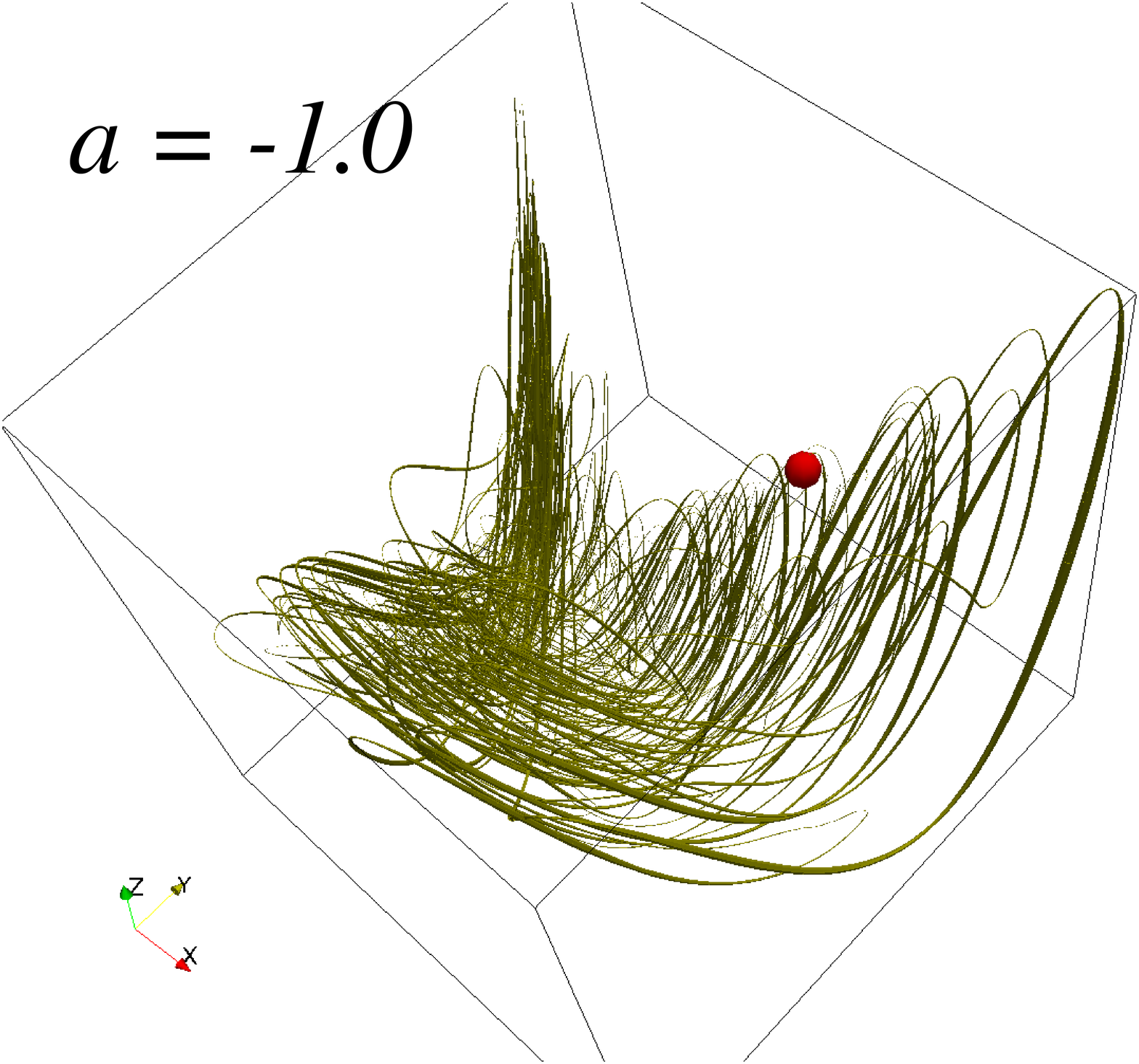}%
\includegraphics[scale=0.11]{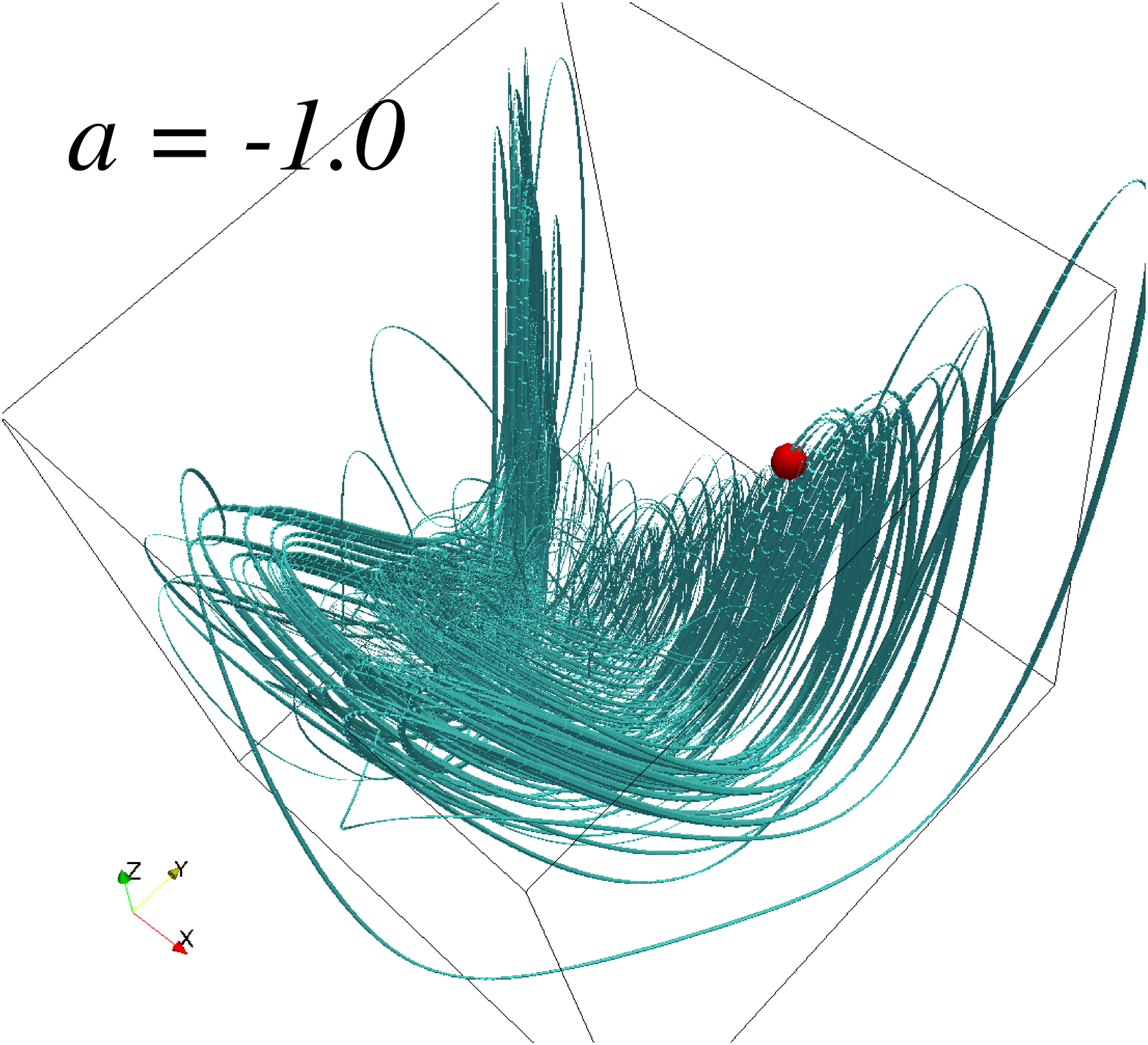}}
\centerline{%
\includegraphics[scale=0.11]{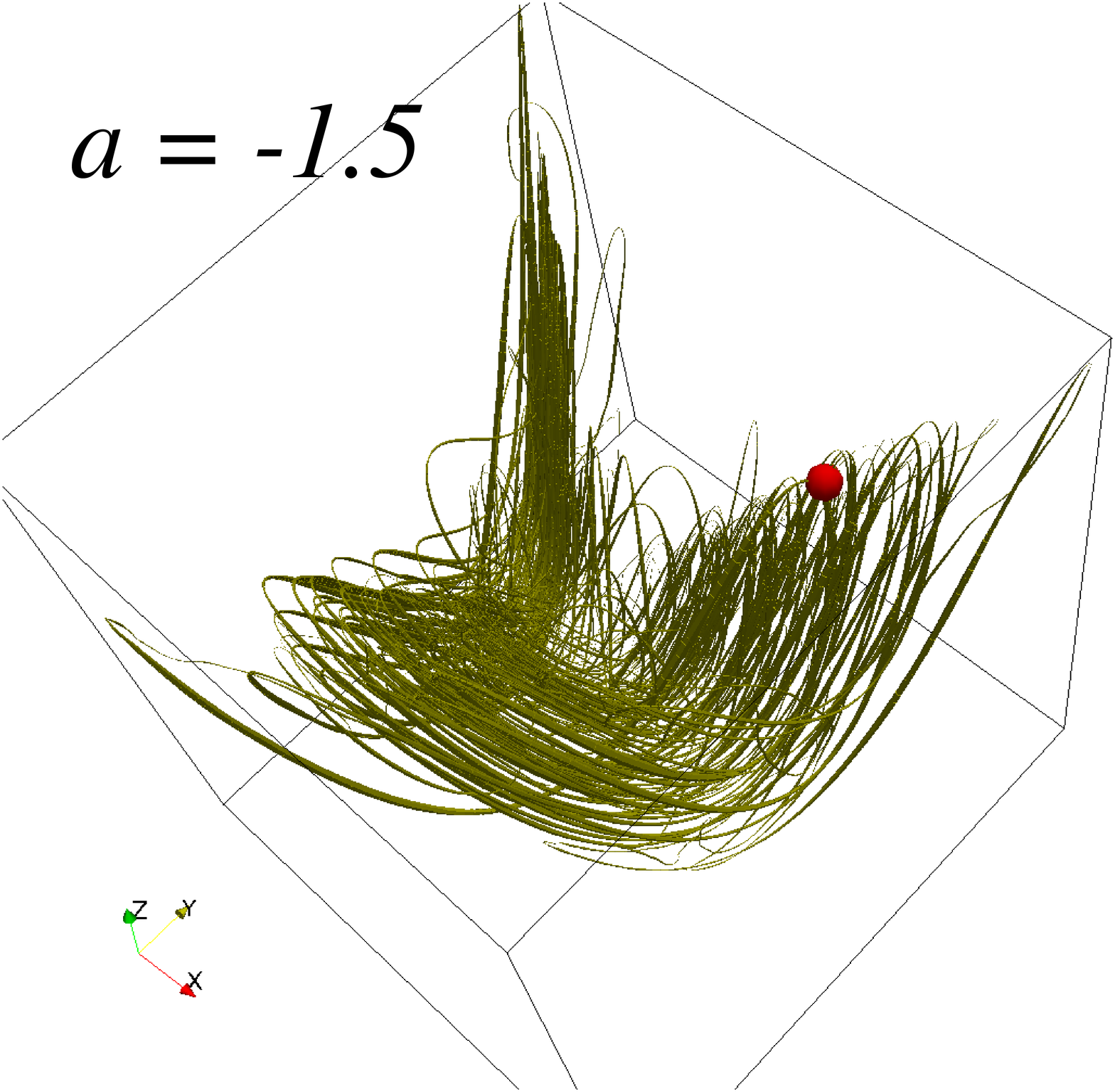}%
\includegraphics[scale=0.11]{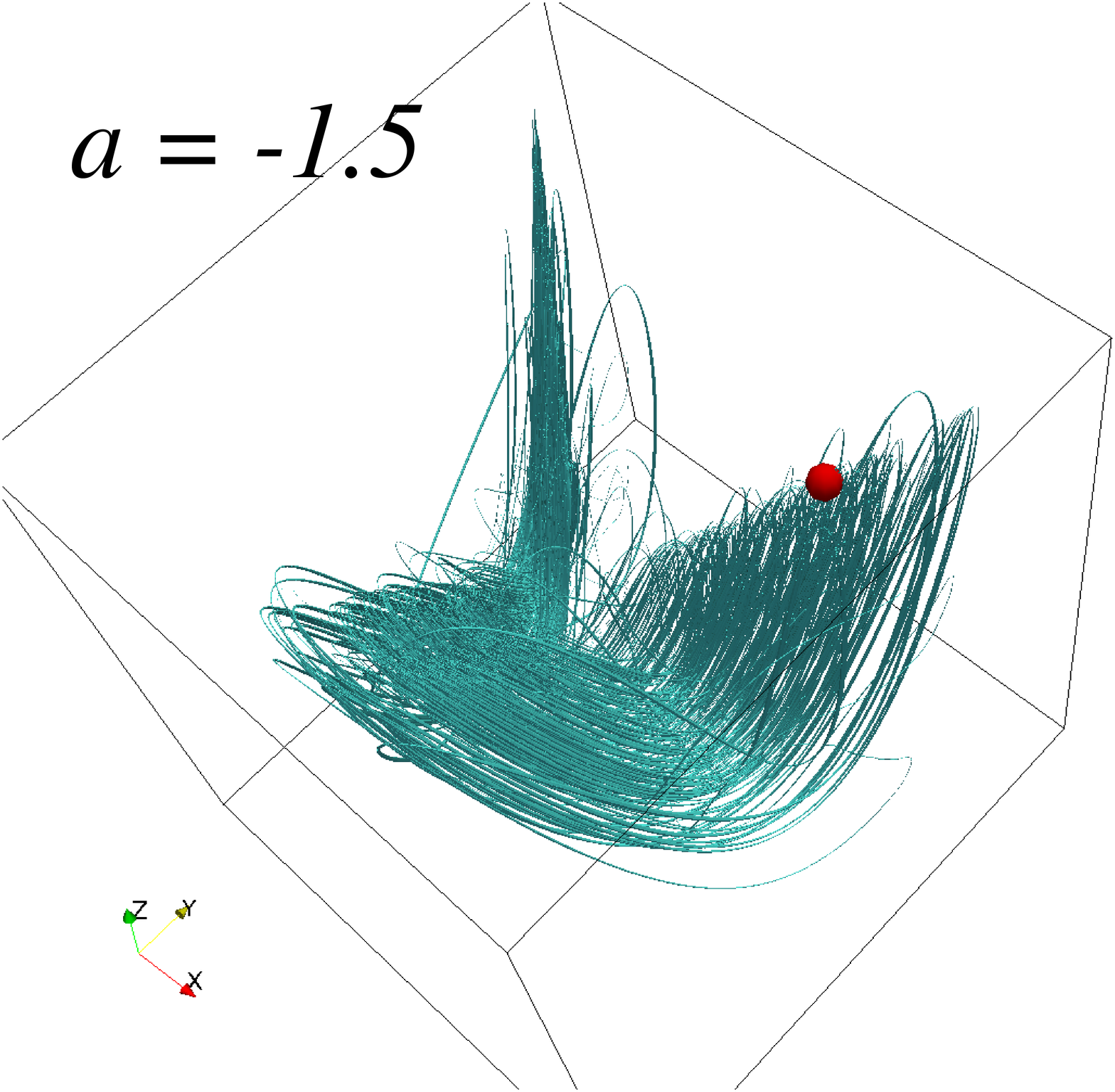}}
\centerline{%
\includegraphics[scale=0.11]{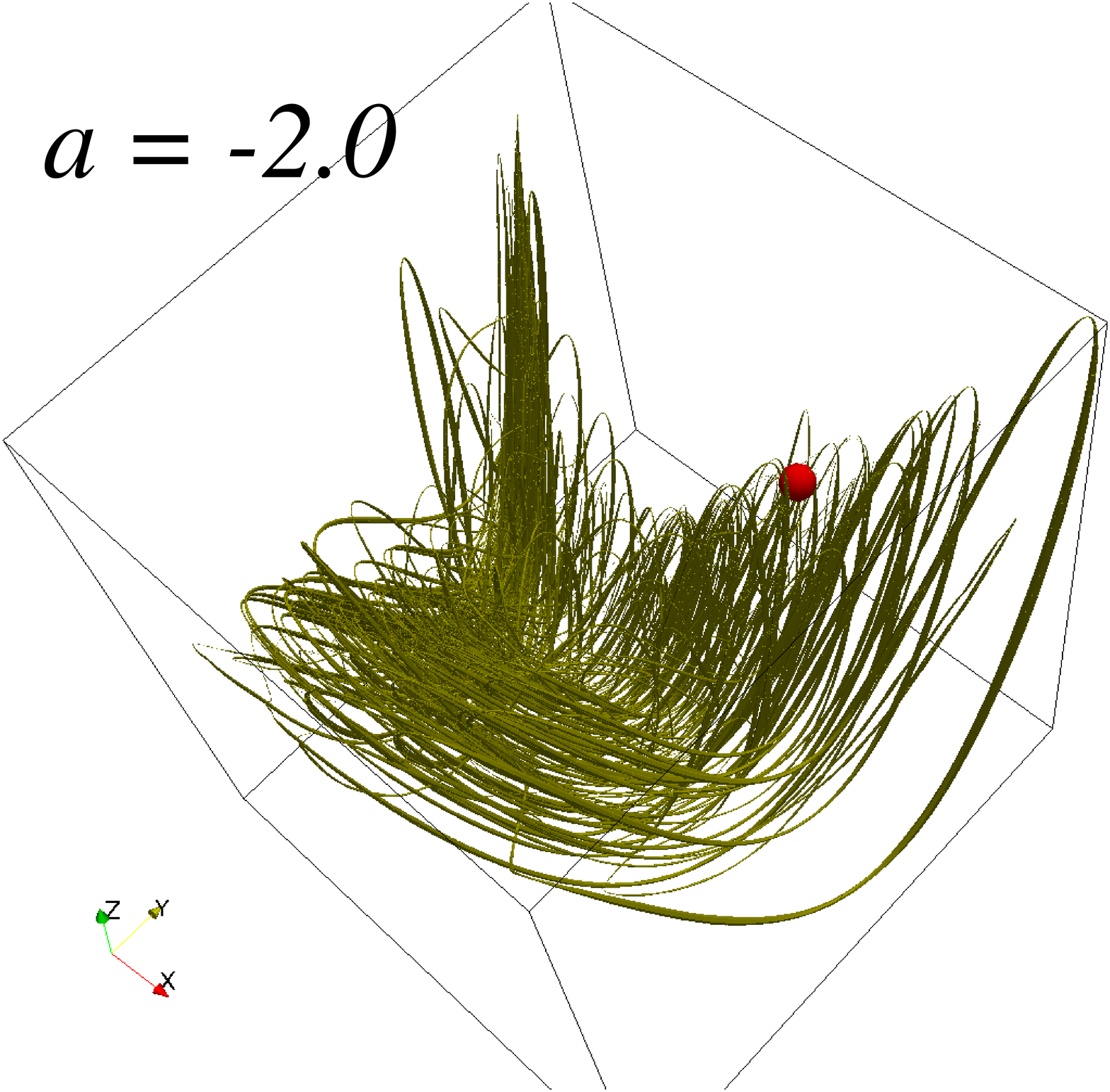}%
\includegraphics[scale=0.11]{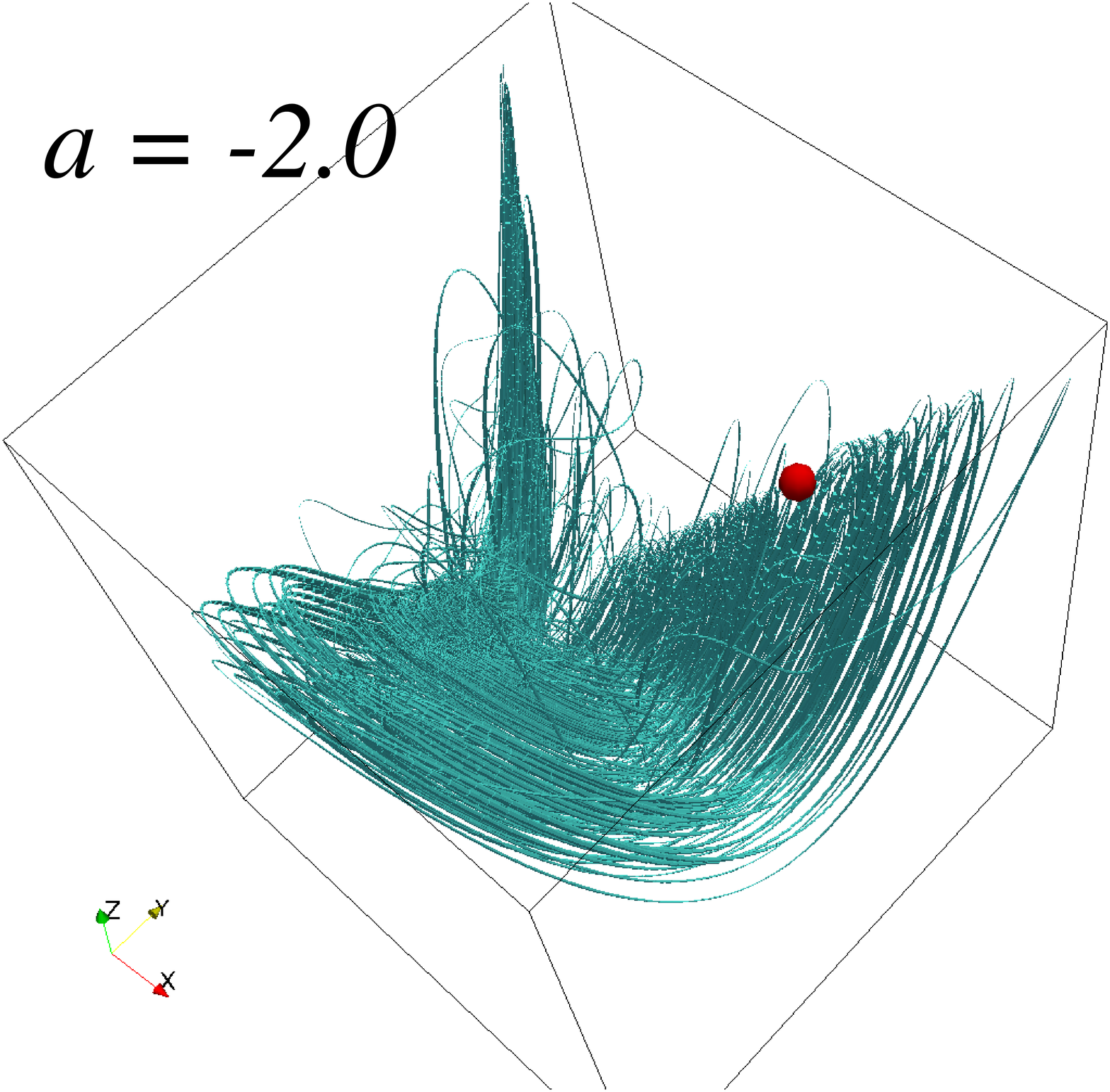}}
\centerline{%
\includegraphics[scale=0.11]{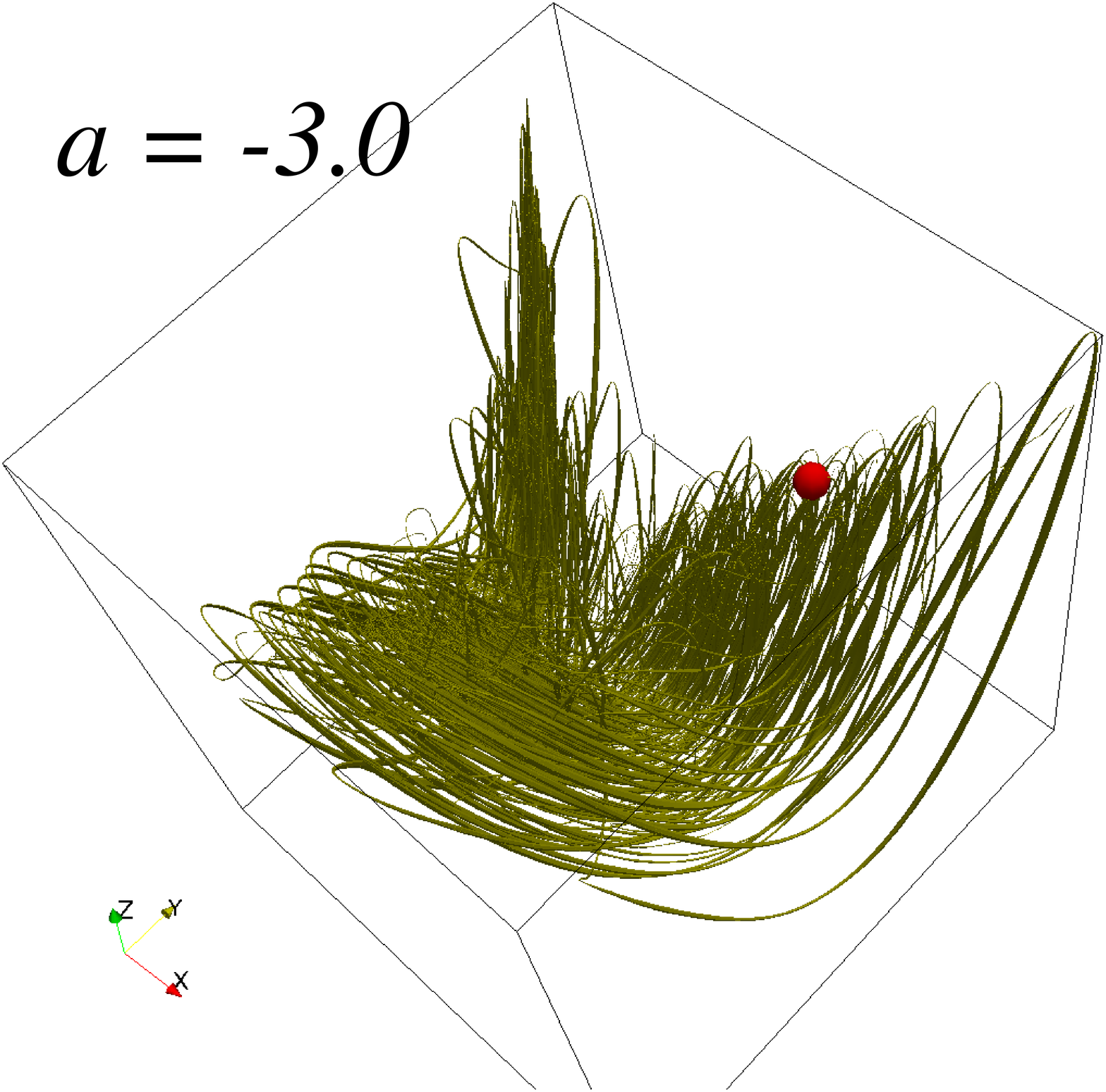}%
\includegraphics[scale=0.11]{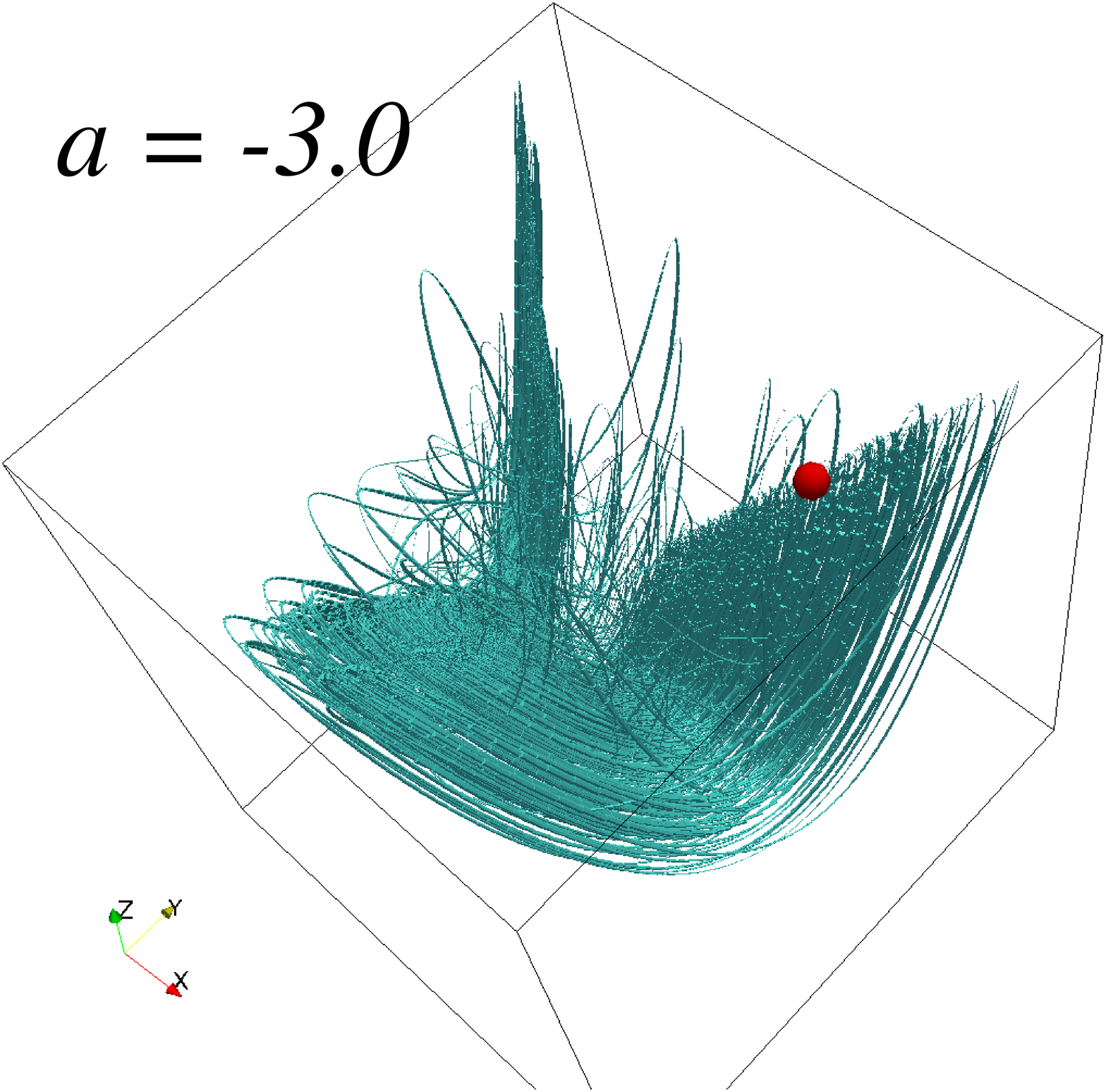}}
\centerline{%
\includegraphics[scale=0.11]{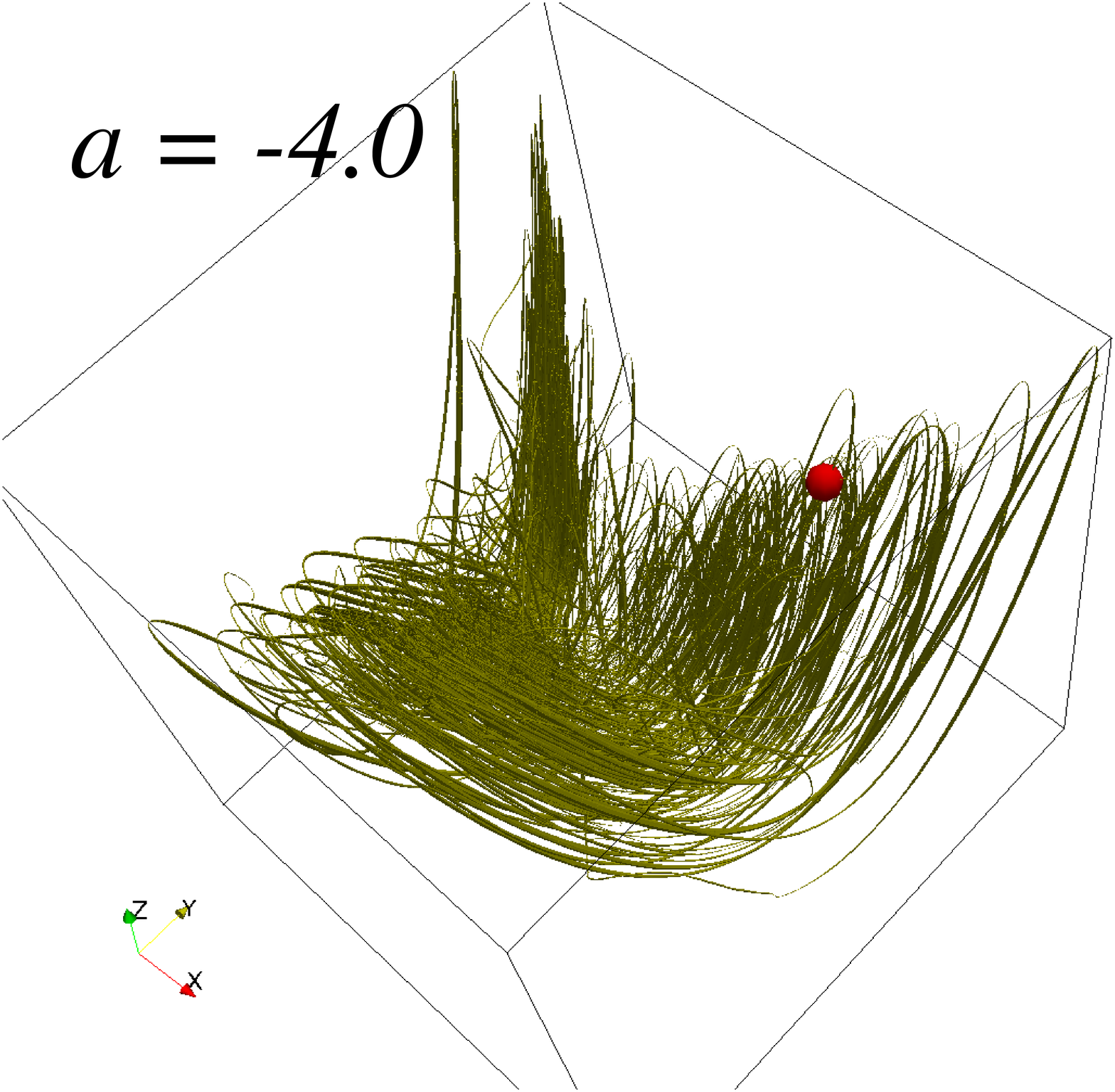}%
\includegraphics[scale=0.11]{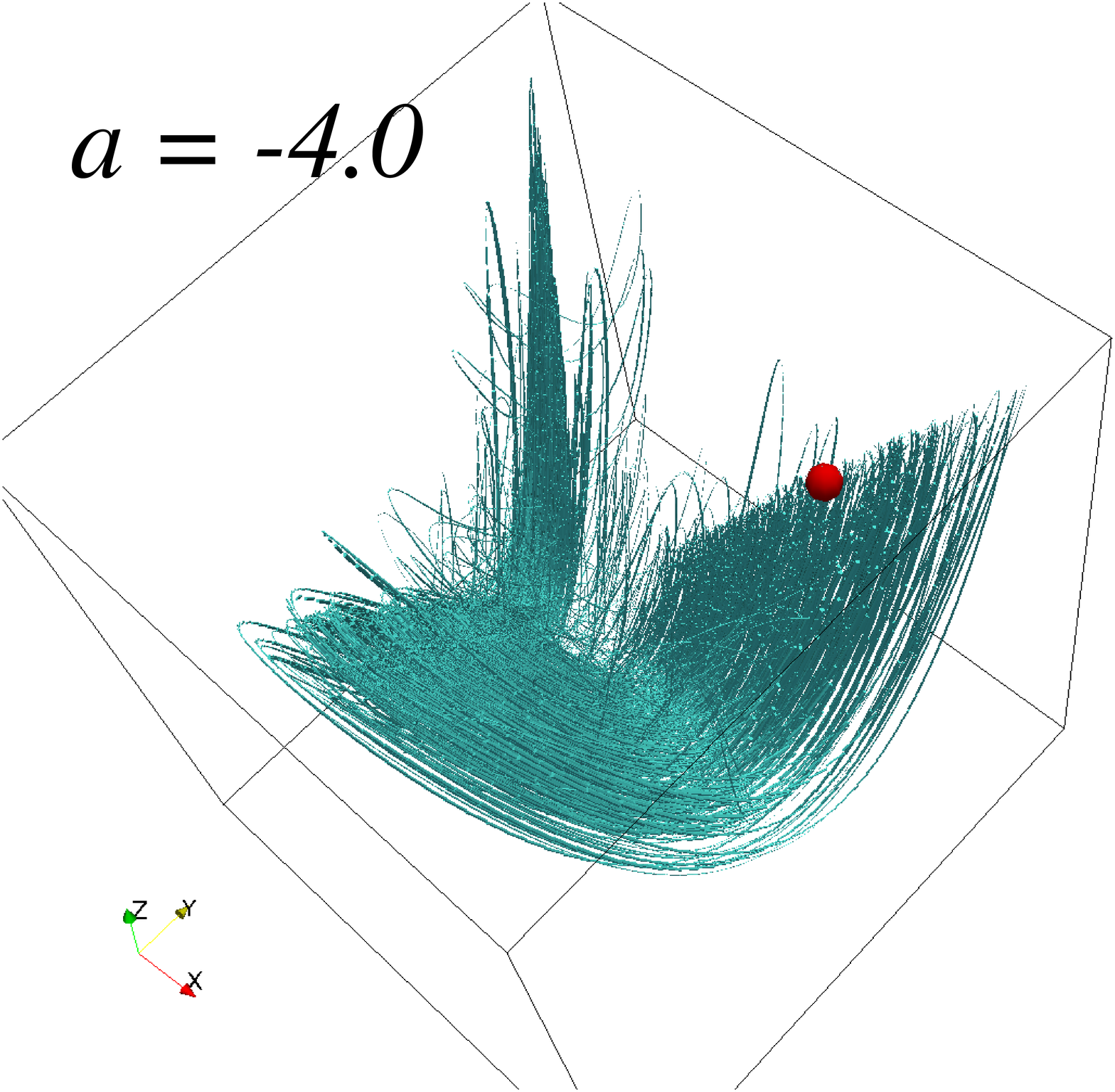}}
 \caption{\label{phase} Phase-space orbit, $\vec{X}(\vec{\kappa}, t)$, of
 the turbulent solution under the random forcing for various $a$'s 
with $\vec{\kappa} = (k_1, k_2, k_3) = (4, 8, 16)$ (left panels) 
and $(32, 64, 128)$ (right panels). The numerical solutions are the same
as those in Sec.\ref{s:random} with $2^{14}$ grid points.
The point (sphere) corresponds to the stationary solution under 
the deterministic forcing. The outline box in each panel 
is $-3 \le X_j \le 1 ~(j = 1, 2, 3)$.
The duration of the orbit shown here 
is $T = 1600$ for the former $\vec{\kappa}$ and $T = 400$ for the latter.
The figure of the latter looks denser since the typical time scale of the 
variation of the orbit is smaller.
}
\end{figure}

The orbit in the $\vec{X}$-space for each $a$ is shown in Fig.~\ref{phase}. 
Qualitative observations are now in order. 
The orbit of the turbulent solution meanders a certain surface with a thickness, which
is called here the \textit{attracting set}. Its overall shape is the same for different $a$'s.
An interesting question would be whether the thickness of the attracting set 
goes to zero as the amplitude of the random forcing tends to zero.
The stationary solution, which is visualized as a big sphere in Fig.~\ref{phase},
is located at one edge of the attracting set, not in the middle. 
This implies that the precise form of the time-averaged $E(k)$ of the turbulent
solution in the inertial range can be slightly different from the energy spectrum of the stationary 
solution.
Comparing the orbits between the two scale ranges, $\vec{\kappa}_1$ and $\vec{\kappa}_2$, we observe
that the orbits $\vec{X}(\vec{\kappa}_1)$ and $\vec{X}(\vec{\kappa}_2)$ appears almost the same,
which may be a manifestation of the near self-similarity of the energy spectrum within the inertial range.
From this, it is tempting to seek a three-variable modeling of the inertial-range dynamics
of the gCLMG turbulence.

\section{\label{s:sc}Summary and Concluding Discussion} 

\subsection{Summary}

We have numerically studied solutions of the viscous gCLMG eq.
under two kinds of large-scale monoscale forcing for certain range of 
negative $a$'s. 
Solutions strongly depend on the nature of the forcing. 
However the common characteristic structure independent on the forcing
is vorticity pulses developed
around stagnation points (velocity null point) with negative velocity
gradient.

When the forcing is random, the solution become turbulent, which
were analyzed with standard tools of studying NS turbulence.
We observed that the energy spectra of the gCLMG turbulence appear
to have power-law behaviors in the intermediate wavenumber range. 
However their scaling exponents are different from the dimensional 
prediction of the cascade of the inviscid invariant, Eq.(\ref{abmo}).

We then looked for direct evidence for or against the cascade with 
the filtering flux method 
for the five cases $a = -1.0, -1.5, -2.0, -3.0$ and $-4.0$.
We found that the invariant, Eq.(\ref{abmo}), cascades down to smaller scales 
except for the $a = -4.0$ case. 
It showed that turbulent cascade occurs for non-quadratic conservative quantities.
We then considered the K\'arm\'an-Howarth-Monin relation of the gCLMG turbulence
and discussed possible dissipative weak solution of the inviscid gCLMG eq.

Through the structure functions of the vorticity in the inertial range,
we observed that they were not simple power-law functions, supporting
non power-law type correction seen in the energy spectra. 
Although, if we assume that the leading behavior of the structure function was power-law,
the data indicated the negative H\"older exponent of the vorticity increment 
and hence blowup of the vorticity (of the vorticity pulses) in the inviscid 
limit. 

When the forcing is deterministic and stationary, the solution becomes
stationary. 
This nonlinear stationary solutions have almost identical
energy spectra with those of the turbulent solutions. By increasing
the resolution of the numerical simulation, we parametrized possible form
of the asymptotic (as $\nu \to 0$) energy spectra of the stationary solutions
as Eq.(\ref{e}) in the inertial range.
This parametrization supported the presence of the correction 
to the dimensionally predicted $k^{-3}$ due to the cascade 
of the inviscid invariant. However the functional form of the correction
is different from the Kraichnan's log-correction proposed for the 2D
enstrophy-cascade turbulence. 

The stationary solution has single vorticity pulse. We found that its
height and width scales with certain powers of the viscosity. 
We argued this viscous scaling with a boundary-layer analysis and 
obtained the scaling exponents as a function of $a$ in Eq.(\ref{ve}).
This viscous scaling also indicated blowup of the vorticity in the inviscid 
limit. Next we showed that the energy spectra of the stationary solution
in the intermediate wavenumbers (which corresponds to the inertial range 
in the turbulent solution) is also close to that of the inviscid and unforced 
solution of the gCLMG eq.

Finally, by normalizing the turbulent solution with the nonlinear stationary solution,
we found that the phase-space orbit of the turbulent solution is self-similar
in the inertial range.
This self-similarity is observed for all the five cases of $a$ studied here.
An important message here is that, not only this self-similarity, but all the 
other properties of the gCLMG solutions are qualitatively the same for all the cases
of $a$.

\subsection{Concluding discussion}
Our motivation of studying the gCLMG equation is to obtain insights
on the statistical laws of the NS turbulence and
a possible role of singular behavior of the solutions to the 
Navier-Stokes or Euler equations. 

With the suitable range of the parameter $a$, the randomly forced cases
exhibited certain similarities to the NS turbulence case, which
are the cascade of the inviscid conservative quantities 
and the broad energy spectra. 
Due to the dimension of the dissipation rate of the conservative quantity,
$(-a + 1)$-th power of time, the statistical laws were compared
to those of the 2D enstrophy-cascade NS turbulence, in particular,
the energy spectrum with the logarithmic correction and the vorticity
structure functions. 

One insight obtained here empirically, which may be useful 
to the 2D NS turbulence, is the expression with a high-order logarithmic correction of 
the energy spectrum, Eq.(\ref{e}). This spectrum around $k^{-3}$ implies
that the cascade of the gCLMG turbulence is not local as argued for 
the 2D enstrophy-cascade turbulence, see, e.g., \cite{e05, ccfs}.  
By non-locality it is meant that effect of the large-scale motions on the flux
is not diminished even if we have a sufficient scale separation.
It can be shown that, for the $a = -2$ case, a large-scale effect on 
the cascade of the gCLMG turbulence is not negligible (infrared non-local
in the language of \cite{e05}).
However for other $a$'s, there is a possibility
of local cascade. 
For example, for $a = -3$, let us now assume that 
the flux shown in Fig.\ref{fluxodd} becomes flat and that 
the second order structure function is a power law, $S_2(r) \propto r^{0.6}$ 
(which is contrary to our conclusion though).
Then it can be inferred that the cascade of $\tilde{C}_{-3}$ becomes infrared local from
although its energy spectrum is around $k^{-3}$. 

Concerning the relation between the statistical laws and the singularity,
the vorticity structure functions of the gCLMG turbulence
indicated blowup of the vorticity as $\nu \to 0$. This blowup of the vorticity
is also supported by the behavior of the nonlinear stationary solution.
This implies that the qualitative aspect of the even-order structure functions allows
us to detect signature of the inviscid-limit blowup as discussed already in, 
for example, \cite{f}. 
We also speculate that the negative exponent of the sixth order 
structure function, provided that the non-power-law correction is small, 
corresponds to the spatial decay of the vorticity in the neighborhood of the pulse.
However we are not able to identify quantitatively
relation between the statistics and the singularity, such as
the power-law exponent of the structure function and the scaling exponent of the blowup. 

We have shown the strong dependence of numerical solutions of the gCLMG eq.
on the forcing. It appears that this has nothing to do with the NS turbulence.
However recently it became known that a numerical solution to the 3D NS equations
with a large-scale deterministic forcing in the periodic cube reaches a near-stationary laminar state
after an extremely long duration of the turbulent state \cite{lm}. 
If the large-scale forcing is random, our simulation 
shows that a solution to the NS equations does not become laminar at least 
in the same duration of the simulation of the deterministic forcing case.
For the (2D) Kolmogorov flows, where the single Fourier-mode stationary forcing
is added, it is found that stable stationary solutions exist at large Reynolds numbers \cite{kimoka}
in certain cases.
Therefore this sort of the forcing dependence is not limited to a small class
of nonlinear PDE's with periodic boundary condition. 
Furthermore, it may imply that a random dynamical system approach is fruitful
when studying large-time asymptotic behavior of turbulent state.

We have seen that a number of the subtle properties of turbulent flows were
realized in the gCLMG solutions. This suggests its role as a unique testing 
ground worth further rigorous and theoretical studies. Specifically, turbulent 
statistical laws can be understood via singularities characterized with vorticity pulse,
which are present both in the nonlinear stationary solution and also the inviscid solution.

\section*{Acknowledgments}
We acknowledge stimulating discussions
with Koji Ohkitani, Hisashi Okamoto, Yukio Kaneda and Michio Yamada
and the support by Grants-in-Aid for Scientific Research 
KAKENHI (B) No.~26287023 from JSPS. 

\appendix
\section{Incomplete self-similarity analysis of the energy spectrum}

Here we argue that the functional form of the energy spectrum, Eq.(\ref{e}) with $c_0 = 3$ and $\theta = -1$,
can be obtained with the incomplete self-similarity (see, e.g., Sec.~8.3 of \cite{scaling}).

First we assume that the energy spectrum in the inertial range 
has the following form 
of correction to the dimensional result Eq.(\ref{spcform}):
\begin{eqnarray}
 E(k) 
&=& \beta_a^{3/(1 - a)} k^{-3} \Phi(k, \nu, k_f)  \nonumber \\
&=& \beta_a^{3/(1 - a)} k^{-3} \Phi\left(\frac{k}{k_f}, \frac{k_f}{k_d}\right).
\end{eqnarray}
Here $k_d$ is the dissipation wavenumber, 
which can be either the Kolmogorov-Kraichnan dissipation wavenumber 
$1/\eta_a = \beta_a^{-1/[2(a - 1)]} \nu^{-1/2}$
or the inverse of the boundary-layer width $1/\delta \propto \nu^{-\mu} = \nu^{-(1-a)/(1-2a)}$ that is 
the viscous length scale discussed in Sec.\ref{vp}.

Since the inertial-range property emerges in the intermediate asymptotics, namely
$k_f \to 0$ and $k_d \to \infty ~(\nu \to 0)$, we second assume that the correction
can be expanded with a small parameter $\varepsilon$.
Specifically, the leading behavior of the correction is assumed to be
\begin{eqnarray}
 \Phi\left(\frac{k}{k_f}, \frac{k_f}{k_d}\right)
\simeq
\left(\frac{k_f}{k_d}\right)^{\alpha_0 + \varepsilon \alpha_1}
\left[A_0  + \varepsilon A_1 \right],
\end{eqnarray}
where $\alpha_0, \alpha_1$ and $A_0$ are constants; 
$A_1$ can be a function of $k / k_f$.
In fact, the standard assumption on $\Phi$ would be 
$\Phi = (k / k_f)^{\alpha_0 + \varepsilon \alpha_1} [A_0 + \varepsilon A_1]$. 
We do not follow this since our intention is to obtain $\Phi$
as a non-power-law function of $k / k_f$.

Third, we assume $\alpha_0 = 0$ in order to have a nonvanishing limit of $\Phi$ as $k_d \to \infty$.
Then the $\alpha_1$ part can be written as
\begin{eqnarray}
 \left(\frac{k_f}{k_d}\right)^{\varepsilon \alpha_1}
 &=& \exp\left[\varepsilon \alpha_1 \left(\log \frac{k_f}{k} + \log \frac{k}{k_d} \right)\right].
\end{eqnarray}
Fourth, based on this form, 
we assume that the small parameter is $\varepsilon = \log^{-1} (k_f / k)$. 
This yields
\begin{eqnarray}
 E(k) \simeq
\beta_a^{3/(1 - a)} 
k^{-3} 
\left[A_0 + \frac{A_1}{\log \frac{k_f}{k}} \right]
\exp\left[\alpha_1 \frac{\log \frac{k_f}{k_d} }{-\log{\frac{k}{k_f}}} \right] \nonumber\\
\simeq
A_0 \beta_a^{3/(1 - a)} 
k^{-3} 
\exp\left[ - \alpha_1 \left(\log\frac{k_f}{k_d}\right)  \left(\log\frac{k}{k_f}\right)^{-1} \right],
\end{eqnarray}
which corresponds to Eq.(\ref{e}) with $\theta = -1$. 
Notice that the constant $c_2$ in Eq.(\ref{e}) is now proportional to $\log (k_f/ k_d)$.  

\section{Locally self-similar analysis of the inviscid blowup solution}

\begin{figure} 
\vspace*{0.5cm}
\includegraphics[scale = 0.6]{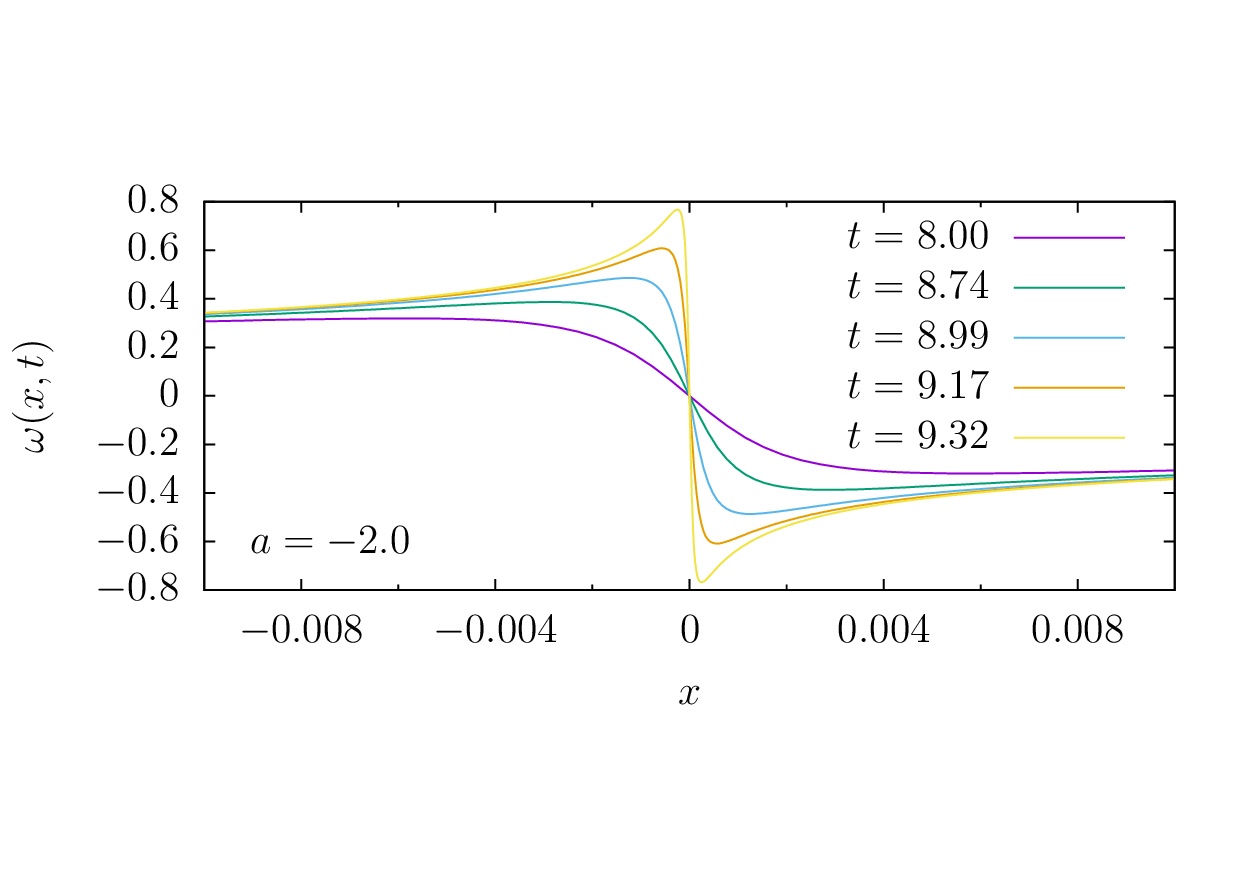}
\includegraphics[scale = 0.6]{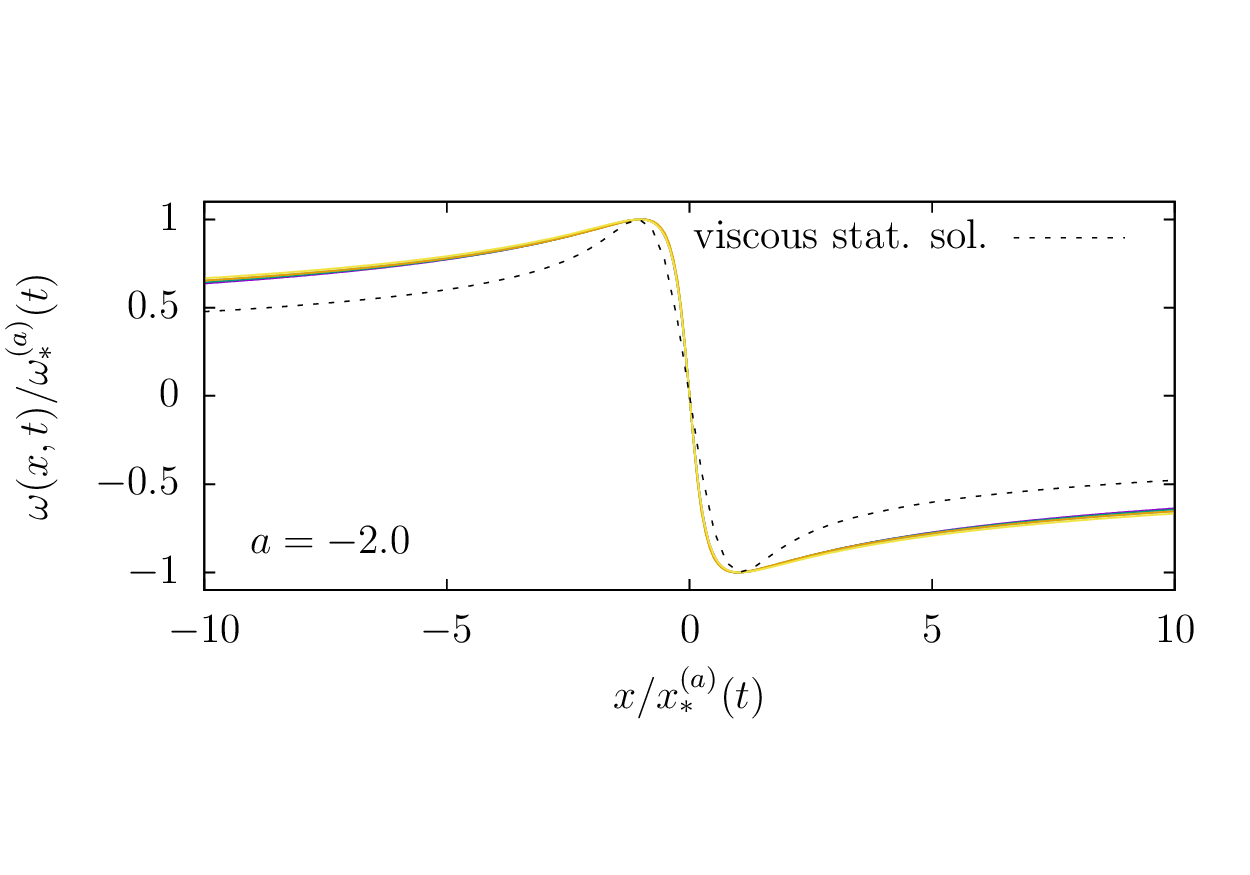}
\caption{\label{profinv} Left: Time evolution of the vorticity pulse of the inviscid unforced gCLMG eq. with $a = -2.0$  (the coordinate is shifted so that the center of the pulse is at the origin). Right: scaled pulses together with the scaled nonlinear stationary solution of the viscous and forced gCLMG eq., which were shown in Fig.\ref{profscale} for comparison. The functional form of the inviscid self-similar solution is different from that of 
the nonlinear stationary solution discussed in Sec.\ref{vp}}
\end{figure}

With the initial data $\omega(x, t = 0) = 0.1 \sin x$, the solution
of the inviscid unforced gCLMG eq. has a single vorticity pulse 
at $x = \pi$. Let us shift the coordinate so that the center of the 
pulse is at the origin. The time variation of the pulse profile is shown 
in Fig.\ref{profinv} for the case of $a = -2.0$ as an example 
(for other $a$'s the results are similar).
Now we write the maximum of the vorticity as $\omega_*^{(a)}(t)$
and the location of the maximum of the vorticity as $-x_*^{(a)}(t) ~(< 0)$.
As shown in Fig.\ref{profinv}, the pulse profiles at different times 
can be collapsed to a single curve
by scaling the solution with $x_*^{(a)}(t)$ and $\omega_*^{(a)}(t)$. 
This suggests that the inviscid pulse can be described with a locally 
self-similar solution.

\begin{figure} 
\includegraphics[scale = 0.6]{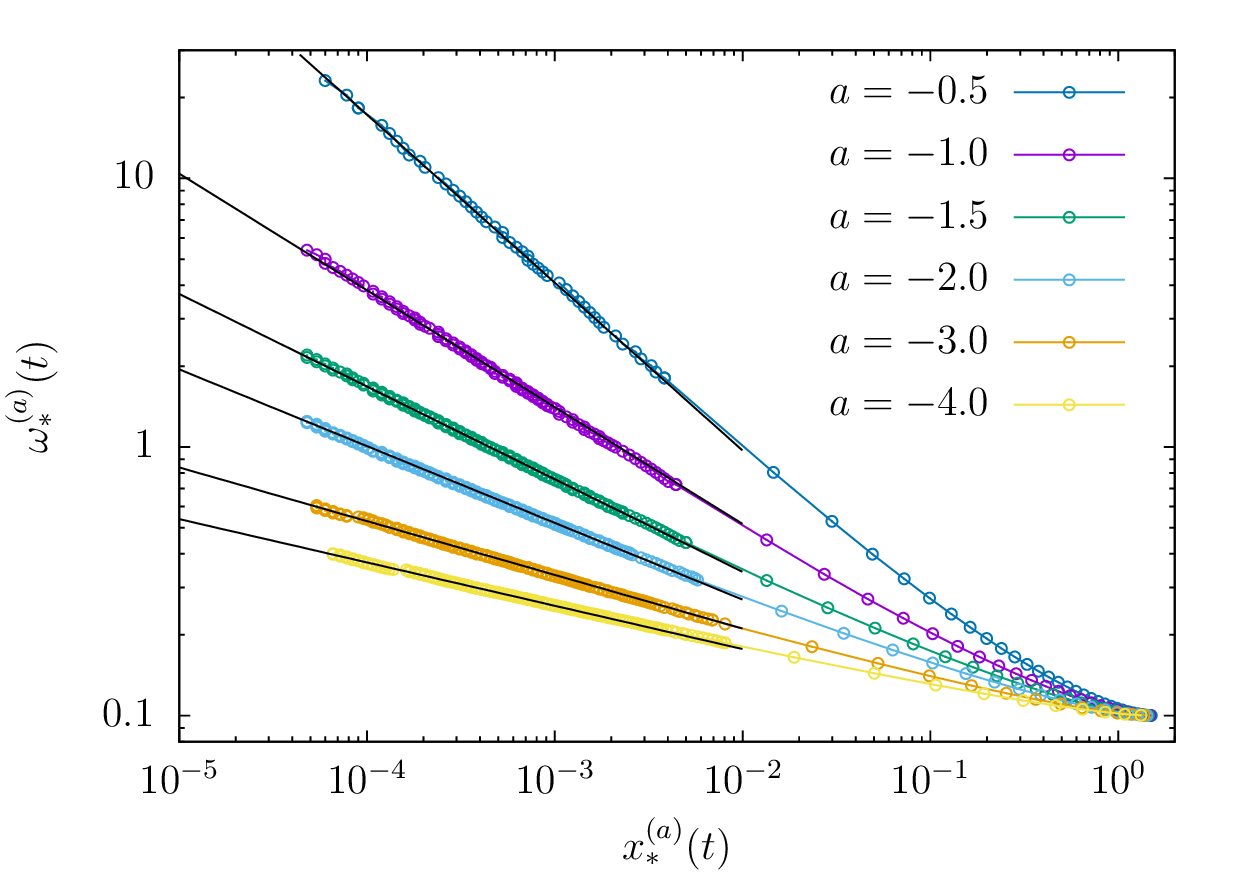}
 \caption{\label{invscaling} Power-law relation between the pulse height $\omega_*^{(a)}(t)$
 and the pulse width $x_*^{(a)}(t)$. The solid lines represent empirically fitted power-laws whose
 exponents are $\xi_a / \zeta_a = -1/1.6, -1/2.3, -1/2.9, -1/3.5, -1/5.0, -1/6.2$ from top to bottom.
 Different data points correspond to data at different times.
 The case $a = -0.5$ is included although the conservation  of $\tilde{C}_{-0.5}$ is not guaranteed.}
\end{figure}

To analyze the rate of the blowup, let us here assume that the local 
self-similar form is
\begin{equation}
\omega(x, t) = \omega_*^{(a)}(t) F^{(a)}(x / x_*^{(a)}(t)) 
\end{equation}
(the scaled profile depends on the parameter $a$) and that
the height and the width of the pulse have the following power-law dependence
\begin{equation}
  \omega_*^{(a)}(t) \simeq (t_*^{(a)} - t)^{\xi_a}, \quad 
   x_*^{(a)}(t) \simeq (t_*^{(a)} - t)^{\zeta_a}
\label{ipower}
\end{equation}
with the finite blowup time $t_*^{(a)}$.
The assumption (\ref{ipower}) is supported by the numerical data which
shows the power-law behavior $\omega_*^{(a)}(t) \propto  [x_*^{(a)}(t)]^{\xi_a / \zeta_a}$,
plotted in Fig.\ref{invscaling}. 

Now we argue that the exponent $\xi_a$ can be determined 
as $\xi_a = -1$ based on the conservation law.
The local conservation law $\partial_t [|\omega|^{-a}/(-a)] = \partial_x [ u |\omega|^{-a}]$ can be
integrated in $0 \le x \le x_*^{(a)}(t)$. The result is
\begin{equation}
 \partial_t \left[ \int_0^{x_*^{(a)}(t)} \frac{|\omega|^{-a}}{-a} dx \right] 
 = \left. (u |\omega|^{-a}) \right|_{x = x_*^{(a)}(t)},
\label{invcons}
\end{equation}
where we need some assumption on the behavior of the velocity at $x = x_*^{(a)}(t)$.
The numerical data indicates that the velocity around $x_*^{(a)}(t)$
is not locally self similar with $\omega_*^{(a)}(t)$ and $x_*^{(a)}(t)$.
This is as expected from the shape of the energy spectrum close to $k^{-3}$ 
since the velocity is dominated by the large-scale modes.
Nevertheless the data shows that the time variation of the velocity is very close
to the dimensional analysis result:
$u(x_*^{(a)}(t), t) \simeq \omega_*^{(a)}(t) x_*^{(a)}(t) \simeq (t_*^{(a)} - t)^{\xi_a + \zeta_a}$
(a small discrepancy in the exponent is seen for $a = -3.0$ and $-4.0$, though).
With this scaling of the velocity, Eq.(\ref{invcons}) yields the relation 
among the exponents as 
$-a \xi_a + \zeta_a - 1 = \xi_a + \zeta_a - a \xi_a$. Hence $\xi_a = -1$.

\begin{figure} 
\includegraphics[scale = 0.6]{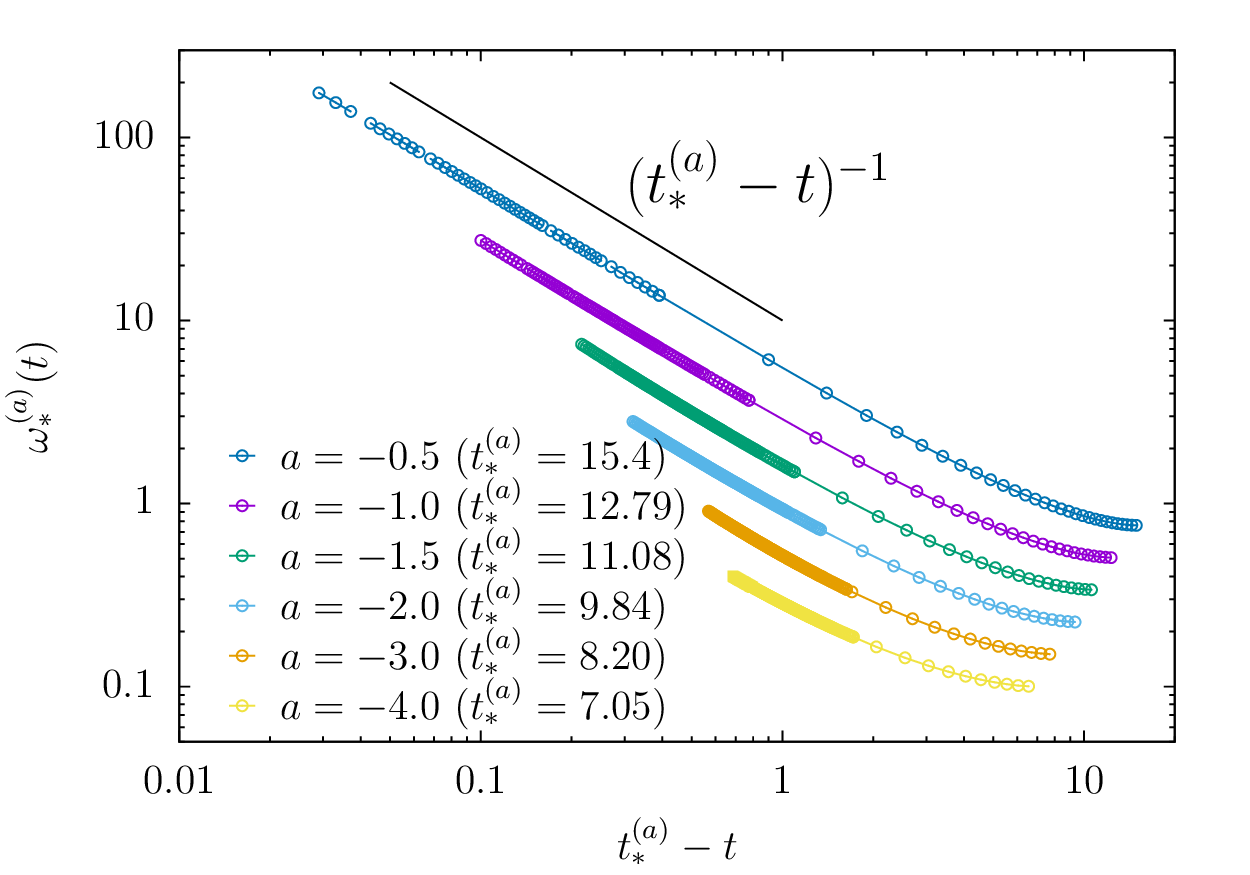}
\includegraphics[scale = 0.6]{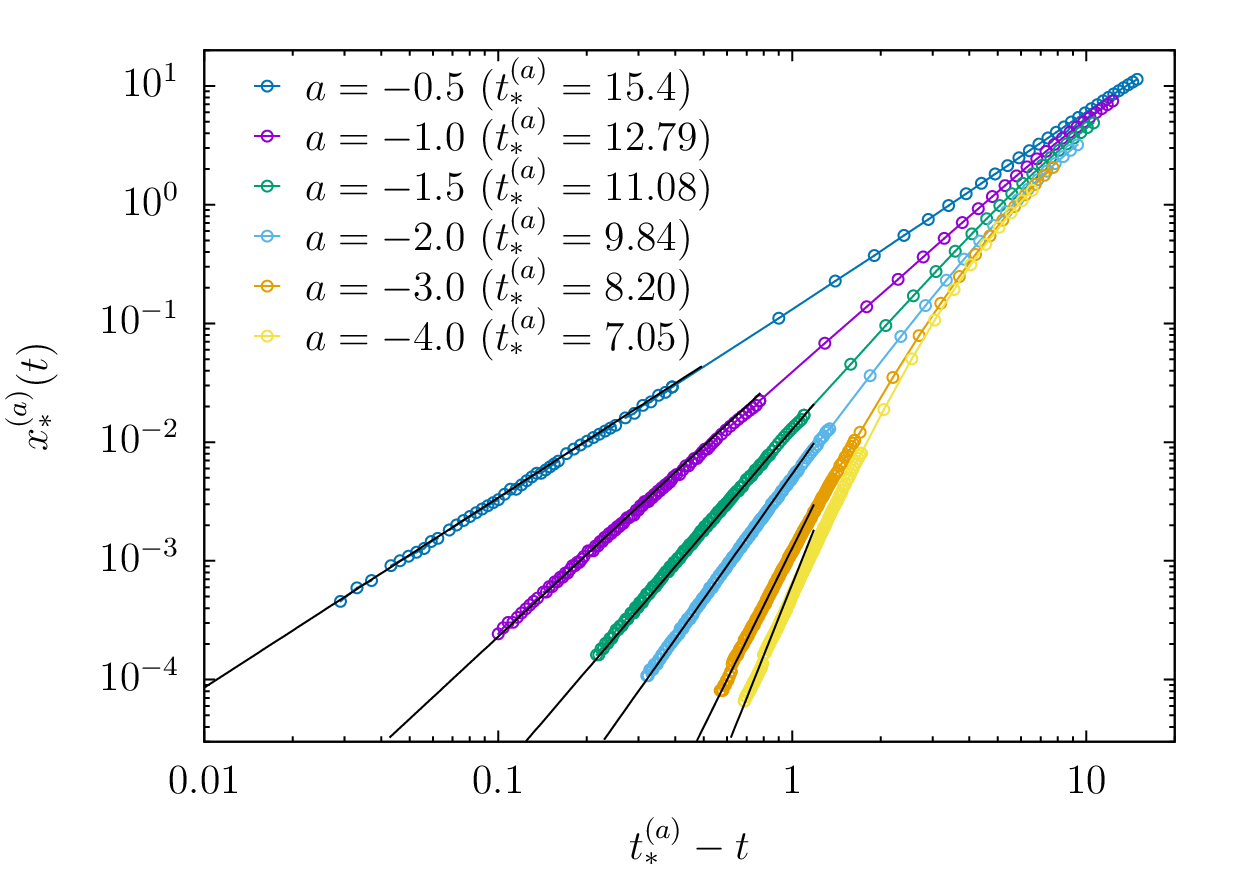}
\caption{\label{invscalingos} Power-law behavior of $\omega^{(a)}_*(t) \simeq (t_*^{(a)} - t)^{-1}$ (Top) and of $x_*^{(a)} \simeq (t_*^{(a)} - t)^{\zeta_a}$ (Bottom) with an empirical choice of $t_*^{(a)}$. The curves are shifted vertically for clarity. The solid lines in the bottom panel represent power laws with the exponents $\zeta_a= 1.6, 2.3, 2.9, 3.5, 5.0, 6.2$, which are determined in Fig.\ref{invscaling} independently on the choice of the blowup time $t_*^{(a)}$.}
\end{figure}

This scaling $\omega_*^{(a)}(t) \propto (t_*^{(a)} - t)^{-1}$ can be observed 
as shown in Fig.\ref{invscalingos}, provided that somewhat subjective choice
of the unknown blowup time, $t_*^{(a)}$, is made 
(recall that the power-law exponent emerged in such a figure is quite sensitive
to choice of the origin $t_*^{(a)}$).
About the maximum of $H(\omega)$, the same temporal scaling $(t_*^{(a)} - t)^{-1}$ 
is observed with the same choice of $t_*^{(a)}$'s (figure not shown), 
indicating that the blowup criterion obtained in \cite{oswn}, 
$\int_0^{t_*^{(a)}}\max_{x} H(\omega)(t) dt = \infty$, is satisfied. 

The numerical data about the scaling of the width of the pulse,
$x_*^{(a)}(t) \simeq (t_*^{(a)} - t)^{\zeta_a}$, 
is shown in Fig.\ref{invscalingos}.
One heuristic assumption leading to determination of
the exponent $\zeta_a$ is that the left hand side of Eq.(\ref{invcons}),
which is the rate of change of the part of the conservative quantity
contained in the pulse, is time-independent.
This gives us $-a \xi_a + \zeta_a - 1 = 0$. Together with $\xi_a = -1$,
the assumptions yields $\zeta_a = 1 - a$. 
However this does not agree well with the numerical results.

In \cite{cc10}, a particular self-similar solution of the inviscid, unforced
gCLMG eq. for any $a$ was found, which corresponds formally to the exponents 
$\xi_a = -1$ and $\zeta_a = -a$. 
The particular solution is different from the locally self-similar form 
analyzed here.

\end{document}